\documentclass[aps,prb, superscriptaddress,twocolumn,longbibliography]{revtex4-1}
\usepackage{amsmath}
\usepackage{amssymb}
\usepackage{graphicx}
\usepackage[usenames,dvipsnames]{xcolor}
\usepackage{tikz}
\usepackage{pgffor}
\usepackage{verbatim}
\usepackage{bbm}
\usepackage{float}
\usepackage{mathrsfs}
\usepackage[colorlinks, breaklinks=true,linkcolor=red, citecolor=blue, linktocpage=true]{hyperref}

\newcommand{\<}{\langle}
\renewcommand{\>}{\rangle}
\newcommand{\be}{\begin{equation} }
\newcommand{\ee}{\end{equation} }
\newcommand{\ba}{\begin{eqnarray} }
\newcommand{\ea}{\end{eqnarray} }

\newcommand{\down}{\downarrow}

\newcommand{\bpm}{\begin{pmatrix}}
\newcommand{\epm}{\end{pmatrix}}
\newcommand{\bmm}{\begin{matrix}}
\newcommand{\emm}{\end{matrix}}
\newcommand{\up}{\uparrow}

\newcommand{\la}{\label}
\newcommand{\bea}{\begin{eqnarray}}
\newcommand{\eea}{\end{eqnarray}}
\renewcommand{\v}[1]{\boldsymbol{#1}}

\begin{document}

\title{Formalism for the solution of quadratic Hamiltonians with large cosine terms}

 \author{Sriram Ganeshan}
 \affiliation{Condensed Matter Theory Center and Joint Quantum Institute, Department of Physics, University of Maryland, College Park, MD 20742, USA}
\author{Michael Levin}
\affiliation{Department of Physics, James Frank Institute, University of Chicago, Chicago, IL  USA}


\begin{abstract}
We consider quantum Hamiltonians of the form $H = H_0 - U \sum_j \cos(C_j)$ where $H_0$ is a quadratic function 
of position and momentum variables $\{x_1, p_1, x_2, p_2,...\}$ and the $C_j$'s are linear in these variables. 
We allow $H_0$ and $C_j$ to be completely general with only two restrictions: we require that (1) the $C_j$'s
are linearly independent and (2) $[C_j, C_k]$ is an integer multiple of $2\pi i$ for all $j,k$ 
so that the different cosine terms commute with one another. Our main result is a recipe for solving these Hamiltonians and 
obtaining their exact low energy spectrum in
the limit $U \rightarrow \infty$. This recipe involves constructing creation and annihilation operators and is 
similar in spirit to the procedure for diagonalizing quadratic Hamiltonians. 
In addition to our exact solution in the infinite $U$ limit, we also discuss how to analyze these systems
when $U$ is large but finite. Our results are relevant to a number of different physical systems, but one of the most natural 
applications is to understanding the effects of electron scattering on quantum Hall edge modes. 
To demonstrate this application, we use our formalism to solve a toy model for a fractional quantum spin Hall edge 
with different types of impurities. 
\end{abstract}


\maketitle


\section{Introduction}
In this paper we study a general class of quantum Hamiltonians which are relevant to a number of different
physical systems. The Hamiltonians we consider are defined on a 
$2N$ dimensional phase space $\{x_1,...,x_N,p_1,...,p_N\}$ with $[x_i, p_j] = i \delta_{ij}$. They take the form
\begin{align}
H = H_0 - U \sum_{i=1}^M \cos(C_i)
\label{genHamc}
\end{align}
where $H_0$ is a \emph{quadratic} function of the phase space variables $\{x_1,...,x_N,p_1,...,p_N\}$, and $C_i$ is \emph{linear} 
in these variables. The $C_i$'s can be chosen arbitrarily except for two restrictions: 
\begin{enumerate}
\item{$\{C_1,...,C_M\}$ are linearly independent.}
\item{$[C_i, C_j]$ is an integer multiple of $2\pi i$ for all $i,j$.}
\end{enumerate}
Here, the significance of the second condition is that it guarantees that the cosine terms commute with one another: $[\cos(C_i),\cos(C_j)] = 0$ for all $i,j$.

For small $U$, we can straightforwardly analyze these Hamiltonians by treating the cosine terms as perturbations to $H_0$. But how can
we study these systems when $U$ is large? The most obvious approach is to expand around $U = \infty$, just as in the
small $U$ case we expand around $U=0$. But in order to make such an expansion, we first need to be able to solve these Hamiltonians exactly in the
infinite $U$ limit. The purpose of this paper is to describe a systematic procedure for obtaining such a solution, at least at low energies.

The basic idea underlying our solution is that when $U$ is very large, the cosine terms act as
\emph{constraints} at low energies. Thus, the low energy spectrum of $H$ can be described by an effective Hamiltonian $H_{\text{eff}}$
defined within a constrained Hilbert space $\mathcal{H}_{\text{eff}}$. This effective Hamiltonian $H_{\text{eff}}$ is quadratic in 
$\{x_1,...,x_N,p_1,...,p_N\}$ since $H_0$ is quadratic and the constraints are linear in these variables. We can therefore diagonalize
$H_{\text{eff}}$ and in this way determine the low energy properties of $H$.

Our main result is a general recipe for finding the exact low energy spectrum of $H$ in the limit
$U \rightarrow \infty$. This recipe consists of two steps and is only slightly more complicated than what is required 
to solve a conventional quadratic Hamiltonian. The first step involves finding creation and annihilation
operators for the low energy effective Hamiltonian $H_{\text{eff}}$ (Eq. \ref{auxeqsumm}). The second step of the recipe involves finding integer linear 
combinations of the $C_i$'s that have simple commutation relations with one another. In practice, this step amounts to finding 
a change of basis that puts a particular integer skew-symmetric matrix into
canonical form (Eq. \ref{zprime}). Once these two steps are completed, the low energy spectrum can be written 
down immediately (Eq. \ref{energyspectsumm}).

In addition to our exact solution in the infinite $U$ limit, we also discuss how to analyze these systems
when $U$ is large but finite. In particular, we show that in the finite $U$ case, we need to add small 
(non-quadratic) corrections to the effective Hamiltonian $H_{\text{eff}}$ in order to reproduce the low energy physics of $H$. 
One of our key results is a discussion of the general form of 
these finite $U$ corrections, and how they scale with $U$.

Our results are useful because there are a number of physical systems where 
one needs to understand the effect of cosine-like interactions on a quadratic Hamiltonian. 
An important class of examples are the edges of Abelian fractional quantum Hall (FQH) states. Previously it has been argued 
that a general Abelian FQH edge can be modeled as collection of $p$ chiral Luttinger 
liquids with Hamiltonian~\cite{wenedge, 1991-FrohlichKerler, WenReview, WenBook}
\begin{equation*}
H_0 = \int dx \frac{1}{4\pi} (\partial_x \Phi)^T V (\partial_x \Phi)
\end{equation*}
Here $\Phi(x) = (\phi_1(x),...,\phi_p(x))$, with each component $\phi_I$ 
describing a different (bosonized) edge mode while $V$ is a $p \times p$ matrix that describes the velocities 
and density-density interactions between the different edge modes. The commutation 
relations for the $\phi_I$ operators are $[\phi_I(x), \partial_y \phi_J(y)] = 2\pi i (K^{-1})_{IJ} \delta(x-y)$ where 
$K$ is a symmetric, integer $p \times p$ matrix which is determined by the bulk FQH state. 

The above Hamiltonian $H_0$ is quadratic and hence exactly soluble, but in many cases it is unrealistic because it 
describes an edge in which electrons do not scatter between the different edge modes. In order to incorporate scattering
into the model, we need to add terms to the Hamiltonian of the form 
\begin{equation*}
H_{\text{scat}} = \int U(x) \cos(\Lambda^T K \Phi - \alpha(x)) dx
\end{equation*}
where $\Lambda$ is a $p$-component integer vector that describes the number of electrons scattered from each edge mode~\cite{kanefisher}. However,
it is usually difficult to analyze the effect of these cosine terms beyond the small $U$ limit 
where perturbative techniques are applicable. (An important exception is when $\Lambda$ is a 
null vector~\cite{Haldane}, i.e. $\Lambda^T K \Lambda = 0$: in this case, the fate of the edge modes can be determined 
by mapping the system onto a Sine-Gordon model~\cite{wang2013weak}).

Our results can provide insight into this class of systems because they allow us to construct exactly soluble toy models 
that capture the effect of electron scattering at the edge. Such toy models can be obtained by replacing 
the above continuum scattering term $H_{\text{scat}}$ by a collection of
discrete impurity scatterers, $U \sum_i \cos(\Lambda^T K \Phi(x_i) - \alpha_i)$, and then taking the limit $U \rightarrow \infty$.
It is not hard to see that the latter cosine terms obey conditions (1) and (2) from above, so we can solve the resulting
models exactly using our general recipe. Importantly, this construction is valid for \emph{any} choice of $\Lambda$, whether or not $\Lambda$
is a null vector.

Although the application to FQH edge states is one of the most interesting aspects of our results, 
our focus in this paper is on the general formalism rather than the implications for specific
physical systems. Therefore, we will 
only present a few simple examples involving a fractional quantum spin Hall edge with 
different types of impurities. The primary purpose of these examples is to 
demonstrate how our formalism works rather than to obtain novel results.

We now discuss the relationship with previous work. One paper that explores some related ideas is Ref.~\onlinecite{gottesman2001encoding}. In that paper, 
Gottesman, Kitaev, and Preskill discussed Hamiltonians similar to (\ref{genHamc}) for the case 
where the $C_i$ operators do not commute, i.e. $[C_i, C_j] \neq 0$. They showed that these Hamiltonians can have degenerate ground states and proposed using these degenerate states 
to realize qubits in continuous variable quantum systems. 

Another line of research that has connections to the present work involves the problem
of understanding constraints in quantum mechanics. In particular, a number of previous works have studied the problem of a 
quantum particle that is constrained to move on a surface by a strong confining potential~\cite{JensenKoppe,da1981quantum}. 
This problem is similar in spirit to one we 
study here, particularly for the special case where $[C_i, C_j] = 0$: in that case, if we identify $C_i$ as position coordinates $x_i$, then the Hamiltonian 
(\ref{genHamc}) can be thought of as describing a particle that is constrained to move on a periodic array of hyperplanes. 

Our proposal to apply our formalism to FQH edge states also has connections to the previous literature. In particular, 
it has long been known that the problem of an impurity in a non-chiral
Luttinger liquid has a simple exact solution in the limit of infinitely strong backscattering~\cite{kane1992transmission, kane1992transport,chamon, chamon0}. The infinite
backscattering limit for a single impurity has also been studied for more complicated Luttinger liquid systems~\cite{ChklovskiiHalperin,chamon1997distinct, PhysRevLett.99.066803, ganeshan2012fractional}. The advantage of our approach to these systems is that our methods allow us to study not just single impurities but also multiple coherently coupled impurities, and to obtain 
the full quantum dynamics not just transport properties.

The paper is organized as follows. In section \ref{summsect} we summarize our formalism and main results. In section \ref{examsect} 
we illustrate our formalism with some examples involving fractional quantum spin Hall edges with impurities. We discuss directions for future work in the conclusion.
The appendices contain the general derivation of our formalism as well as other technical results.


\section{Summary of results}\la{summsect}

\subsection{Low energy effective theory}
\label{effsummsect}
Our first result is that we derive an \emph{effective theory} that describes the low energy spectrum of 
\begin{align*}
H = H_0 - U \sum_{i=1}^M \cos(C_i)
\end{align*}
in the limit $U \rightarrow \infty$. This effective theory consists of an effective Hamiltonian $H_{\text{eff}}$ and an effective 
Hilbert space $\mathcal{H}_{\text{eff}}$. Conveniently, we find a simple algebraic expression for $H_{\text{eff}}$ and $\mathcal{H}_{\text{eff}}$ that
holds in the most general case. Specifically, the effective Hamiltonian is given by
\begin{equation}
H_{\text{eff}} =  H_0 - \sum_{i,j=1}^{M} \frac{(\mathcal{M}^{-1})_{ij}}{2} \cdot \Pi_i \Pi_j
\label{Heffgenc}
\end{equation}
where the operators $\Pi_1,..., \Pi_{M}$ are defined by
\begin{equation}
\Pi_{i} = \frac{1}{2\pi i} \sum_{j=1}^{M} \mathcal{M}_{ij} [C_j, H_0]
\end{equation}
and where $\mathcal{M}_{ij}$ is an $M \times M$ matrix defined by
\begin{equation}
\mathcal{M} = \mathcal{N}^{-1}, \quad \mathcal{N}_{ij} = -\frac{1}{4\pi^2}[C_i,[C_j,H_0]]
\end{equation}
This effective Hamiltonian is defined on an effective Hilbert space $\mathcal{H}_{\text{eff}}$, which is a \emph{subspace} of the original Hilbert space $\mathcal{H}$,
and consists of all states $|\psi\>$ that satisfy
\begin{equation}
\cos(C_i)|\psi\> = |\psi\>, \quad i=1,...,M
\label{Hilbeff}
\end{equation}

A few remarks about these formulas: first, notice that $\mathcal{M}$ and $\mathcal{N}$ are matrices of \emph{$c$-numbers} 
since $H_0$ is quadratic and the $C_i$'s are linear combinations of $x_j$'s and $p_j$'s. Also notice that 
the $\Pi_i$ operators are linear functions of $\{x_1,...,x_N,p_1,...,p_N\}$. These observations imply 
that the effective Hamiltonian $H_{\text{eff}}$ is always \emph{quadratic}. 
Another important point is that the $\Pi_i$ operators are conjugate to the $C_i$'s:
\begin{equation}
[C_i, \Pi_j] = 2\pi i \delta_{ij}
\end{equation}
This means that we can think of the $\Pi_i$'s as generalized momentum operators. Finally, notice that
\begin{equation}
[C_i, H_{\text{eff}}] = 0
\label{CiH0}
\end{equation}
The significance of this last equation is that it shows that the Hamiltonian $H_{\text{eff}}$ can be naturally defined within the above 
Hilbert space (\ref{Hilbeff}).

We can motivate this effective theory as follows. First, it is natural to expect that the lowest energy states in the 
limit $U \rightarrow \infty$ are those that minimize the cosine terms. This leads to the effective Hilbert space given in Eq. \ref{Hilbeff}.
Second, it is natural to expect that the dynamics in the $C_1,...,C_M$ directions freezes out at low energies. Hence, the terms that 
generate this dynamics, namely $\sum_{ij} \frac{\mathcal{M}^{-1}_{ij} }{2}\Pi_i \Pi_j$, should be removed from the effective Hamiltonian. This
leads to Eq. \ref{Heffgenc}. Of course this line of reasoning is just an intuitive picture; for a formal derivation of 
the effective theory, we refer the reader to appendix \ref{derivsect}.

At what energy scale is the above effective theory valid? We show that $H_{\text{eff}}$ correctly reproduces the energy spectrum of $H$ for
energies less than $\sqrt{U/m}$ where $m$ is the maximum eigenvalue of $\mathcal{M}_{ij}$. One implication of this result is that our
effective theory is only valid if $\mathcal{N}$ is \emph{non-degenerate}: if $\mathcal{N}$ were degenerate than $\mathcal{M}$ would
have an infinitely large eigenvalue, which would mean that there would be no energy scale below which our theory is valid. Physically,
the reason that our effective theory breaks down when $\mathcal{N}$ is degenerate is that in this case, the dynamics in the $C_1,...,C_M$ 
directions does not completely freeze out at low energies.

To see an example of these results, consider a one dimensional harmonic oscillator with a cosine term:
\begin{equation}
H = \frac{p^2}{2m} + \frac{K x^2}{2} - U \cos(2\pi x)
\end{equation}
In this case, we have $H_0 =  \frac{p^2}{2m} + \frac{K x^2}{2}$ and $C = 2\pi x$. If we substitute
these expressions into Eq. \ref{Heffgenc}, a little algebra gives
\begin{equation*}
H_{\text{eff}} = \frac{K x^2}{2}
\end{equation*}
As for the effective Hilbert space, Eq. \ref{Hilbeff} tells us that $\mathcal{H}_{\text{eff}}$
consists of position eigenstates 
\begin{equation*}
\mathcal{H}_{\text{eff}} = \{|x=q\>, \quad q = \text{(integer)}\}
\end{equation*}
If we now diagonalize the effective Hamiltonian within the effective Hilbert space, we obtain eigenstates 
$|x=q\>$ with energies $E = \frac{K q^2}{2}$. Our basic claim is that these eigenstates and energies should match the low energy
spectrum of $H$ in the $U \rightarrow \infty$ limit. In appendix \ref{Heffex1sect}, we analyze this example in detail and we confirm that 
this claim is correct (up to a constant shift in the energy spectrum). 
 
To see another illustrative example, consider a one dimensional harmonic oscillator with \emph{two} cosine terms,
\begin{equation}
H = \frac{p^2}{2m} + \frac{K x^2}{2} - U \cos(d p) - U \cos(2\pi x)
\la{example2summ}
\end{equation}
where $d$ is a positive integer. This example is fundamentally different from the previous one because the arguments
of the cosine do not commute: $[x,p] \neq 0$. This property leads to some new features, such as degeneracies
in the low energy spectrum. To find the effective theory in this case, we note that
$H_0 =  \frac{p^2}{2m} + \frac{K x^2}{2}$ and $C_1 = dp$, $C_2 = 2\pi x$. With a little algebra,
Eq. \ref{Heffgenc} gives
\begin{equation*}
H_{\text{eff}} = 0
\end{equation*}
As for the effective Hilbert space, Eq. \ref{Hilbeff} tells us that $\mathcal{H}_{\text{eff}}$
consists of all states $|\psi\>$ satisfying
\begin{equation*}
\cos(2\pi x) |\psi\> = \cos(dp) |\psi\> = |\psi\>.
\end{equation*}
One can check that there are $d$ linearly independent states obeying the above conditions; hence if we diagonalize the effective 
Hamiltonian within the effective Hilbert space, we obtain $d$ exactly
degenerate eigenstates with energy $E = 0$. The prediction of our formalism is therefore that $H$ has a $d$-fold ground state degeneracy in the 
$U \rightarrow \infty$ limit. In appendix \ref{Heffex2sect}, we analyze this example and confirm this prediction.

\subsection{Diagonalizing the effective theory}
\label{diagsummsect}
We now move on to discuss our second result, which is a recipe for diagonalizing the effective Hamiltonian $H_{\text{eff}}$. Note that 
this diagonalization procedure is unnecessary for the two examples discussed above, since $H_{\text{eff}}$ is very simple in these cases. However, in
general, $H_{\text{eff}}$ is a complicated quadratic Hamiltonian which is defined within a complicated Hilbert space $\mathcal{H}_{\text{eff}}$,
so diagonalization is an important issue. In fact, in practice, the results in this section are more useful than those in the previous section
because we will see that we can diagonalize $H_{\text{eff}}$ without explicitly evaluating the expression in Eq. \ref{Heffgenc}. 

Our recipe for diagonalizing $H_{\text{eff}}$ has three steps. The first step is to find creation and annihilation operators for $H_{\text{eff}}$. 
Formally, this amounts to finding all operators $a$ that are linear combinations of $\{x_1,...,x_N, p_1,...,p_N\}$, 
and satisfy
\begin{align}
[a, H_{\text{eff}}] &= E a, \nonumber \\
[a, C_i] &= 0, \quad i=1,...,M
\label{commeq}
\end{align}
for some scalar $E \neq 0$. While the first condition is the usual definition of creation and annihilation
operators, the second condition is less standard; the motivation for this condition is that 
$H_{\text{eff}}$ commutes with $C_i$ (see Eq. \ref{CiH0}). As a result, we can impose the requirement $[a,C_i] = 0$ and we will still have
enough quantum numbers to diagonalize $H_{\text{eff}}$ since we can use the $C_i$'s in addition to the $a$'s.

Alternatively, there is another way to find creation and annihilation operators which is often more convenient: instead of 
looking for solutions to (\ref{commeq}), one can look for solutions to
\begin{align}
[a,H_0] &= E a + \sum_{j=1}^M \lambda_{j} [C_j,H_0], \nonumber \\
[a, C_i] &= 0, \quad i=1,...,M
\label{auxeqsumm}
\end{align}
for some scalars $E,\lambda_j$ with $E \neq 0$. Indeed, we show in appendix \ref{altmethsect} that every solution to (\ref{commeq}) is also a solution to 
(\ref{auxeqsumm}) and vice versa, so these two sets of equations are equivalent. In practice, it is easier to work with 
Eq. \ref{auxeqsumm} \textcolor{black}{than} Eq. \ref{commeq} because Eq. \ref{auxeqsumm} is written in terms of $H_0$, 
and thus it does not require us to work out the expression for $H_{\text{eff}}$.

The solutions to (\ref{commeq}), or equivalently (\ref{auxeqsumm}), can
be divided into two classes: ``annihilation operators'' with $E > 0$, and ``creation operators'' with $E < 0$. 
Let $a_1,...,a_K$ denote a complete set of linearly independent annihilation operators. We will denote the corresponding 
$E$'s by $E_1,...,E_K$ and the creation operators by $a_1^\dagger,...,a_K^\dagger$. The creation/annihilation operators 
should be normalized so that
\begin{align*}
[a_k, a_{k'}^\dagger] =\delta_{kk'}, \quad [a_k, a_{k'}] = [a_k^\dagger, a_{k'}^\dagger] = 0
\end{align*}

We are now ready to discuss the second step of the recipe. This step involves searching for linear combinations
of $\{C_1,...,C_M\}$ that have simple commutation relations with one another. The idea behind this step is that we
ultimately need to construct a complete set of quantum numbers for labeling the eigenstates of $H_{\text{eff}}$.
Some of these quantum numbers will necessarily involve the $C_i$ operators since these operators play 
a prominent role in the definition of the effective Hilbert space, $\mathcal{H}_{\text{eff}}$. However, the $C_i$'s are
unwieldy because they have complicated commutation relations with one another. 
Thus, it is natural to look for linear combinations of $C_i$'s that have simpler commutation
relations. 

With this motivation in mind, let $\mathcal{Z}_{ij}$ be the $M \times M$ matrix defined by
\begin{equation}
\mathcal{Z}_{ij} = \frac{1}{2\pi i} [C_i, C_j]
\end{equation}
The matrix $\mathcal{Z}_{ij}$ is integer and skew-symmetric, but otherwise arbitrary. Next, let
\begin{align}
C_i' = \sum_{j=1}^M \mathcal{V}_{ij} C_j \textcolor{black}{+ \chi_i}
\end{align}
for some matrix $\mathcal{V}$ \textcolor{black}{and some vector $\chi$.} Then, $[C_i', C_j'] = 2\pi i \mathcal{Z}_{ij}'$ where
$\mathcal{Z}' = \mathcal{V} \mathcal{Z} \mathcal{V}^T$. The second step of the recipe is to
find a matrix $\mathcal{V}$ with integer entries and determinant $\pm 1$, such that $\mathcal{Z}'$
takes the simple form
\begin{equation}
\mathcal{Z}' = \bpm 0_I & -\mathcal{D} & 0 \\
          \mathcal{D} & 0_I & 0 \\
          0 & 0 & 0_{M-2I} \epm, \quad \mathcal{D} = \bpm d_1 & 0 & \dots & 0 \\
                                                  0 & d_2 & \dots & 0 \\
                                                  \vdots & \vdots & \vdots & \vdots \\
                                                   0 & 0 & \dots & d_I \epm
\label{zprime}
\end{equation}
Here $I$ is some integer with $0 \leq I \leq M/2$ and $0_I$ denotes an $I \times I$ matrix of zeros. 
In mathematical language, $\mathcal{V}$ is an integer change of basis that puts $\mathcal{Z}$ into 
\emph{skew-normal} form. It is known that such a change of basis always exists, though it is not unique.\cite{NewmanBook} 
\textcolor{black}{After finding an appropriate $\mathcal{V}$, the offset $\chi$ should then be chosen so that
\begin{equation}
\chi_i = \pi \cdot \sum_{j < k} \mathcal{V}_{ij} \mathcal{V}_{ik} \mathcal{Z}_{jk} \pmod{2\pi}
\label{chicond}
\end{equation}
The reason for choosing $\chi$ in this way is that it ensures that $e^{iC_i'}|\psi\> = |\psi\>$ 
for any $|\psi\> \in \mathcal{H}_{\text{eff}}$, as can be easily seen from the Campbell-Baker-Hausdorff formula.}

Once we perform these two steps, we can obtain the complete energy spectrum of $H_{\text{eff}}$ with the help of a few results that
we prove in appendix \ref{diagsect}.\footnote{These results rely on a small technical assumption, see Eq. \ref{Rassumption}.} Our first result 
is that $H_{\text{eff}}$ can always be written in the form
\begin{equation}
H_{\text{eff}} = \sum_{k=1}^K E_k a_k^\dagger a_k + F \left(C_{2I+1}',...,C_M'\right)
\label{Heffdiagsumm}
\end{equation}
where $F$ is some (a priori unknown) quadratic function. Our second result (which is really just an observation) 
is that the following operators all commute with each other:
\begin{equation}
\{e^{i C_{1}'/d_1},...,e^{i C_{I}'/d_I},e^{i C_{I+1}'},...,e^{i C_{2I}'},C_{2I+1}',...,C_M'\}
\label{commop}
\end{equation}
Furthermore, these operators commute with the occupation number operators $\{a_1^\dagger a_1,...,a_K^\dagger a_K\}$. Therefore,
we can simultaneously diagonalize (\ref{commop}) along with $\{a_k^\dagger a_k\}$. We denote the simultaneous eigenstates by
\begin{align*}
|\theta_1,...,\theta_I, \varphi_1,...,\varphi_I, x_{I+1}',...,x_{M-I}', n_1,...,n_K\>
\end{align*}
or, in more abbreviated form, $|\v{\theta}, \v{\varphi},\v{x}',\v{n}\>$. Here the different quantum numbers are defined by 
\begin{align}
&e^{i C_i'/d_{i}} |\v{\theta}, \v{\varphi},\v{x}',\v{n}\> = e^{i\theta_{i}} |\v{\theta}, \v{\varphi},\v{x}',\v{n}\>, \ i = 1,...,I \nonumber \\
&e^{i C_i'} |\v{\theta}, \v{\varphi},\v{x}',\v{n}\> = e^{i\varphi_{i-I}} |\v{\theta}, \v{\varphi},\v{x}',\v{n}\>, \ i = I+1,...,2I \nonumber\\
&C_i' |\v{\theta}, \v{\varphi},\v{x}',\v{n}\> = 2\pi x_{i-I}'  |\v{\theta}, \v{\varphi},\v{x}',\v{n}\>, \ i = 2I+1, ..., M \nonumber \\
&a_k^\dagger a_k |\v{\theta}, \v{\varphi},\v{x}',\v{n}\> = n_k  |\v{\theta}, \v{\varphi},\v{x}',\v{n}\>, \  k = 1,...,K  
\label{thetaphixn}
\end{align}
where $0 \leq \theta_i, \varphi_i < 2\pi$, while $x_i'$ is real valued and $n_k$ ranges over 
non-negative integers. 

By construction the $|\v{\theta}, \v{\varphi},\v{x}',\v{n}\>$ states form a complete basis for the Hilbert space $\mathcal{H}$. 
Our third result is that a \emph{subset} of these states form a 
complete basis for the effective Hilbert space $\mathcal{H}_{\text{eff}}$. This subset consists of all  
$|\v{\theta}, \v{\varphi},\v{x}',\v{n}\>$ for which 
\textcolor{black}{\begin{enumerate}
\item{$\v{\theta} = (2\pi \alpha_1/d_1,..., 2\pi \alpha_I/d_I)$ with $\alpha_i =0,1,...,d_i-1$.}
\item{$\v{\varphi} = (0,0,...,0)$.}
\item{$(x_{I+1}',...,x_{M-I}') = (q_1,...,q_{M-2I})$ for some integers $q_i$.}
\end{enumerate}}
We will denote this subset of eigenstates by $\{|\v{\alpha},\v{q},\v{n}\>\}$. 

Putting this together, we can see from equations (\ref{Heffdiagsumm}) and (\ref{thetaphixn}) that the $|\v{\alpha},\v{q},\v{n}\>$ are eigenstates of $H_{\text{eff}}$,
with eigenvalues
\begin{equation}
E = \sum_{k=1}^K n_k E_k + F(2\pi q_1,...,2\pi q_{M-2I})
\label{energyspectsumm}
\end{equation}
We therefore have the full eigenspectrum of $H_{\text{eff}}$ --- up to the determination of the function $F$. With a bit more work, 
one can go further and compute the function $F$ (see appendix \ref{eigensect}) but we will not discuss this issue here because in many cases of 
interest it is more convenient to find $F$ using problem-specific approaches.

To see examples of this diagonalization procedure, we refer the reader to section \ref{examsect}. As for the general derivation of this procedure, see appendix \ref{diagsect}.

\subsection{Degeneracy}
\label{degsummsect}
One implication of Eq. \ref{energyspectsumm} which is worth mentioning is that the energy $E$ is independent of the quantum numbers 
$\alpha_1,...,\alpha_I$. Since $\alpha_i$ ranges from $0 \leq \alpha_i < d_i - 1$, it follows that every eigenvalue of $H_{\text{eff}}$ has a 
degeneracy of (at least)
\begin{equation}
D = \prod_{i=1}^I d_i
\la{Deg}
\end{equation}
In the special case where $\mathcal Z_{ij}$ is non-degenerate (i.e. the case where $M = 2I$), this degeneracy can be conveniently written as
\begin{equation}
D = \sqrt{\det(\mathcal{Z})}
\end{equation}
since
\begin{equation*}
\det(\mathcal{Z}) = \det(\mathcal{Z}') = \prod_{i=1}^I d_i^2
\end{equation*}

For an example of this degeneracy formula, consider the Hamiltonian (\ref{example2summ}) discussed in section \ref{effsummsect}. In this 
case, $C_1 = dp$ while $C_2 = 2\pi x$ so 
\begin{equation*}
\mathcal Z_{ij} = \frac{1}{2\pi i} [C_i, C_j] = \bpm 0 & -d \\ d & 0 \epm
\end{equation*}
Thus, the above formula predicts that the degeneracy for this system is $D = \sqrt{\det(\mathcal Z)} = d$, which is consistent with our previous discussion.

\subsection{Finite $U$ corrections}
\label{finUsummsect}
We now discuss our last major result. To understand this result, note that while $H_{\text{eff}}$ gives the exact low energy spectrum of $H$ in the
infinite $U$ limit, it only gives \emph{approximate} results when $U$ is large but finite. Thus, to complete our
picture we need to understand what types of corrections we need to add to $H_{\text{eff}}$ to obtain an exact effective theory
in the finite $U$ case. 

It is instructive to start with a simple example: 
$H = \frac{p^2}{2m} + \frac{K x^2}{2}  - U \cos(2\pi x)$. As we discussed in section \ref{effsummsect}, 
the low energy effective Hamiltonian in the infinite $U$ limit 
is $H_{\text{eff}} = \frac{K x^2}{2}$ while the low energy Hilbert space $\mathcal{H}_{\text{eff}}$ 
is spanned by position eigenstates $\{|q\>\}$ where $q$ is an integer. 

Let us consider this example in the case where $U$ is large but finite. In this case, we expect that there is some small amplitude for the system
to tunnel from one cosine minima $x=q$ to another minima, $x = q-n$. Clearly we need to add correction terms to $\mathcal{H}_{\text{eff}}$ 
that describe these tunneling processes. But what are these correction terms? It is not hard to see that the most general possible correction terms
can be parameterized as
\begin{equation}
\sum_{n=-\infty}^\infty e^{inp} \cdot \epsilon_n(x)
\end{equation}
where $\epsilon_n(x)$ is some unknown function which also depends on $U$. Physically, each term $e^{inp}$ describes a tunneling process $|q\> \rightarrow |q-n\>$ since $e^{inp} |q\> = |q-n\>$. 
The coefficient $\epsilon_n(x)$ describes the amplitude for this process, which may depend on $q$ in general. (The one exception is the $n=0$ term, which does not describe tunneling at all, but rather describes corrections to the onsite energies for each minima).

Having developed our intuition with this example, we are now ready to describe our general result. Specifically, in the general case we show that the finite $U$ corrections can be written in the form
\begin{equation}
\sum_{\v{m}} e^{i \sum_{j=1}^M m_j \Pi_j} \cdot
\epsilon_{\v{m}}(\{a_k, a_k^\dagger, C_{2I+i}'\})
\label{finUsumm}
\end{equation}
with the sum running over $M$ component integer vectors $\v{m} = (m_1,...,m_M)$. Here, the $\epsilon_{\v{m}}$ are
unknown functions of $\{a_1,...,a_k,a_1^\dagger,...,a_k^\dagger,C_{2I+1'},...,C_M'\}$ which also depend on $U$. 
We give some examples of these results in section \ref{examsect}. For a derivation of the finite $U$ corrections, see appendix \ref{finUsect}. 

\subsection{Splitting of ground state degeneracy}
\label{degsplitsummsect}
One application of Eq. \ref{finUsumm} is to determining how the ground state degeneracy of $H_{\text{eff}}$ splits at finite $U$.
Indeed, according to standard perturbation theory, we can find the splitting of the ground state
degeneracy by projecting the finite $U$ corrections onto the ground state subspace and then diagonalizing the resulting
$D \times D$ matrix. The details of this diagonalization problem are system dependent,
so we cannot say much about it in general. However, we would like to mention a related result that is useful in
this context. This result applies to any system in which the commutator matrix
$\mathcal{Z}_{ij}$ is non-degenerate. Before stating the result, we first need to define
some notation: let $\Gamma_1,...,\Gamma_M$ be operators defined by
\begin{equation}
\Gamma_i = \sum_j (\mathcal{Z}^{-1})_{ji} C_j
\label{Gammadef}
\end{equation}
Note that, by construction, the $\Gamma_i$ operators obey the commutation relations
\begin{align}
[C_i,\Gamma_j] = 2\pi i \delta_{ij}, \quad [a_k,\Gamma_j] = [a_k^\dagger, \Gamma_j] = 0
\end{align}
With this notation, our result is that
\begin{equation}
\<\alpha'|e^{i \sum_{j=1}^M m_j \Pi_j} \cdot \epsilon_{\v{m}} |\alpha\>
= u_{\v{m}} \cdot \<\alpha'|e^{i \sum_{j=1}^M m_j \Gamma_j}|\alpha\>
\label{finUidsumm}
\end{equation}
where $|\alpha\>, |\alpha'\>$ are ground states and $u_{\v{m}}$ is some unknown proportionality constant. This result is useful because it
is relatively easy to compute the matrix elements of $e^{i \sum_{j=1}^M m_j \Gamma_j}$; hence the above relation allows us to 
compute the matrix elements of the finite $U$ corrections (up to the constants $u_{\v{m}}$) without much work. We derive this result in appendix \ref{finUsect}.


\section{Examples}
\label{examsect}
In this section, we illustrate our formalism with some concrete examples. These examples involve
a class of two dimensional electron systems in which the spin-up and spin-down electrons 
form $\nu = 1/k$ Laughlin states with opposite chiralities~\cite{BernevigZhang}. These states are known as ``fractional quantum spin Hall insulators.'' 
We will be primarily interested in the \emph{edges} of fractional quantum spin Hall (FQSH) insulators~\cite{LevinStern}. Since the edge of the 
Laughlin state can be modeled as a single chiral Luttinger liquid, the edge of a FQSH insulator consists of two chiral Luttinger 
liquids with opposite chiralities --- one for each spin direction (Fig. \ref{fig:cleanedge}). 

The examples that follow will explore the physics of the FQSH edge in the presence of impurity-induced scattering. More specifically, 
in the first example, we consider a FQSH edge with a single magnetic impurity; in the second example we consider a FQSH edge with 
multiple magnetic impurities; in the last example we consider a FQSH edge with alternating magnetic and superconducting 
impurities. In all cases, we study the impurities in the infinite scattering limit, which corresponds to $U \rightarrow \infty$ in (\ref{genHamc}).
Then, in the last subsection we discuss how our results change when the scattering strength $U$ is large but finite.

We emphasize that the main purpose of these examples is to illustrate our formalism rather than to derive novel results. In particular, 
many of our findings regarding these examples are known previously in the literature in some form. 

Of all the examples, the last one, involving magnetic and superconducting impurities, is perhaps most interesting: we find that this system 
has a ground state degeneracy that grows exponentially with the
number of impurities. This ground state degeneracy is closely related to the previously known topological degeneracy that appears 
when a FQSH edge is proximity coupled to alternating ferromagnetic and superconducting strips~\cite{lindner2012fractionalizing, cheng2012superconducting, barkeshli2013twist,vaezi2013fractional, clarke2013exotic}.

Before proceeding, we need to explain what we mean by ``magnetic impurities'' and ``superconducting impurities.'' At a formal level, a magnetic
impurity is a localized object that scatters spin-up electrons into spin-down electrons. Likewise a superconducting impurity is a localized
object that scatters spin-up electrons into spin-down \emph{holes}. More physically, a magnetic impurity can be realized by placing the tip of a ferromagnetic
needle in proximity to the edge while a superconducting impurity can be realized by placing the tip of a superconducting needle in proximity to
the edge. 

\subsection{Review of edge theory for clean system}
\begin{figure}[tb]
  \centering
\includegraphics[width=4.5cm,height=4.0cm]{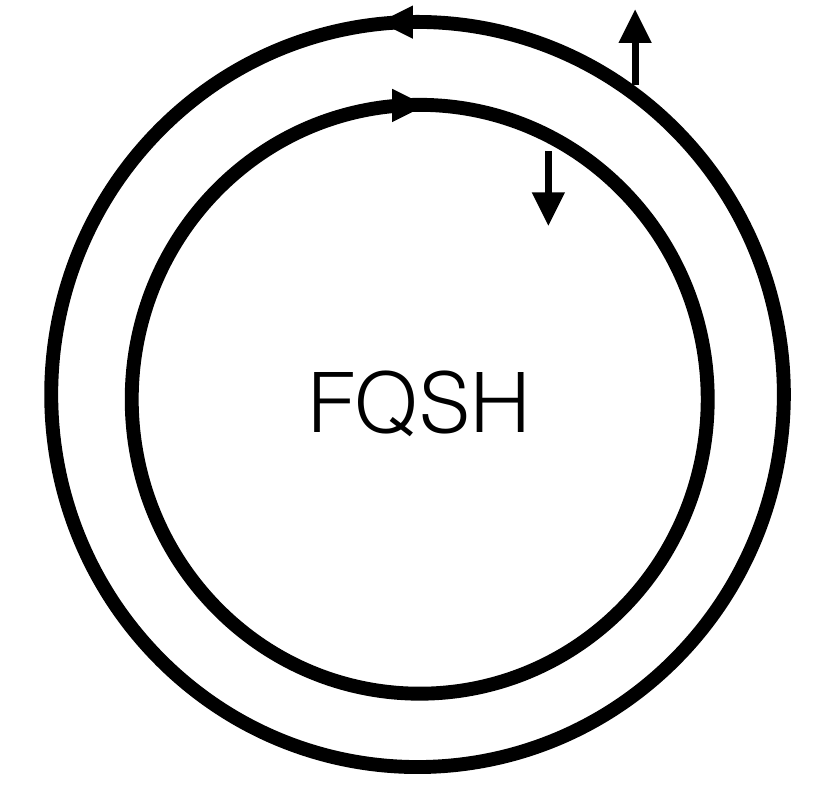}
    \caption{The fractional quantum spin Hall edge consists of two counter-propagating chiral Luttinger Liquids --- one for each spin direction ($\up,\down$)}
  \la{fig:cleanedge}
\end{figure}
As discussed above, the edge theory for the $\nu = 1/k$ fractional quantum spin Hall state consists of
two chiral Luttinger liquids with opposite chiralities --- one for each spin direction (Fig. \ref{fig:cleanedge}). The purpose of this section 
is to review the Hamiltonian formulation of this edge theory.\cite{WenReview,WenBook,LevinStern} More specifically we will discuss the edge theory for a disk geometry where the
circumference of the disk has length $L$. Since we will work in a Hamiltonian formulation, in order to define the edge theory we 
need to specify the Hamiltonian, the set of physical observables, and the canonical commutation relations.

We begin with the set of physical observables. The basic physical observables in the edge theory are a collection of operators 
$\{\partial_y \phi_\up(y), \partial_y \phi_\down(y)\}$ along with two
additional operators $\phi_\up(y_0)$, $\phi_\down(y_0)$ where $y_0$ is an arbitrary, but fixed, point on the boundary of the disk. 
The $\{\partial_y \phi_\up(y), \partial_y \phi_\down(y),\phi_\up(y_0),\phi_\down(y_0)\}$ operators can be thought of as the 
fundamental phase space operators in this system, i.e. the analogues of the $\{x_1,...,x_N,p_1,...,p_N\}$
operators in section \ref{summsect}. Like $\{x_1,...,x_N,p_1,...,p_N\}$, all other
physical observables can be written as functions/functionals of 
$\{\partial_y \phi_\up(y), \partial_y \phi_\down(y),\phi_\up(y_0),\phi_\down(y_0)\}$. Two important examples are the 
operators $\phi_\up(y)$ and $\phi_\down(y)$ which are defined by
\begin{equation}
\phi_\sigma(y) \equiv \phi_\sigma(y_0) + \int_{y_0}^y \partial_x \phi_\sigma dx, \quad \sigma = \up, \down
\label{phidef}
\end{equation}
where the integral runs from $y_0$ to $y$ in the clockwise direction. 

The physical meaning of these operators is as follows: the density of spin-up electrons at position $y$ is given
by $\rho_\up(y) = \frac{1}{2\pi} \partial_y \phi_\up$ while the density of spin-down electron is
$\rho_\down(y) = \frac{1}{2\pi} \partial_y \phi_\down$. The total charge $Q$ and total spin $S^z$ on the edge are 
given by $Q = Q_\up + Q_\down$ and $S^z = 1/2(Q_\up - Q_\down)$ with
\begin{align*}
Q_\sigma = \frac{1}{2\pi} \int_{-L/2}^{L/2} \partial_y \phi_\sigma dy, \quad \sigma = \up,\down
\end{align*}
Finally, the spin-up and spin-down electron creation operators take the form
\begin{align*}
\psi_\up^\dagger = e^{i k \phi_\up}, \quad \psi_\down^\dagger = e^{-i k \phi_\down}
\end{align*}

In the above discussion, we ignored an important subtlety: $\phi_\up(y_0)$ and $\phi_\down(y_0)$ are actually \emph{compact} degrees of freedom 
which are only defined modulo $2\pi/k$. In other words, strictly speaking, $\phi_\up(y_0)$ and $\phi_\down(y_0)$ are \emph{not}
well-defined operators: only $e^{i k \phi_\up(y_0)}$ and $e^{ik \phi_\down(y_0)}$ are 
well-defined. (Of course the same also goes for $\phi_\up(y)$ and $\phi_\down(y)$, 
in view of the above definition). Closely related to this fact, the conjugate ``momenta'' $Q_\up$ and $Q_\down$ are actually 
discrete degrees of freedom which can take only integer values. 

The compactness of $\phi_\up(y_0),\phi_\down(y_0)$ and discreteness of $Q_\up, Q_\down$ is inconvenient for us
since the machinery discussed in section \ref{summsect} is designed for systems in which all the phase space operators 
are real-valued, rather than systems in which some operators are angular valued and some are integer valued.
To get around this issue, we will initially treat $\phi_\up(y_0)$ and $\phi_\down(y_0)$ and the conjugate
momenta $Q_\up$, $Q_\down$ as \emph{real valued} operators. We will then use a trick (described in the next section) 
to dynamically generate the compactness of $\phi_\up(y_0), \phi_\down(y_0)$ as well as the discreteness of $Q_\up, Q_\down$.

Let us now discuss the commutation relations for the $\{\partial_y \phi_\up(y), \partial_y \phi_\down(y),\phi_\up(y_0),\phi_\down(y_0)\}$ 
operators. Like the usual phase space operators $\{x_1,...,x_N,p_1,...,p_N\}$, the commutators of
$\{\partial_y \phi_\up(y), \partial_y \phi_\down(y),\phi_\up(y_0),\phi_\down(y_0)\}$ are $c$-numbers.
More specifically, the basic commutation relations are
\begin{align}
[\partial_x\phi_\up(x),\partial_{y}\phi_{\up}(y)] &=\frac{2\pi i}{k} \partial_x\delta(x-y) \nonumber \\
[\partial_x\phi_{\down}(x),\partial_{y}\phi_{\down}(y)] &=-\frac{2\pi i}{k} \partial_x\delta(x-y) \nonumber \\
[\phi_\up(y_0), \partial_y \phi_\up(y)] &= \frac{2\pi i}{k}\delta(y-y_0) \nonumber \\
[\phi_\down(y_0), \partial_y \phi_\down(y)] &= -\frac{2\pi i}{k}\delta(y-y_0)
\label{phicommrel0}
\end{align}
with the other commutators vanishing:
\begin{align*}
[\phi_\up(y_0), \partial_y \phi_\down(y)] &= [\phi_\down(y_0), \partial_y \phi_\up(y)] =0 \\
[\phi_\up(y_0), \phi_\down(y_0)] &= [\partial_x\phi_\up(x),\partial_{y}\phi_{\down}(y)] = 0
\end{align*}
Using these basic commutation relations, together with the definition of $\phi_\sigma(y)$ (\ref{phidef}), one can 
derive the more general relations
\begin{align}
[\phi_\up(x),\partial_{y}\phi_{\up}(y)] &=\frac{2\pi i}{k}\delta(x-y) \nonumber \\
[\phi_{\down}(x),\partial_{y}\phi_{\down}(y)] &=-\frac{2\pi i}{k}\delta(x-y) \nonumber \\
[\phi_\up(x), \partial_y \phi_{\down}(y)] &= 0
\label{phicommrel}
\end{align}
as well as
\begin{align}
[\phi_\up(x), \phi_\up(y)] &= \frac{\pi i}{k} \text{sgn}(x,y) \nonumber \\
[\phi_\down(x), \phi_\down(y)] &= -\frac{\pi i}{k} \text{sgn}(x,y) \nonumber \\
[\phi_\up(x), \phi_\down(y)] &= 0
\label{genphicommrel}
\end{align}
where the $\text{sgn}$ function is defined by $\text{sgn}(x,y) = +1$ if $y_0 < x < y$ and $\text{sgn}(x,y) = -1$ if $y_0 < y < x$, with
the ordering defined in the clockwise direction. The latter commutation relations (\ref{phicommrel}) and (\ref{genphicommrel}) will be
particularly useful to us in the sections that follow.

Having defined the physical observables and their commutation relations, the last step is to define the Hamiltonian for the edge theory.
The Hamiltonian for a perfectly clean, homogeneous edge is
\begin{equation}
H_0 = \frac{kv}{4\pi}\int_{-L/2}^{L/2}[(\partial_{x}\phi_{\up}(x))^{2}+(\partial_{x}\phi_{\down}(x))^{2}]dx
\label{Hclean}
\end{equation}
where $v$ is the velocity of the edge modes.

At this point, the edge theory is complete except for one missing element: we have not given an explicit definition 
of the Hilbert space of the edge theory. There are two different
(but equivalent) definitions that one can use. The first, more abstract, definition is that the Hilbert space is the unique 
irreducible representation of the operators 
$\{\partial_y \phi_\up, \partial_y \phi_\down,\phi_\up(y_0),\phi_\down(y_0)\}$ and the commutation relations
(\ref{phicommrel0}). (This is akin to defining the Hilbert space of the 1D harmonic oscillator as the irreducible representation
of the Heisenberg algebra $[x,p] = i$). The second definition, which is more concrete but also more complicated, 
is that the Hilbert space is spanned by 
the complete orthonormal basis $\{|q_\up, q_\down, \{n_{p\up}\}, \{n_{p\down}\}\>\}$
where the quantum numbers $q_\up, q_\down$ range over all integers
\footnote{Actually, \unexpanded{$q_\uparrow, q_\downarrow$} 
range over arbitrary real numbers in our fictitious representation of the edge: as explained above, we initially pretend that 
\unexpanded{$Q_\uparrow, Q_\downarrow$} are not quantized, and then introduce quantization later on using a trick.}
while $n_{p\up}, n_{p\down}$ range over all nonnegative integers
for each value of $p = 2\pi/L, 4\pi/L,...$. These basis states have a simple physical meaning: 
$|q_\up, q_\down, \{n_{p\up}\}, \{n_{p\down}\}\>$ corresponds to a state with
charge $q_\up$ and $q_\down$ on the two edge modes, and with $n_{p \up}$ and $n_{p \down}$ phonons with momentum $p$ on the two edge
modes. 

\subsection{Example 1: Single magnetic impurity}
\label{fmsisect}

With this preparation, we now proceed to study a fractional quantum spin Hall edge with a single magnetic
impurity in a disk geometry
of circumference $L$ (Fig. \ref{fig:si}a). We 
assume that the impurity, which is located at $x = 0$, generates a backscattering term of the form 
$\frac{U}{2}(\psi_\up^\dagger(0) \psi_\down(0) + h.c)$. Thus, in the
bosonized language, the system with an impurity is described by the Hamiltonian
\begin{align}
&H=H_0-U\cos(C), \quad C = k (\phi_\up(0) + \phi_\down(0))
\label{fmsi}
\end{align}
where $H_0$ is defined in Eq. \ref{Hclean}. Here, we temporarily ignore the question of how we regularize the
cosine term; we will come back to this point below.

Our goal is to
find the low energy spectrum of $H$ in the strong backscattering limit, $U \rightarrow \infty$. We will accomplish this
using the results from section \ref{summsect}. Note that, in using these results, we implicitly assume that
our formalism applies to systems with infinite dimensional phase spaces, even though we only derived it in the finite
dimensional case.

\begin{figure}[tb]
  \centering
\includegraphics[width=8.0cm,height=4.0cm]{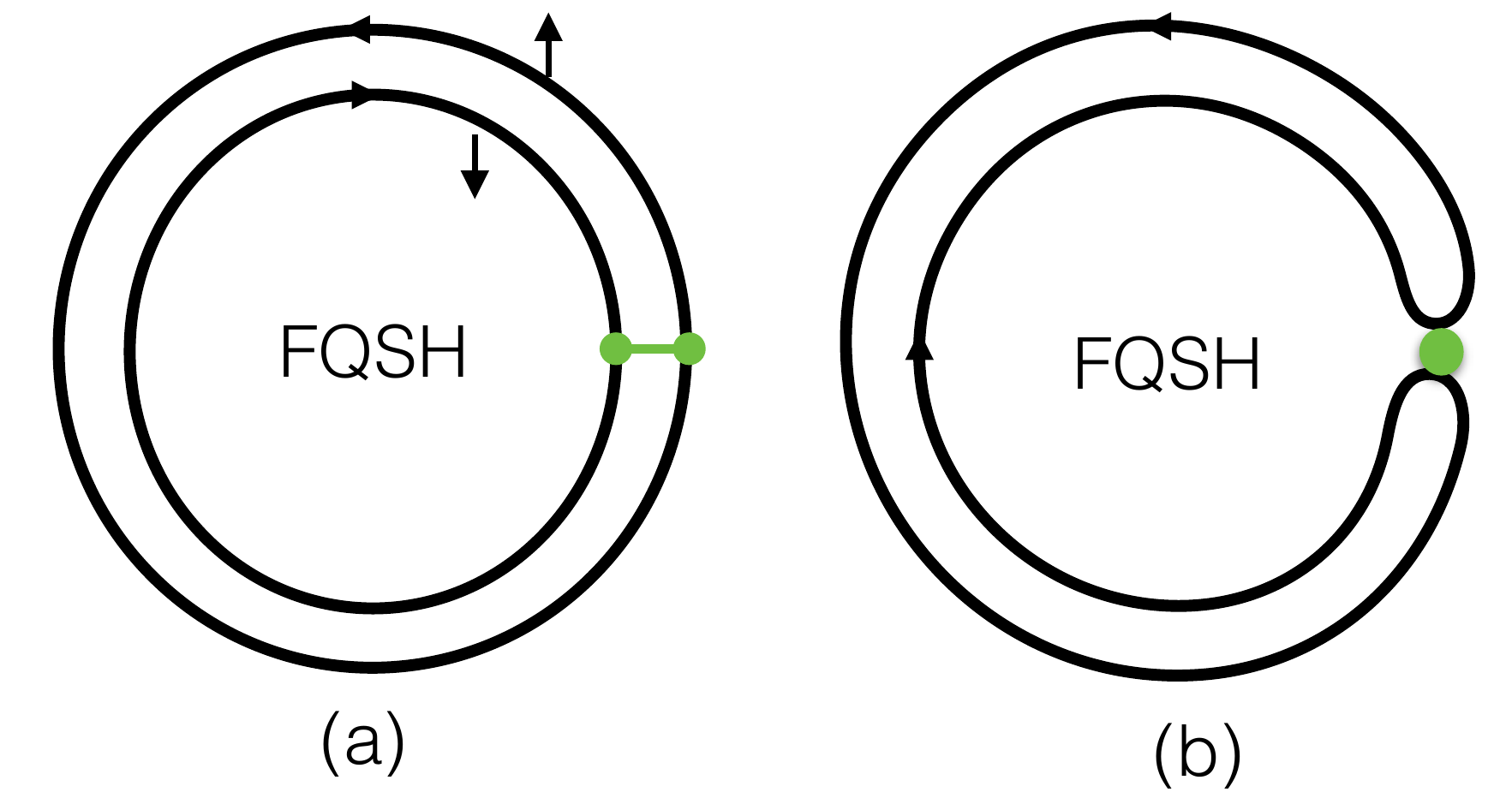}
    \caption{(a) A magnetic impurity on a fractional quantum spin Hall edge causes spin-up electrons to backscatter
into spin-down electrons. (b) In the infinite backscattering limit, the impurity effectively
reconnects the edge modes.}
  \la{fig:si}
\end{figure}

First we describe a trick for correctly accounting for the compactness of
$\phi_\up(y_0), \phi_\down(y_0)$ and the quantization of $Q_\up, Q_\down$. The idea is simple: we initially treat these variables as if they
are real valued, and then we introduce compactness \emph{dynamically} by adding two additional cosine terms to our Hamiltonian:
\begin{equation}
H = H_0 - U \cos(C) - U \cos(2\pi Q_\up) - U \cos(2\pi Q_\down)
\end{equation}
These additional cosine terms effectively force $Q_\up$ and $Q_\down$ to be quantized at low energies, thereby generating the compactness
that we seek.\footnote{Strictly speaking we also need to add (infinitesimal) quadratic terms to the Hamiltonian of the form 
\unexpanded{$\epsilon( \phi_\up(y_0)^2 + \phi_\down(y_0)^2)$} so that the $\mathcal{N}$ matrix is non-degenerate. However, these terms play no role in 
our analysis so we will not include them explicitly.}
We will include all three cosine terms in our subsequent analysis.

The next step is to calculate the low energy effective Hamiltonian $H_{\text{eff}}$ and low energy Hilbert space $\mathcal{H}_{\text{eff}}$.
Instead of working out the expressions in Eqs. \ref{Heffgenc}, \ref{Hilbeff}, we will skip this computation and proceed directly to
finding creation and annihilation operators for $H_{\text{eff}}$ using Eq. \ref{auxeqsumm}. (This approach works because equation (\ref{auxeqsumm}) 
does not require us to find the explicit form of $H_{\text{eff}}$). 

According to Eq. \ref{auxeqsumm}, we can find the creation and annihilation
operators for $H_{\text{eff}}$ by finding all operators $a$ such that (1) $a$ is a linear combination of our fundamental phase space 
operators $\{\partial_y \phi_\up, \partial_y \phi_\down, \phi_\up(y_0), \phi_\down(y_0)\}$ and (2) $a$ obeys
\begin{align}
[a, H_{0}] &= E a + \lambda [C, H_0] + \lambda_\up [Q_\up, H_0] + \lambda_\down [Q_\down, H_0] \nonumber \\
[a, C] &= [a,Q_\up] = [a,Q_\down] = 0 
\label{auxeqsi}
\end{align}
for some scalars $E, \lambda, \lambda_\up, \lambda_\down$ with $E \neq 0$. 

To proceed further, we note that the constraint $[a, Q_\up] = [a, Q_\down] = 0$, implies that $\phi_\up(y_0), \phi_\down(y_0)$ cannot appear in the expression 
for $a$. Hence, $a$ can be written in the general form
\begin{equation}
a=\int_{-L/2}^{L/2} [f_{\up}(y)\partial_{y}\phi_{\up}(y)+f_{\down}(y)\partial_{y}\phi_{\down}(y)] dy
\la{genmode}
\end{equation}
Substituting this expression into the first line of Eq.~\ref{auxeqsi} we obtain the differential equations
\begin{align*}
-ivf_{\up}'(y) & =Ef_{\up}(y)+\lambda kiv\delta(y) \nonumber \\
ivf_{\down}'(y) & = Ef_{\down}(y)-\lambda kiv\delta(y)
\end{align*}
(The $\lambda_\up, \lambda_\down$ terms drop out of these equations since $Q_\up, Q_\down$ commute with $H_0$).
These differential equations can be solved straightforwardly. The most general solution takes the form
\begin{align}
f_{\up}(y) & = e^{ipy}[A_{1}\Theta(-y)+A_{2}\Theta(y)] \nonumber \\
f_{\down}(y) & = e^{-ipy}[B_{1}\Theta(-y)+B_{2}\Theta(y)]
\la{solsi}
\end{align}
where $p = E/v$, and 
\begin{equation}
A_{2}=A_{1}-\lambda k, B_{2}=B_{1}-\lambda k
\label{bcsi}
\end{equation}
Here $\Theta$ is the Heaviside step function defined as
\begin{equation*}
\Theta(x) = \begin{cases} 0 & -L/2 \leq x \leq 0 \\
			  1 & 0 \leq x \leq L/2 \end{cases}
\end{equation*}
(Note that the above expressions (\ref{solsi}) for $f_\up, f_\down$ do not obey periodic boundary conditions 
at $x = \pm L/2$; we will not impose these boundary conditions until later in our calculation). Eliminating $\lambda$ 
from (\ref{bcsi}) we see that
\begin{equation}
A_2 - A_1 = B_2 - B_1
\label{constraintAB1}
\end{equation}

We still have to impose one more condition on $a$, namely $[a,C] = 0$. This condition leads to a second constraint on $A_1, A_2, B_1, B_2$,
but the derivation of this constraint is somewhat subtle. The problem is that if we simply substitute (\ref{genmode}) into $[a,C] = 0$, we find
\begin{equation}
f_{\up}(0) = f_{\down}(0) 
\la{conssi}
\end{equation}
It is unclear how to translate this relation into one involving $A_1, A_2, B_1, B_2$ since $f_\up, f_\down$ are discontinuous at $x=0$ 
and hence $f_\up(0), f_\down(0)$ are ill-defined. The origin of this puzzle is that the cosine term in Eq. \ref{fmsi} contains short-distance 
singularities and hence is not well-defined. To resolve this issue we regularize the argument of the cosine term, replacing 
$C = k(\phi_\up(0)+\phi_\down(0))$ with
\begin{equation}
C \rightarrow \int_{-L/2}^{L/2} k ( \phi_\up(x) + \phi_\down(x)) \tilde{\delta}(x) dx
\label{Cregsi}
\end{equation}
where $\tilde{\delta}$ is a narrowly peaked function with $\int \tilde{\delta}(x) dx =1$. Here, we can think of 
$\tilde{\delta}$ as an approximation to a delta function. Note that $\tilde{\delta}$ effectively introduces a short-distance cutoff
and thus makes the cosine term non-singular. After making this replacement, it is straightforward to repeat the 
above analysis and solve the differential equations for $f_\up, f_\down$. In appendix \ref{regapp}, we work out this 
exercise, and we find that with this regularization, the condition $[a,C] = 0$ leads to the constraint
\begin{equation}
\frac{A_1 + A_2}{2} = \frac{B_1 + B_2}{2} 
\label{constraintAB2}
\end{equation}
Combining our two constraints on $A_1,B_1, A_2, B_2$ (\ref{constraintAB1},\ref{constraintAB2}), we obtain the relations
\begin{align}
A_1 = B_1, \quad A_2 = B_2
\end{align}

So far we have not imposed any restriction on the momentum $p$. The momentum constraints come from the periodic boundary conditions on $f_\up, f_\down$:
\begin{align*}
f_\up(-L/2) = f_\up(L/2), \quad f_\down(-L/2) = f_\down(L/2)
\end{align*}
Using the explicit form of $f_\up, f_\down$, these boundary conditions give
\begin{equation*}
A_1 e^{-ipL/2} = A_2 e^{ipL/2}, \quad B_1 e^{ipL/2} = B_2 e^{-ipL/2}
\end{equation*}
from which we deduce
\begin{equation}
e^{2ipL} = 1, \quad A_2 = A_1 e^{-ipL}
\end{equation}
Putting this all together, we see that the most general possible creation/annihilation operator for $H_{\text{eff}}$ is given by
\begin{align*}
a_p =  A_1 \int_{-L/2}^{L/2} &(e^{ipy} \partial_y \phi_\up + e^{-ipy} \partial_y \phi_\down) \Theta(-y) \nonumber \\
+ e^{-ipL}&(e^{ipy} \partial_y \phi_\up + e^{-ipy} \partial_y \phi_\down) \Theta(y) dy
\end{align*}
where $p$ is quantized as $p = \pm \pi/L, \pm 2\pi/L,...$ and $E_p = vp$. (Note that $p=0$
does not correspond to a legitimate creation/annihilation operator according to the definition given above, since we
require $E \neq 0$).

Following the conventions from \textcolor{black}{section \ref{diagsummsect}}, we will refer to the operators with $E_p > 0$ --- or equivalently $p > 0$ ---
as ``annihilation operators'' and the other operators as ``creation operators.'' Also, we will choose the normalization 
constant $A_1$ so that $[a_p, a_{p'}^\dagger] = \delta_{pp'}$ for $p,p' > 0$. This gives the expression
\begin{align}
a_p = \sqrt{\frac{k}{4\pi|p|L}} \int_{-L/2}^{L/2} &(e^{ipy} \partial_y \phi_\up + e^{-ipy} \partial_y \phi_\down) \Theta(-y) 
\nonumber \\
+ e^{-ipL}&(e^{ipy} \partial_y \phi_\up + e^{-ipy} \partial_y \phi_\down) \Theta(y) dy
\end{align}

The next step is to compute the commutator matrix $\mathcal{Z}_{ij}$. In the case at hand, we have three cosine terms
$\{\cos(C_1), \cos(C_2), \cos(C_3)\}$, where 
\begin{equation*}
C_1 = C, \quad C_2 = 2\pi Q_\up, \quad C_3= 2\pi Q_\down 
\end{equation*}
Therefore $\mathcal{Z}_{ij}$ is given by
\begin{equation*}
\mathcal{Z}_{ij} = \frac{1}{2\pi i} [C_i, C_j] = \bpm 0 & 1 & -1 \\ -1 & 0 & 0 \\ 1 & 0 & 0 \epm
\end{equation*}

\textcolor{black}{To proceed further we need to find an appropriate change of variables of the form $C_i' = \sum_{j=1}^{3} \mathcal{V}_{ij} C_j + \chi_i$.
Here, $\mathcal V$ should an integer matrix with determinant $\pm 1$ with the property 
that $ \mathcal{Z}_{ij}' = \frac{1}{2\pi i} [C_i', C_j']$ is in skew-normal form, while $\chi$ should be a real
vector satisfying Eq. \ref{chicond}.} It is easy to see that the following change of variables does the job:
\begin{equation*}
C_1' = C_1, \quad C_2' = -2\pi Q_\up, \quad C_3' = 2\pi Q_\up + 2\pi Q_\down
\end{equation*}
Indeed, for this change of variables,
\begin{equation*}
\mathcal{Z}' = \bpm 0 & -1 & 0 \\ 1 & 0 & 0 \\ 0 & 0 & 0 \epm
\end{equation*}
We can see that this is in the canonical skew normal form shown in Eq. \ref{zprime}, with the parameters $M=3$, $I = 1$,
$d_1 = 1$.     

We are now in a position to write down the low energy effective Hamiltonian $H_{\text{eff}}$: according to Eq. 
\ref{Heffdiagsumm},
$H_{\text{eff}}$ must take the form
\begin{equation}
H_{\text{eff}} = \sum_{p > 0} v p a_p^\dagger a_p + F \cdot (C_3')^2
\label{Heffsi1}
\end{equation}
where $F$ is some (as yet unknown) constant. To determine the constant $F$, we make two observations. First,
we note that the first term in Eq. \ref{Heffsi1} can be rewritten as $\sum_{p \neq 0} \frac{v |p|}{2} (a_{-p} a_p)$.
Second, we note that $C_3' = 2\pi Q$ is proportional to $a_{p =0}$. Given these observations, it is natural to interpret
the $F (C_3')^2$ term as the missing $p = 0$ term in the sum. This suggests that we can fix the coefficient $F$ using 
continuity in the $p \rightarrow 0$ limit. To this end, we observe that
\begin{equation*}
\lim_{p \rightarrow 0} \frac{v|p|}{2} a_{-p} a_p =  \frac{vk}{8 \pi L} \cdot (C_3')^2  
\end{equation*}
We conclude that $F = \frac{vk}{8 \pi L}$. Substituting this into (\ref{Heffsi1}),
we derive 
\begin{equation}
H_{\text{eff}} = \sum_{p > 0} v p a_p^\dagger a_p + \frac{vk}{8 \pi L} \cdot (C_3')^2
\label{Heffsi2}
\end{equation}
where the sum runs over $p = \pi/L, 2\pi/L, ...$.

In addition to the effective Hamiltonian, we also need to discuss the effective Hilbert space $\mathcal{H}_{\text{eff}}$
in which this Hamiltonian is defined. According to the results of \textcolor{black}{section \ref{diagsummsect}}, the effective Hilbert space 
$\mathcal{H}_{\text{eff}}$ is spanned by states $\{|q,\{n_p\}\>\}$ where $|q,\{n_p\}\>$ is the unique simultaneous eigenstate of the form
\begin{align*}
e^{iC_1'}|q,\{n_p\}\> &= |q,\{n_p\}\>, \\
e^{iC_2'}|q,\{n_p\}\> &= |q,\{n_p\}\>, \\
C_3' |q,\{n_p\}\> &= 2\pi q |q,\{n_p\}\>, \\
a_p^\dagger a_p |q,\{n_p\}\> &= n_p |q,\{n_p\}\>
\end{align*}
Here $n_p$ runs over non-negative integers, while $q$ runs over all integers. 
Note that we do not need to label the $\{|q,\{n_p\}\>\}$ basis states with $\alpha$ quantum numbers since $d_1 = 1$ so there is no 
degeneracy.

Having derived the effective theory, all that remains is to diagonalize it. Fortunately we can accomplish this without any
extra work: from (\ref{Heffsi2}) it is clear that 
the $\{|q,\{n_p\}\>\}$ basis states are also eigenstates of $H_{\text{eff}}$ with energies given by
\begin{equation}
E = \sum_{p > 0} v p n_p +  \frac{\pi vk}{2L} \cdot q^2
\label{Esi}
\end{equation}
We are now finished: the above equation gives the complete energy spectrum of $H_{\text{eff}}$, and thus the complete
low energy spectrum of $H$ in the limit $U \rightarrow \infty$. 

To understand the physical interpretation of this energy spectrum, we
can think of $n_p$ as describing the number of phonon excitations with momentum $p$, while $q$ describes the
total charge on the edge. With these identifications, the first term in (\ref{Esi}) describes the total
energy of the phonon excitations --- which are linearly dispersing with velocity $v$ --- while the second term
describes the charging/capacitative energy of the edge. 

It is interesting that at low energies, 
our system has only \emph{one} branch of phonon modes and one charge degree of freedom, while the clean edge theory (\ref{Hclean}) 
has two branches of phonon modes and two charge degrees of freedom --- one for each spin direction. The explanation for
this discrepancy can be seen in Fig. \ref{fig:si}b: in the infinite $U$ limit, the impurity induces
perfect backscattering which effectively reconnects the edges to form a single chiral edge of length $2L$.

\subsection{Example 2: Multiple magnetic impurities}
\label{fmmisect}
We now consider a fractional quantum spin Hall edge in a disk geometry 
with $N$ magnetic impurities located at positions $x_1,...,x_N$ (Fig. \ref{fig:mi}a). 
Modeling the impurities in the same way as in the previous section, the Hamiltonian is
\begin{align}
H=H_0-U\sum_{i=1}^{N}\cos(C_i), \quad C_{i}=k (\phi_{\up}(x_i)+\phi_{\down}(x_i))
\la{fmmi}
\end{align}
where $H_0$ is defined in Eq. \ref{Hclean}.

As in the single impurity case, our goal is to understand the low energy physics of $H$ in the limit $U \rightarrow \infty$.
We can accomplish this using the same approach as before.
The first step is to take account of the compactness of $\phi_\up, \phi_\down$ and the discrete nature of 
$Q_\up, Q_\down$ by adding two additional cosine terms to our Hamiltonian:
\begin{equation*}
H = H_0 - U\sum_{i=1}^{N}\cos(C_i) - U \cos(2\pi Q_\up) - U \cos(2\pi Q_\down)
\end{equation*}

\begin{figure}[tb]
  \centering
\includegraphics[width=8cm,height=4.5cm]{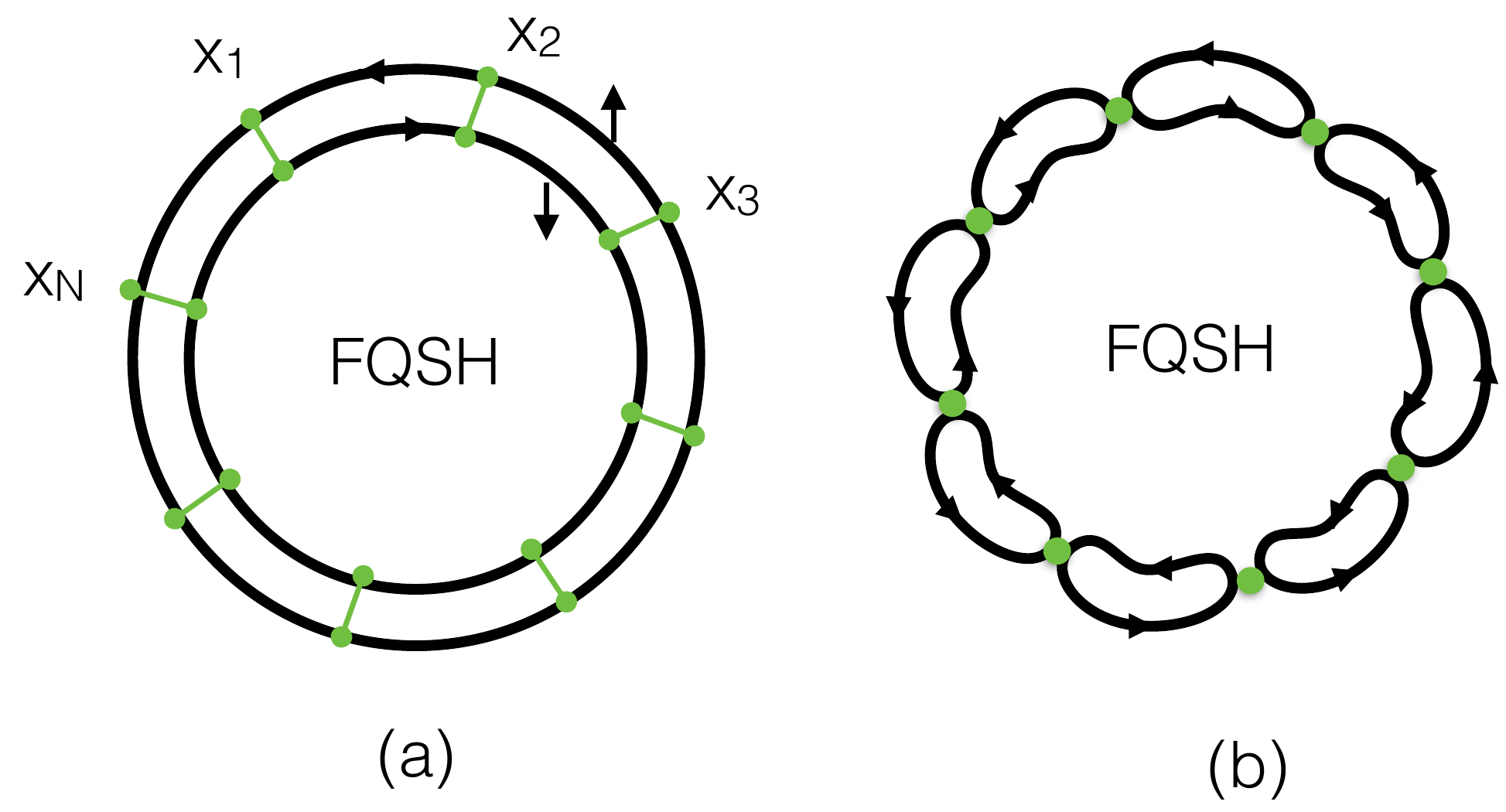}
    \caption{(a) A collection of $N$ magnetic impurities
on a fractional quantum spin Hall edge. The impurities are located at positions $x_1,...,x_N$. (b) In the infinite backscattering
limit, the impurities effectively reconnect the edge modes, breaking the edge into $N$ disconnected components.}
  \la{fig:mi}
\end{figure}

Next, we find the creation and annihilation operators for $H_{\text{eff}}$ using Eq. \ref{auxeqsumm}. That is, we search for all 
operators $a$ such that (1) $a$ is a linear combination of our fundamental phase 
space operators $\{\partial_y \phi_\up, \partial_y \phi_\down, \phi_\up(y_0), \phi_\down(y_0)\}$ and (2) $a$ obeys
\begin{align}
[a, H_{0}] &= E a + \sum_{i=1}^N \lambda_i [C_i, H_0] + \lambda_\up [Q_\up, H_0] + \lambda_\down [Q_\down, H_0] \nonumber \\
[a, C_j] &= [a,Q_\up] = [a,Q_\down] = 0
\label{auxeqfm}
\end{align}
for some $E, \textcolor{black}{\lambda_i}, \lambda_\up, \lambda_\down$ with $E \neq 0$.

Given that $[a, Q_\up] = [a, Q_\down] = 0$, we know that $\phi_\up(y_0), \phi_\down(y_0)$ cannot appear in the expression 
for $a$. Hence, $a$ can be written in the general form
\begin{equation*}
a=\int_{-L/2}^{L/2} [f_{\up}(y)\partial_{y}\phi_{\up}(y)+f_{\down}(y)\partial_{y}\phi_{\down}(y)] dy
\end{equation*}
Substituting these expressions into the first line of Eq.~\ref{auxeqfm}, we obtain
\begin{align*}
-ivf_{\up}'(y) & =Ef_{\up}(y)+ kiv \sum_{j=1}^N \lambda_j \delta(y - x_j) \nonumber \\
ivf_{\down}'(y) & = Ef_{\down}(y)- kiv \sum_{j=1}^N \lambda_j \delta(y-x_j)
\end{align*}
Solving these differential equations gives
\begin{align*}
f_{\up}(y) &=\sum_{j=1}^{N}A_{j}e^{ipy}\Theta(x_{j-1}<y<x_{j})] \nonumber \\
f_{\down}(y) &=\sum_{j=1}^{N}B_{j}e^{-ipy}\Theta(x_{j-1}<y<x_{j})]
\end{align*}
where $p = E/v$ and where
\begin{equation*}
A_{j+1}=A_{j}-\lambda_{j}k e^{-ipx_{j}}, \quad B_{j+1}=B_{j}-\lambda_{j}k e^{ipx_{j}}
\end{equation*}
Here, $\Theta(a < y < b)$ is defined to take the value $1$ if $y$ is in the interval $[a,b]$ and
$0$ otherwise. Also, we use a notation in which $x_{0}$ is identified with $x_N$. Eliminating $\lambda_j$, we derive
\begin{equation}
(A_{j+1} - A_{j}) e^{ipx_j} = (B_{j+1} - B_{j}) e^{-ipx_j}
\label{constraintABfm1}
\end{equation}

We still have to impose the condition $[a,C_j] = 0$, which gives an additional set of constraints on $\{A_j, B_j\}$. As 
in the single impurity case, we regularize the cosine terms to derive these constraints. That is, we replace
$C = k(\phi_\up(x_j)+\phi_\down(x_j))$ with
\begin{equation}
C \rightarrow \int_{-L/2}^{L/2} k ( \phi_\up(x) + \phi_\down(x)) \tilde{\delta}(x-x_j) dx
\label{Cregmi}
\end{equation}
where $\tilde{\delta}$ is a narrowly peaked function with $\int \tilde{\delta}(x) dx =1$, i.e., an approximation to a 
delta function. With this regularization, it is not hard to show that $[a,C_j] = 0$ gives the constraint 
\begin{equation}
\frac{1}{2}(A_{j} + A_{j+1}) e^{ipx_j} = \frac{1}{2}(B_{j} + B_{j+1}) e^{-ipx_j}
\label{constraintABfm2}
\end{equation}
Combining (\ref{constraintABfm1}), (\ref{constraintABfm2}) we derive
\begin{align}
A_j e^{ipx_j} = B_j e^{-ipx_j}, \quad A_{j+1} e^{ipx_j} = B_{j+1} e^{-ipx_j}
\label{constraintABfm3}
\end{align}

Our task is now to find all $\{A_j, B_j,p\}$ that satisfy (\ref{constraintABfm3}). For simplicity, we will specialize
to the case where the impurities are uniformly spaced with spacing $s$, i.e. $x_{j+1} - x_j = s = L/N$ for all $j$.
In this case, equation (\ref{constraintABfm3}) implies that $e^{2ips} = 1$,
so that $p$ is quantized in integer multiples of $\pi/s$. For any such $p$, (\ref{constraintABfm3}) has $N$ linearly independent solutions
of the form
\begin{align*}
B_j &= A_j = 0 \ \ \text{for } j \neq m \\
B_j &= A_j e^{2i p x_j} \neq 0 \ \ \text{for } j =  m 
\end{align*}
with $m=1,...,N$. 
Putting this all together, we see that the most general possible creation/annihilation operator for $H_{\text{eff}}$ is given by
\begin{align*}
a_{pm} = \sqrt{\frac{k}{4\pi|p|s}} \int_{-L/2}^{L/2} &[(e^{ipy}\partial_y \phi_\up + e^{2ip x_m} e^{-ipy} \partial_y \phi_\down) \\
         & \cdot \Theta(x_{m-1}<y<x_{m})]dy
\end{align*}
with $E_{pm} = vp$. Here the index $m$ runs over $m=1,...,N$ while $p$ takes values $\pm \pi/s, \pm 2\pi/s,...$.
(As in the single impurity case, $p=0$ does not correspond to a legitimate creation/annihilation operator, since 
we require that $E \neq 0$).

Following the conventions from \textcolor{black}{section \ref{diagsummsect}}, 
we will refer to the operators with $E > 0$ --- or equivalently $p > 0$ ---
as ``annihilation operators'' and the other operators as ``creation operators.'' Note that we have normalized the 
$a$ operators so that $[a_{pm}, a_{p'm'}^\dagger] = \delta_{pp'} \delta_{mm'}$ for $p,p' > 0$.

The next step is to compute the commutator matrix $\mathcal{Z}_{ij} = \frac{1}{2\pi i} [C_i, C_j]$. 
Let us denote the $N+2$ cosine terms as $\{\cos(C_1),...,\cos(C_{N+2})\}$ where 
$C_{N+1} = 2\pi Q_\up$, $C_{N+2} = 2\pi Q_\down$. Using (\ref{genphicommrel}) we find that
 $\mathcal{Z}_{ij}$ takes the form
\begin{eqnarray*}
\mathcal{Z}_{ij}
&=& \bpm 0 & \cdots & 0 & 1 & -1 \\ \vdots & \vdots & \vdots & \vdots & \vdots \\ 0 & \cdots & 0 & 1 & -1  \\
-1 & \cdots & -1 & 0 & 0 \\ 1 & \cdots & 1 & 0 & 0 \epm
\end{eqnarray*}

\textcolor{black}{To proceed further we need to find an appropriate change of variables of the form $C_i' = \sum_{j=1}^{N+2} \mathcal{V}_{ij} C_j + \chi_i$.
Here, $\mathcal V$ should be chosen so that $ \mathcal{Z}_{ij}' = \frac{1}{2\pi i} [C_i', C_j']$ is in skew-normal form, while $\chi$ 
should be chosen so that it obeys Eq. \ref{chicond}.}
It is easy to see that the following change of variables does the job:
\begin{align*}
&C_1' = C_1, \quad C_2' = -2\pi Q_\up, \quad C_3' = 2\pi Q_\up + 2\pi Q_\down \\
&C_m' = C_{m-2} - C_{m-3}, \quad m= 4,...,N+2 
\end{align*}
Indeed, it is easy to check that 
\begin{eqnarray*}
\mathcal{Z}_{ij}' &=& \frac{1}{2\pi i} [C_i', C_j'] \nonumber \\
&=& \bpm 0 & -1 & 0 & \cdots & 0 \\ 1 & 0 & 0 & \cdots & 0 \\ 0 & 0 & 0 & \cdots & 0 \\ \vdots & \vdots & \vdots & \vdots & \vdots \\
0 & 0 & 0 & \cdots & 0 \epm
\end{eqnarray*}
We can see that this is in the canonical skew-normal form shown in Eq. \ref{zprime}, with the parameters $M = N+2$, $I = 1$, $d_1 = 1$.

We are now in a position to write down the low energy effective Hamiltonian $H_{\text{eff}}$: according to Eq. \ref{Heffdiagsumm},
$H_{\text{eff}}$ must take the form
\begin{equation}
H_{\text{eff}} = \sum_{m=1}^{N} \sum_{p > 0} v p a_{pm}^\dagger a_{pm} + F(C_3',..., C_{N+2}')
\label{Hefffm}
\end{equation}
where the sum runs over $p = \pi/s, 2\pi/s,...$ and
where $F$ is some quadratic function of $N$ variables. To determine $F$, we first need to work out more concrete expressions for 
$C_m'$. 
The $m=3$ case is simple: $C_3' = 2\pi Q$. On the other hand, for $m = 4,...,N+2$, we have
\begin{eqnarray*}
C_m' &=& k(\phi_\up(x_{m-2}) + \phi_\down(x_{m-2})) \\
&-& k(\phi_\up(x_{m-3}) + \phi_\down(x_{m-3}) \\
&=& k \int_{x_{m-3}}^{x_{m-2}} (\partial_y \phi_\up + \partial_y \phi_\down) dy \\
\end{eqnarray*}
where the second line follows from the definition of $\phi_\up, \phi_\down$ (\ref{phidef}) along with
the assumption that the impurities are arranged in the order $y_0 < x_1 < ... < x_N$ in the clockwise direction.

With these expressions we can now find $F$. We use the same trick as in the single impurity case: 
we note that the first term in Eq. \ref{Hefffm} can be rewritten as as $\sum_m \sum_{p \neq 0} \frac{v |p|}{2} (a_{-pm} a_{pm})$,
and we observe that
\begin{equation*}
\lim_{p \rightarrow 0} \frac{v|p|}{2} a_{-pm} a_{pm} =  \frac{v}{8 \pi k s} \cdot (C_{m+2}')^2 
\end{equation*}
for $m=2,...,N$, while
\begin{equation*}
\lim_{p \rightarrow 0} \frac{v|p|}{2} a_{-p1} a_{p1} =  
\frac{v}{8 \pi k s} \cdot (k C_3' - \sum_{m=4}^{N+2} C_{m'})^2
\end{equation*}
Assuming that $F$ reproduces the missing $p=0$ piece of the first term in Eq. \ref{Hefffm}, we deduce that
\begin{eqnarray}
F(C_3',...,C_{N+2}') &=&  \frac{v}{8 \pi k s} \cdot \sum_{m=4}^{N+2} (C_m')^2 \nonumber \\
&+& \frac{v}{8 \pi k s} \cdot (k C_3' - \sum_{m=4}^{N+2} C_{m}')^2 
\end{eqnarray}

In addition to the effective Hamiltonian, we also need to discuss the effective Hilbert space $\mathcal{H}_{\text{eff}}$. 
Applying the results of \textcolor{black}{section \ref{diagsummsect}}, we see that $\mathcal{H}_{\text{eff}}$ is spanned by
states $\{|\v{q},\{n_{pm}\}\>\}$ where $|\v{q},\{n_{pm}\}\>$ is the unique simultaneous eigenstate of the form
\begin{align*}
e^{iC_1'} |\v{q},\{n_{pm}\}\> &= |\v{q},\{n_{pm}\}\>, \\
e^{iC_2'} |\v{q},\{n_{pm}\}\> &= |\v{q},\{n_{pm}\}\>, \\
C_i' |\v{q},\{n_{pm}\}\> &= 2\pi q_{i-2} |\v{q},\{n_{pm}\}\>, \ i = 3,...,N+2\\
a_{pm}^\dagger a_{pm} |\v{q},\{n_{pm}\}\> &= n_{pm} |\v{q},\{n_{pm}\}\>
\end{align*}
Here $n_{pm}$ runs over non-negative integers, while $\v{q}$ is an $N$ component vector, $\v{q} = (q_1,...,q_N)$ where each component $q_i$ runs over all integers.
As in the single impurity case, we do not need to label the $\{|\v{q},\{n_{pm}\}\>\}$ basis states with $\alpha$ quantum numbers 
since $d_1 = 1$ and thus there is no degeneracy.

Now that we have derived the effective theory, all that remains is to diagonalize it. To do this, we note that
the $\{|\v{q},\{n_{pm}\}\>\}$ basis states are also eigenstates of $H_{\text{eff}}$ with energies given by
\begin{eqnarray}
E = \sum_{m=1}^N \sum_{p > 0} v p n_{pm} &+& \frac{\pi v}{2 k s} \sum_{m=2}^{N} q_m^2 \label{Efm} \\
&+& \frac{\pi v}{2 k s}  (k q_1 - q_2 -...-q_N)^2 \nonumber
\end{eqnarray}
The above equation gives the complete low energy spectrum of $H$ in the limit $U \rightarrow \infty$.

Let us now discuss the physical interpretation of these results. As in the single impurity case, when 
$U \rightarrow \infty$ the impurities generate perfect backscattering, effectively reconnecting the edge modes.
The result, as shown in Fig. \ref{fig:mi}b, is the formation of $N$ disconnected chiral modes 
living in the $N$ intervals, $[x_N,x_1], [x_1, x_2], ...,[x_{N-1}, x_N]$. 

With this picture in mind, the $n_{pm}$ quantum numbers have a natural interpretation as the 
number of phonon excitations with momentum $p$ on the $m$th disconnected component of the edge. Likewise, if
we examine the definition of $q_m$, we can see that $q_m/k$ is equal to the total charge in the $m$th 
component of the edge, i.e. the total charge in the interval $[x_{m-1},x_m]$, for $m=2,...,N$.
On the other hand, the quantum number $q_1$ is slightly different: $q_1$ is equal to the total charge on
the entire boundary of the disk $[-L/2, L/2]$. Note that since $q_m$ is quantized to be an integer for all $m=1,...,N$, it follows
that the charge in each interval $[x_{m-1},x_m]$ is quantized in integer multiples of $1/k$ while the total charge on the whole
edge is quantized as an \emph{integer}. These quantization laws are physically sensible: indeed, the
fractional quantum spin Hall state supports quasiparticle excitations with charge $1/k$, so it makes sense that disconnected
components of the edge can carry such charge, but at the same time we also know that the \emph{total} charge on 
the boundary must be an integer.

Putting this all together, we see that the first term in (\ref{Efm}) can be interpreted as the energy of the phonon 
excitations, summed over all momenta and all disconnected components of the edge. Similarly the second term can be interpreted
as the charging energy of the disconnected components labeled by $m=2,...,N$, while the third term can be interpreted as the
charging energy of the first component labeled by $m=1$.

So far in this section we have considered magnetic impurities which backscatter spin-up electrons into spin-down electrons. These impurities explicitly break time reversal symmetry.
However, one can also consider non-magnetic impurities which preserve time reversal symmetry and backscatter \emph{pairs} of spin-up electrons into pairs of spin-down electrons. When the scattering strength $U$ is sufficiently strong these impurities can cause a \emph{spontaneous} breaking of time reversal symmetry, leading to a two-fold degenerate ground state.~\cite{XuMoore,WuBernevigZhang,LevinStern} This physics can also be captured by an appropriate toy model and we provide an example in Appendix \ref{ssb}.

\subsection{Example 3: Multiple magnetic and superconducting impurities}
\label{fmscsect}
We now consider a fractional quantum spin Hall edge in a disk geometry of circumference $L$
with $2N$ alternating magnetic and superconducting impurities. We take the magnetic impurities
to be located at positions $x_1,x_3,...,x_{2N-1}$ while the superconducting impurities are located
at positions $x_2,x_4,...,x_{2N}$ (Fig. \ref{fig:fmscmi}a).  
We assume that the magnetic impurities generate a backscattering term of the form
$\frac{U}{2}(\psi_\up^\dagger(0) \psi_\down(0) + h.c)$, while the superconducting impurities
generate a pairing term of the form $\frac{U}{2}(\psi_\up^\dagger(0) \psi_\down^\dagger(0) + h.c)$.
The Hamiltonian is then
\begin{align}
&H=H_0-U\sum_{i=1}^{2N}\cos(C_i) \\
&C_{i}=k (\phi_{\up}(x_i)+(-1)^{i+1} \phi_{\down}(x_i))
\la{Hfmsc}
\end{align}
where $H_0$ is defined in Eq. \ref{Hclean}.

As in the previous cases, our goal is to understand the low energy physics of $H$ in the limit $U \rightarrow \infty$.
As before, we take account of the compactness of $\phi_\up, \phi_\down$ and the discrete nature of
$Q_\up, Q_\down$ by adding two additional cosine terms to our Hamiltonian:
\begin{equation}
H = H_0 - U\sum_{i=1}^{2N}\cos(C_i) - U \cos(2\pi Q_\up) - U \cos(2\pi Q_\down)
\end{equation}

\begin{figure}[tb]
  \centering
\includegraphics[width=8cm,height=4.2cm]{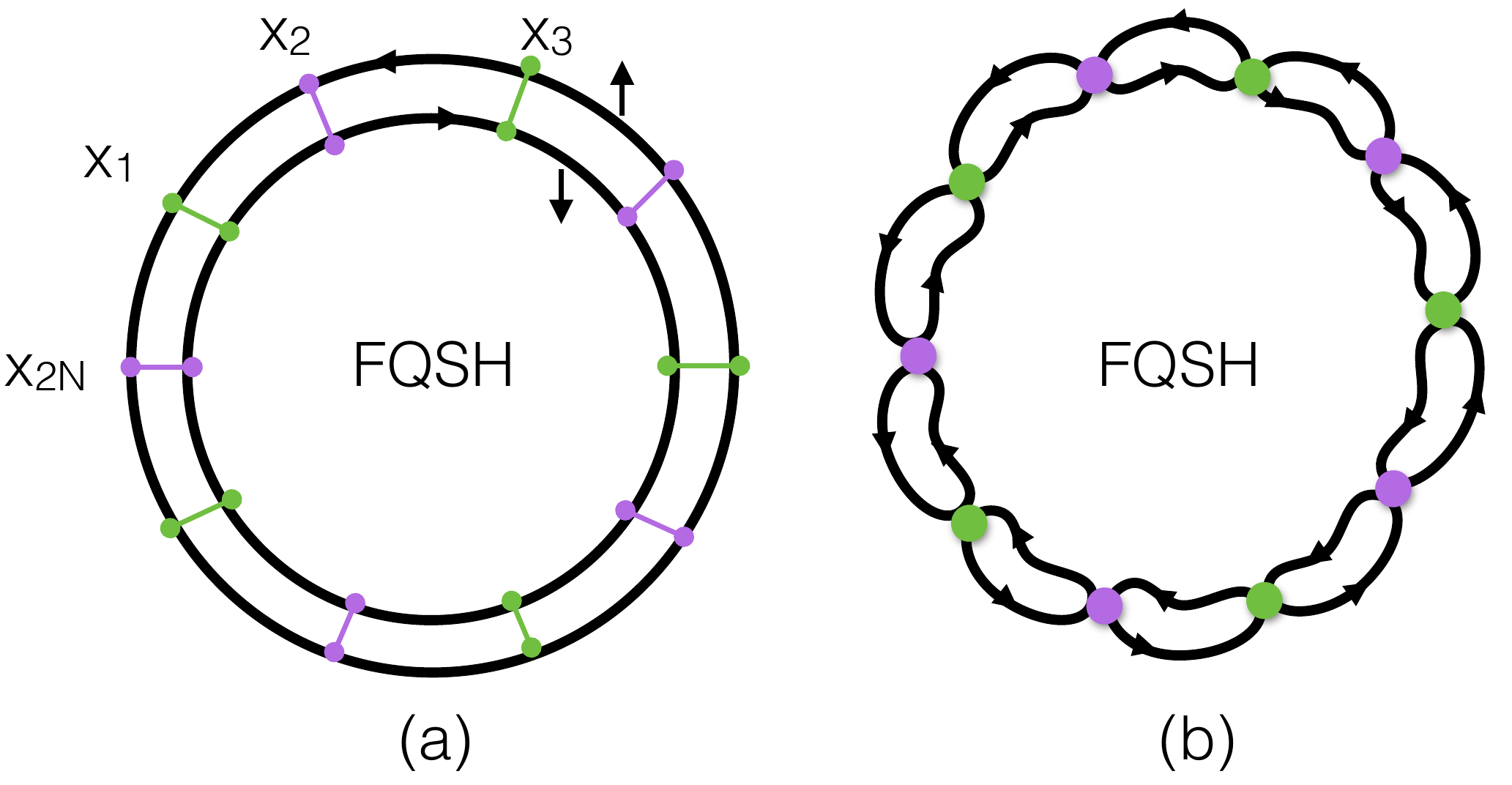}
    \caption{
(a) A collection of $2N$ alternating magnetic and superconducting impurities on a fractional quantum spin Hall edge.
The magnetic impurities are located at positions $x_1,x_3,....,x_{2N-1}$ while the superconducting impurities
are located at positions $x_2,x_4,...,x_{2N}$. The magnetic impurities scatter spin-up electrons into spin-down
electrons while the superconducting impurities scatter spin-up electrons into spin-down holes.
(b) In the infinite $U$
limit, the impurities effectively reconnect the edge modes, breaking the edge into $2N$ disconnected components.}

  \la{fig:fmscmi}
\end{figure}

Next, we find the creation and annihilation operators for $H_{\text{eff}}$ using Eq. \ref{auxeqsumm}. That is, we search for all
operators $a$ such that (1) $a$ is a linear combination of our fundamental phase
space operators $\{\partial_y \phi_\up, \partial_y \phi_\down, \phi_\up(y_0), \phi_\down(y_0)\}$ and (2) $a$ obeys
\begin{align}
[a, H_{0}] &= E a + \sum_{i=1}^{2N} \lambda_i [C_i, H_0] + \lambda_\up [Q_\up, H_0] + \lambda_\down [Q_\down, H_0] \nonumber \\
[a, C_j] &= [a,Q_\up] = [a,Q_\down] = 0
\label{auxeqscfm}
\end{align}
for some $E,\textcolor{black}{\lambda_i}, \lambda_\up, \lambda_\down$ with $E \neq 0$.

As before, since $[a, Q_\up] = [a, Q_\down] = 0$, it follows that $\phi_\up(y_0), \phi_\down(y_0)$ cannot appear in the expression
for $a$. Hence, $a$ can be written in the general form
\begin{equation*}
a=\int_{-L/2}^{L/2} [f_{\up}(y)\partial_{y}\phi_{\up}(y)+f_{\down}(y)\partial_{y}\phi_{\down}(y)] dy
\end{equation*}
Substituting this expression into the first line of Eq.~\ref{auxeqscfm}, we obtain
\begin{align*}
-ivf_{\up}'(y) & = Ef_{\up}+kiv\sum_{j}\lambda_{j}\delta(y-x_{j})\\
ivf_{\down}'(y) & = Ef_{\down}-kiv\sum_{j}(-1)^{j+1}\lambda_{j}\delta(y-x_{j})
\end{align*}
Solving the above first order differential equation we get
\begin{align*}
f_{\up}(y) &=\sum_{j=1}^{N}A_{j}e^{ipy}\Theta(x_{j-1}<y<x_{j})] \nonumber \\
f_{\down}(y) &=\sum_{j=1}^{N}B_{j}e^{-ipy}\Theta(x_{j-1}<y<x_{j})]
\end{align*}
where $p = E/v$ and where
\begin{align*}
A_{j+1}=A_{j}-\lambda_{j}k e^{-ipx_{j}}, B_{j+1}=B_{j}-(-1)^{j+1}\lambda_{j}k e^{ipx_{j}} 
\end{align*}
Here, $\Theta(a < y < b)$ is defined to take the value $1$ if $y$ is in the interval $[a,b]$ and
$0$ otherwise. Also, we use a notation in which $x_0$ is identified with $x_N$. Eliminating $\lambda_j$, we derive
\begin{equation}
(A_{j+1} - A_{j}) e^{ipx_j} = (-1)^{j+1} (B_{j+1} - B_{j}) e^{-ipx_j}
\label{constraintABfmsc1}
\end{equation}

We still have to impose the requirement $[a,C_j] = 0$ and derive the corresponding constraint on $\{A_j,B_j\}$. As in 
the previous cases, the correct way to do this is to regularize the cosine terms, replacing
\begin{equation}
C_j \rightarrow \int_{-L/2}^{L/2} k ( \phi_\up(x) + (-1)^{j+1} \phi_\down(x)) \tilde{\delta}(x-x_j) dx
\label{Cregsc}
\end{equation}
where $\tilde{\delta}$ is a narrowly peaked function with $\int \tilde{\delta}(x) dx =1$, i.e., an approximation to a
delta function. With this regularization, it is not hard to show that $[a,C_j] = 0$ gives the constraint
\begin{equation}
\frac{1}{2}(A_{j} + A_{j+1}) e^{ipx_j} = \frac{1}{2} (-1)^{j+1} (B_{j} + B_{j+1}) e^{-ipx_j}
\label{constraintABfmsc2}
\end{equation}
Combining (\ref{constraintABfmsc1}), (\ref{constraintABfmsc2}) we derive
\begin{align}
A_j e^{ipx_j} &=  (-1)^{j+1} B_j e^{-ipx_j}, \nonumber \\
A_{j+1} e^{ipx_j} &=  (-1)^{j+1} B_{j+1} e^{-ipx_j}
\label{constraintABfmsc3}
\end{align}

Our task is now to find all $\{A_j, B_j,p\}$ that satisfy (\ref{constraintABfmsc3}). For simplicity, we will specialize
to the case where the impurities are uniformly spaced with spacing $s$, i.e. $x_{j+1} - x_j = s = L/N$ for all $j$.
In this case, equation (\ref{constraintABfm3}) implies that $e^{2ips} = -1$,
so that $p$ is quantized in half-odd-integer multiples of $\pi/s$. For any such $p$, (\ref{constraintABfm3}) has $N$ linearly independent solutions
of the form
\begin{align*}
B_j &= A_j = 0 \ \ \text{for } j \neq m \\
B_j &= (-1)^{j+1} A_j e^{2i p x_j} \neq 0 \ \ \text{for } j =  m
\end{align*}
with $m=1,...,2N$.

Putting this all together, we see that the most general possible creation/annihilation operator for $H_{\text{eff}}$ is given by
\begin{align*}
a_{pm} &= \int_{-L/2}^{L/2} [(e^{ipy}\partial_y \phi_\up + (-1)^{m+1} e^{2ip x_m} e^{-ipy} \partial_y \phi_\down) \\
         & \cdot \Theta(x_{m-1}<y<x_{m})]dy \cdot  \sqrt{\frac{k}{4\pi|p|s}} 
\end{align*}
with $E_{pm} = vp$. Here the index $m$ runs over $m=1,...,2N$ while $p$ takes values $\pm \pi/2s, \pm 3\pi/2s,...$.
Note that we have normalized the $a$ operators so that $[a_{pm}, a_{p'm'}^\dagger] = \delta_{pp'} \delta_{mm'}$ for $p,p' > 0$.

The next step is to compute the commutator matrix $\mathcal{Z}_{ij} = \frac{1}{2\pi i} [C_i,C_j]$. Let us
denote the $2N+2$ cosine terms as $\{\cos(C_1),...,\cos(C_{2N+2})\}$ where 
$C_{2N+1} = 2\pi Q_\up$, $C_{2N+2} = 2\pi Q_\down$. Using the commutation relations (\ref{genphicommrel}),
we find
\begin{eqnarray*}
\mathcal{Z}_{ij} 
&=& \bpm 0 & k & 0 & k & \cdots & 0 & k & 1 & -1 \\ 
	-k & 0 & k & 0 & \cdots & k & 0 & 1 & 1 \\
	 0 &-k & 0 & k & \cdots & 0 & k & 1 & -1 \\
	-k & 0 &-k & 0 & \cdots & k & 0 & 1 & 1 \\
	\vdots & \vdots & \vdots & \vdots & \vdots & \vdots & \vdots & \vdots & \vdots \\
	 0 & -k & 0 & -k & \cdots & 0 & k & 1 & -1 \\
	-k & 0 & -k & 0 & \cdots & -k & 0 & 1 & 1 \\
	-1 &-1 & -1 & -1 & \cdots &-1 &-1 & 0 & 0 \\
	1 & -1 &  1 & -1 & \cdots & 1 & -1 & 0 & 0 \epm
\end{eqnarray*}

\textcolor{black}{To proceed further we need to find an appropriate change of variables of the form $C_i' = \sum_{j=1}^{N+2} \mathcal{V}_{ij} C_j + \chi_i$.
Here, $\mathcal V$ should be chosen so that $ \mathcal{Z}_{ij}' = \frac{1}{2\pi i} [C_i', C_j']$ is in skew-normal form, while $\chi$ should be chosen so that it obeys Eq. \ref{chicond}.} 
It is easy to see that the following change of variables does the job:
\begin{align*}
&C_m' = C_{2m+1} - C_{2m-1}, \quad m= 1,...,N-1, \\
&C_N' = 2\pi Q_\up + 2\pi Q_\down, \quad C_{N+1}' = C_1, \\
&C_m' = C_{2m-2N-2} - C_{2N}, \quad m= N+2,...,2N\\
&C_{2N+1}' =  C_{2N}-C_1 -2\pi k Q_\up \textcolor{black}{+ \pi}, \quad C_{2N+2}' =  -2\pi Q_\up, 
\end{align*}
Indeed, it is easy to check that
\begin{eqnarray*}
\mathcal{Z}_{ij}' &=& \frac{1}{2\pi i} [C_i', C_j'] \nonumber \\
&=& \bpm 0_{N+1} & -\mathcal{D} \\
          \mathcal{D} & 0_{N+1} & \epm,
\end{eqnarray*}
where $\mathcal{D}$ is the $N+1$ dimensional diagonal matrix
\begin{equation*}
\mathcal{D} = \bpm 2k & 0 & \cdots & 0 & 0 \\
	            0 & 2k & \cdots & 0 & 0 \\
		  \vdots & \vdots & \vdots & \vdots & \vdots \\
		    0 & 0 & \cdots & 2 & 0 \\
		    0 & 0 & \cdots & 0 & 1 \epm
\end{equation*}
We can see that this is in the canonical skew-normal form shown in Eq. \ref{zprime}, with the parameters $M = 2N+2$, $I = N+1$ and
\begin{equation*} 
d_1 = ... = d_{N-1} = 2k, \quad d_{N} = 2, \quad d_{N+1} = 1
\end{equation*}

With these results we can write down the low energy effective Hamiltonian $H_{\text{eff}}$: according to Eq. \ref{Heffdiagsumm},
$H_{\text{eff}}$ must take the form
\begin{equation}
H_{\text{eff}} = \sum_{m=1}^{2N} \sum_{p > 0} v p a_{pm}^\dagger a_{pm} 
\label{Hefffmsc}
\end{equation}
where the sum runs over $p = \pi/2s, 3\pi/2s,...$. Notice that $H_{\text{eff}}$ does not include a term of the form $F(C_{2I+1}',...,C_M')$
which was present in the previous examples. The reason that this term is not present is that $M = 2I$ in this case --- that is, none of the
$C_i'$ terms commute with all the other $C_j'$. This is closely related to the fact that the momentum $p$ is quantized in half-odd integer multiples
of $\pi/2s$ so unlike the previous examples, we cannot construct an operator $a_{pm}$ with $p=0$ (sometimes called a ``zero mode'' operator).

Let us now discuss the effective Hilbert space $\mathcal{H}_{\text{eff}}$.
According to the results of \textcolor{black}{section \ref{diagsummsect}}, the effective Hilbert space $\mathcal{H}_{\text{eff}}$ is spanned by
states $\{|\v{\alpha},\{n_{pm}\}\>\}$ where $|\v{\alpha},\{n_{pm}\}\>$ is the unique simultaneous eigenstate of the form
\begin{align}
e^{iC_i'/2k} |\v{\alpha},\{n_{pm}\}\> &= e^{i\pi \alpha_{i}/k}|\v{\alpha},\{n_{pm}\}\>, \ i = 1,...,N-1 \nonumber \\
e^{iC_N'/2} |\v{\alpha},\{n_{pm}\}\> &= e^{i\pi \alpha_N}|\v{\alpha},\{n_{pm}\}\>, \nonumber \\
e^{iC_{N+1}'} |\v{\alpha},\{n_{pm}\}\> &= |\v{\alpha},\{n_{pm}\}\>, \nonumber \\
e^{iC_i'} |\v{\alpha},\{n_{pm}\}\> &= |\v{\alpha},\{n_{pm}\}\>, \ i =N+2,...,2N+2 \nonumber \\ 
a_{pm}^\dagger a_{pm} |\v{\alpha},\{n_{pm}\}\> &= n_{pm} |\v{\alpha},\{n_{pm}\}\> \label{quantnumdef3}
\end{align}
Here the label $n_{pm}$ runs over non-negative integers, while $\v{\alpha}$ is an abbreviation for the $N$ component integer vector 
$(\alpha_1,...,\alpha_{N})$ where $\alpha_N$ runs over two values $\{0,1\}$, and the other $\alpha_i$'s run over $\{0,1...,2k-1\}$.

As in the previous cases, we can easily diagonalize the effective theory: clearly the 
$\{|\v{\alpha},\{n_{pm}\}\>\}$ basis states are also eigenstates of $H_{\text{eff}}$ with energies given by
\begin{equation}
E = \sum_{m=1}^N \sum_{p > 0} v p n_{pm}
\label{Efmsc}
\end{equation}
The above equation gives the complete low energy spectrum of $H$ in the limit $U \rightarrow \infty$.

An important feature of the above energy spectrum (\ref{Efmsc}) is that the energy $E$ is independent of $\v{\alpha}$. It follows
that every state, including the ground state, has a degeneracy of
\begin{equation}
D = 2 \cdot (2k)^{N-1}
\label{degfmsc}
\end{equation}
since this is the number of different values that $\v{\alpha}$ ranges over.

We now discuss the physical meaning of this degeneracy. As in the previous examples, when $U \rightarrow \infty$,
the impurities reconnect the edge modes, breaking the edge up into $2N$ disconnected components associated with the intervals
$[x_{2N},x_1],[x_1, x_2],...,[x_{2N-1},x_{2N}]$ (Fig. \ref{fig:fmscmi}b). The $n_{pm}$ quantum numbers describe the number of
phonon excitations of momentum $p$ in the $m$th component of the edge. The $\v{\alpha}$ quantum numbers also have
a simple physical interpretation. Indeed, if we examine the definition of $\v{\alpha}$ (\ref{quantnumdef3}), we can see that 
for $i \neq N$, $e^{i \pi \alpha_i/k} = e^{i \pi q_i}$ where $q_i$ is the total charge in the interval $[x_{2i-1}, x_{2i+1}]$ while
$e^{i \pi \alpha_{N}} = e^{i \pi q}$ where $q$ is the total charge on the edge. Thus, for $i \neq N$, $\alpha_i/k$ is
the total charge in the interval $[x_{2i-1}, x_{2i+1}]$ modulo $2$ while $\alpha_N$ is the total charge on the edge 
modulo $2$. The quantum number $\alpha_N$ ranges over two possible values $\{0,1\}$ since the total charge 
on the edge must be an integer while the other $\alpha_i$'s range over $2k$ values $\{0,1,...,2k-1\}$ since the fractional quantum
spin Hall state supports excitations with charge $1/k$ and hence the charge in the interval 
$[x_{2i-1}, x_{2i+1}]$ can be any integer multiple of this elementary value.

It is interesting to compare our formula for the degeneracy (\ref{degfmsc}) to that of 
Refs.~\onlinecite{lindner2012fractionalizing, cheng2012superconducting, barkeshli2013twist,vaezi2013fractional, clarke2013exotic}. Those papers studied a 
closely related system consisting of a FQSH 
edge in proximity to alternating ferromagnetic and superconducting strips (Fig. \ref{fig:fmscmi2}a). The authors found that this related system has
a ground state degeneracy of $D = (2k)^{N-1}$, which agrees exactly with our result, since 
Refs.~\onlinecite{lindner2012fractionalizing, cheng2012superconducting, barkeshli2013twist,vaezi2013fractional, clarke2013exotic} did not include the
two-fold degeneracy associated with fermion parity. In fact, it is not surprising that the two systems share the same
degeneracy since one can tune from their system to our system by shrinking the size of the ferromagnetic and superconducting
strips while at the same time increasing the strength of the proximity coupling (see Fig. \ref{fig:fmscmi2}b-c).

Although our system shares the same degeneracy as the one studied in Refs.~\onlinecite{lindner2012fractionalizing, cheng2012superconducting, barkeshli2013twist,vaezi2013fractional, clarke2013exotic}, one should keep in mind that there is an important
difference between the two degeneracies: the degeneracy in Refs.~\onlinecite{lindner2012fractionalizing, cheng2012superconducting, barkeshli2013twist,vaezi2013fractional, clarke2013exotic} is 
topologically protected and cannot be split by any local perturbation, while our degeneracy is not protected and splits at any finite value of $U$, as we explain in the
next section. That being said, if we modify our model in a simple way, we \emph{can} capture the physics of a topologically protected degeneracy. In particular, the only modification
we would need to make is to replace each individual magnetic impurity with a long array of many magnetic impurities, and similarly we would replace each individual superconducting impurity with a long array of many superconducting impurities. After making this change, the degeneracy would remain nearly exact even at finite $U$, with a splitting which is exponentially small in the length of the arrays.

\begin{figure}[tb]
  \centering
\includegraphics[width=8.5cm,height=3.5cm]{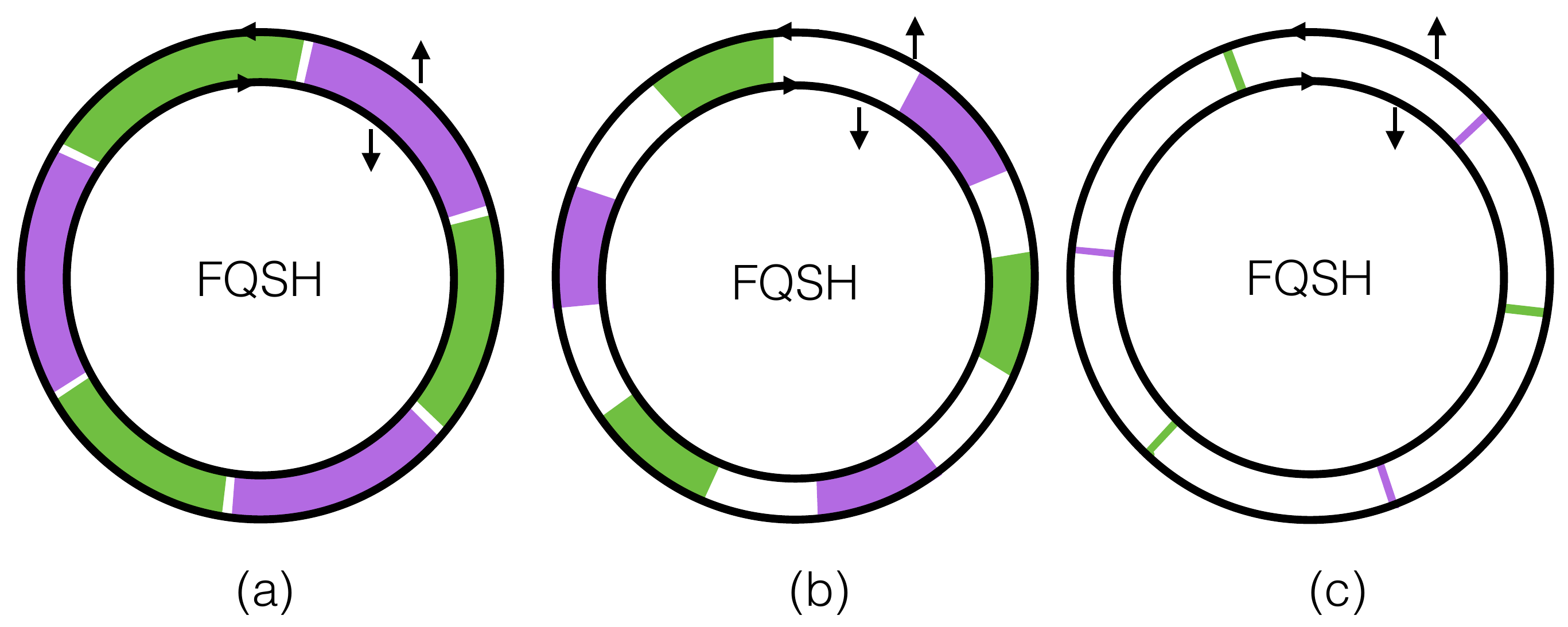}
    \caption{(a) A fractional quantum spin Hall edge in proximity to alternating ferromagnetic
and superconducting strips. (b-c) By shrinking the size of the ferromagnetic and superconducting strips while at 
the same time increasing the strength of the proximity coupling, we can continuously deform the system into a fractional quantum
spin Hall edge with magnetic and superconducting \emph{impurities}.}
  \la{fig:fmscmi2}
\end{figure}

\subsection{Finite \texorpdfstring{$U$}{U} corrections}
In the previous sections we analyzed the low energy physics of three different systems in the limit 
$U \rightarrow \infty$. In this section, we discuss how these results change when $U$ is large but \emph{finite}.

\subsubsection{Single magnetic impurity}
We begin with the simplest case: a single magnetic impurity on the edge of a $\nu = 1/k$ fractional quantum spin Hall
state. We wish to understand how finite $U$ corrections affect the low energy spectrum derived in section \ref{fmsisect}.

We follow the approach outlined in section \ref{finUsummsect}. According to this approach, the first step is to construct the 
operator $\Pi$ which is conjugate to the argument of the cosine term, $C = k (\phi_\up(0) + \phi_\down(0))$. To do this, we regularize 
$C$ as in equation (\ref{Cregsi}), replacing $C \rightarrow \int_{-L/2}^{L/2} k ( \phi_\up(x) + \phi_\down(x)) \tilde{\delta}(x) dx$. For concreteness,
we choose the regulated delta function $\tilde{\delta}$ to be 
\begin{equation*}
\tilde{\delta}(x) = \begin{cases} \frac{1}{b} & |x| \leq b/2 \\
				  0 & |x| > b/2 \end{cases}
\end{equation*}
With this regularization, we find:
\begin{align*}
[C, H_0] &=  \frac{kvi}{b} \int_{-b/2}^{b/2} \textcolor{black}{(\partial_x \phi_\up - \partial_x \phi_\down)} dx \nonumber \\
[C,[C,H_0]] &= -\frac{4\pi k v}{b} 
\end{align*}
so that
\begin{align*}
\mathcal{M} &= \frac{\pi b}{k v} \nonumber \\
\Pi &=  \frac{1}{2}  \int_{-b/2}^{b/2}\textcolor{black}{(\partial_x \phi_\up - \partial_x \phi_\down)} dx
\end{align*}

According to Eq. \ref{finUsumm}, the low energy theory at finite $U$ is obtained by adding terms to $H_{\text{eff}}$ (\ref{Heffsi2}) of the form
$\sum_{n= -\infty}^{n=\infty} e^{i n \Pi} \epsilon_n(\{a_p, a_p^\dagger\}, C_3')$. Here, the $\epsilon_n$ are some unknown functions whose precise
form cannot be determined without more calculation. We should mention that the $\epsilon_n$ functions also depend 
on $U$ --- in fact, $\epsilon_n \rightarrow 0$ as $U \rightarrow \infty$ --- but for notational simplicity we have chosen not to show this 
dependence explicitly. In what follows, instead of computing $\epsilon_n$, we take a more qualitative approach: we simply
assume that $\epsilon_n$ contains all combinations of $a_p, a_p^\dagger, C_3'$ that are not forbidden by locality or other
general principles, and we derive the consequences of this assumption.
 
The next step is to analyze the effect of the above terms on the low energy spectrum. This analysis depends on which parameter regime
one wishes to consider; here, we focus on the limit where $L \rightarrow \infty$, while $U$ is fixed but large. 
In this case, $H_{\text{eff}}$ (\ref{Heffsi2}) has a gapless spectrum, so we cannot use conventional perturbation theory to analyze the effect of the finite $U$ corrections; instead we 
need to use a renormalization group (RG) approach. This RG analysis has been carried out previously~\cite{kane1992transmission, kane1992transport} 
and we will not repeat it here. Instead, we merely summarize a few key results: first, one of the terms generated by finite $U$, namely 
$e^{i \Pi}$, is relevant for $k > 1$ and marginal for $k=1$. Second, the operator $e^{i \Pi}$
can be interpreted physically as describing a quasiparticle tunneling event where
a charge $1/k$ quasiparticle tunnels from one side of the impurity to the other. Third, this operator
drives the system from the $U =\infty$ fixed point to the $U=0$ fixed point. 

These results imply that when $k > 1$, for any finite $U$, the low energy spectrum in the thermodynamic limit 
$L \rightarrow \infty$ is always described by the $U=0$ theory $H_0$. Thus, in this case, the finite $U$ corrections have
an important effect on the low energy physics. We note that these conclusions are consistent with the RG analysis of magnetic impurities 
given in Ref.~\onlinecite{BeriCooper}.

\subsubsection{Multiple magnetic impurities}

We now move on to consider a system of $N$ equally spaced magnetic impurities on an edge of circumference $L$. As in the
single impurity case, the first step in understanding the finite $U$ corrections is to compute the $\Pi_i$ operators that
are conjugate to the $C_i$'s. Regularizing the cosine terms as in the previous case, a straightforward calculation gives
\begin{align*}
\mathcal{M}_{ij} &= \frac{\pi b}{k v} \delta_{ij} \nonumber \\
\Pi_i &=  \frac{1}{2} \int_{x_i-b/2}^{x_i+b/2}\textcolor{black}{(\partial_x \phi_\up - \partial_x \phi_\down)}dx
\end{align*}
where $i,j=1,...,N$. 
\footnote{We do not bother to compute the two remaining $\Pi_i$'s since 
we do not want to include finite $U$ corrections from the corresponding cosine terms 
\unexpanded{$\cos(2\pi Q_\uparrow)$} and 
\unexpanded{$\cos(2\pi Q_\downarrow)$}. Indeed, these terms were introduced as a mathematical trick and so their $U$ coefficients must 
be taken to infinity to obtain physical results.}

According to Eq. \ref{finUsumm}, the finite $U$ corrections contribute additional terms to $H_{\text{eff}}$ (\ref{Hefffm}) of the form
$\sum_{\v{n}} e^{i \sum_{j=1}^N n_j \Pi_j} \epsilon_{\v{n}}$ where $\v{n} =(n_1,...,n_N)$ is an $N$-component
integer vector. Here, $\epsilon_{\v{n}}$ is some unknown function of the operators $\{a_{pm}, a_{pm}^\dagger, C_m'\}$ 
which vanishes as $U \rightarrow \infty$.

We now discuss how the addition of these terms affects the low energy spectrum in two different parameter regimes. 
First we consider the limit where $L \rightarrow \infty$ with $U$ and $N$ fixed. 
This case is a simple generalization of the single impurity system discussed above, and it is easy to see 
that the same renormalization group analysis applies here. Thus, in this limit the finite $U$ corrections have a dramatic
effect for $k > 1$ and cause the low energy spectrum to revert back to the $U=0$ system for any finite value of $U$, no 
matter how large. 

The second parameter regime that we consider is where $L, N \rightarrow \infty$ with $U$ 
and $L/N$ fixed. The case is different from the previous one because $H_{\text{eff}}$ (\ref{Hefffm}) has a finite \emph{energy gap} in this limit
(of order $v/s$ where $s = L/N$). Furthermore, $H_{\text{eff}}$ has a unique ground state. These two properties are stable to
small perturbations, so we conclude that the system will continue to have a unique ground state and an energy gap for finite 
but sufficiently large $U$. 

The presence of this energy gap at large $U$ is not surprising. Indeed, in the above limit, our system can be thought of as a toy model 
for a fractional quantum spin Hall edge that is proximity coupled to a ferromagnetic strip. It is well known that a ferromagnet can 
open up an energy gap at the edge if the coupling is sufficiently strong\cite{LevinStern, BeriCooper}, which is exactly what we have found here.

\subsubsection{Multiple magnetic and superconducting impurities}
\label{finUscsect}
Finally, let us discuss a system of $2N$ equally spaced alternating magnetic and superconducting impurities on an edge
of circumference $L$.  As in the previous cases, the first step in understanding the finite $U$ corrections is to compute 
the $\Pi_i$ operators that are conjugate to the $C_i$'s. Regularizing the cosine terms as in the previous cases, a straightforward calculation gives
\begin{eqnarray*}
\mathcal{M}_{ij} &=& \frac{\pi b}{k v} \delta_{ij} \nonumber \\
\Pi_i &=&  \frac{1}{2}
\int_{x_i-b/2}^{x_i+b/2}\textcolor{black}{(\partial_x \phi_\up - (-1)^{i+1} \partial_x \phi_\down)} dx
\end{eqnarray*}
where $i,j=1,...,2N$. 

As a first step towards understanding the finite $U$ corrections, we consider a scenario in which only one of the impurities/cosine
terms has a finite coupling constant $U$, while the others have a coupling constant which is infinitely large. 
This scenario is easy to analyze because we only have to 
include the corrections associated with a \emph{single} impurity. For concreteness, we assume that the impurity 
in question is superconducting rather than magnetic and we label the corresponding cosine term by $\cos(C_{2j})$ 
(in our notation the superconducting impurities are labeled by even integers).
Having made these choices, we can immediately write down the finite $U$ corrections:
according to Eq. \ref{finUsumm}, these corrections take the form
\begin{equation*}
\textcolor{black}{\sum_{n=-\infty}^\infty} e^{i n \Pi_{2j}} \epsilon_{n}(\{a_{pm},a_{pm}^\dagger\}), 
\end{equation*}
where the $\epsilon_n$ are some unknown functions which vanish as $U \rightarrow \infty$.

Our next task is to understand how these corrections affect the low energy spectrum. The answer to this question depends
on which parameter regime one wishes to study: here we will focus on the regime where $L,N \rightarrow \infty$ with $L/N$ 
and $U$ fixed. In this limit, $H_{\text{eff}}$ (\ref{Hefffmsc}) has a finite energy gap of order $v/s$ where $s = L/N$. At the same time, the ground state is highly
degenerate: in fact, the degeneracy is exponentially large in the system size, growing as $D = 2\cdot(2k)^{N-2}$.
Given this energy spectrum, it follows that at the lowest energy scales, the only effect of the finite $U$ 
corrections is to split the ground state degeneracy. 

To analyze this splitting, we need to compute the matrix elements of the finite $U$ corrections between different 
ground states and then diagonalize the resulting $D \times D$ matrix. Our strategy will be to use the identity 
(\ref{finUidsumm}) which relates the matrix elements of the finite $U$ corrections to the matrix elements of $e^{ni\Gamma_{2j}}$.
Following this approach, the first step in our calculation is to compute $\Gamma_{2j}$. Using the definition (\ref{Gammadef}), we find
\begin{equation*}
\Gamma_{2j} = \frac{1}{2k} (C_{2j+1} - C_{2j-1})
\end{equation*}
assuming $j \neq N$. (The case $j = N$ is slightly more complicated due to our conventions for describing 
the periodic boundary conditions at the edge, so we will assume $j \neq N$ in what follows).

The next step is to find the matrix elements of the operator $e^{in \Gamma_{2j}}$ between different ground states. 
To this end, we rewrite $\Gamma_{2j}$ in terms of the $C_i'$ operators:
$\Gamma_{2j} = \frac{C_j'}{2k}$. The matrix elements of $e^{in \Gamma_{2j}}$ can now be computed straightforwardly 
using the known matrix elements of $C_j'$ (see Eqs. \ref{eciprimemat1}-\ref{eciprimemat2}):
\begin{equation}
\<\v{\alpha}'|e^{in\Gamma_{2j}}|\v{\alpha}\> = e^{\frac{\pi in\alpha_j}{k}} \delta_{\v{\alpha}' \v{\alpha}} 
\label{Gamma2j}
\end{equation}
where $|\v{\alpha}\>$ denotes the ground state $|\v{\alpha}\> \equiv |\v{\alpha},\v{n} = 0\>$.

At this point we apply the identity (\ref{finUidsumm}) which states that the matrix elements of 
$e^{i n \Pi_{2j}} \epsilon_{n}(\{a_{pm},a_{pm}^\dagger\})$ are equal to the matrix elements of
$u_n \cdot e^{i n \Gamma_{2j}}$ where $u_n$ is some unknown proportionality constant. Using this identity together
with (\ref{Gamma2j}), we conclude that the matrix elements of the finite $U$ corrections are given by
\begin{equation*}
f \left(\frac{\alpha_j}{k}\right) \delta_{\v{\alpha}' \v{\alpha}}
\end{equation*}
where $f(x) = \sum_{n= -\infty}^{\infty} u_n e^{\pi inx}$. 

We are now in position to determine the splitting of the ground states. To do this, we note that
while we don't know the values of the $u_n$ constants and therefore we don't know $f(x)$, we expect that generically the function $f$ will have
a unique minimum for $\alpha_j \in \{0,1,...,2k-1\}$. Assuming this is the case, we conclude that the finite $U$ corrections
favor a particular value of $\alpha_j$, say $\alpha_j = 0$. Thus these corrections reduce the ground state degeneracy from 
$D$ to $D/2k$.

So far we have analyzed the case where one of the superconducting impurities is characterized by a finite coupling
constant $U$ while the other impurities are at infinite coupling. Next, suppose that
\emph{all} the superconducting impurities have finite $U$ while all the magnetic impurities have infinite $U$. In this
case, similar analysis as above shows that the matrix elements of the finite $U$ corrections are of the following 
form:
\footnote{In fact, in deriving this expression we used one more piece of information in addition to the general considerations discussed
above, namely that the superconducting impurities are completely disconnected from one another by 
the magnetic impurities and therefore the finite $U$ corrections do not generate any ``interaction'' terms, 
e.g. of the form $g(\alpha_i, \alpha_j) \delta_{\boldsymbol{\alpha}' \boldsymbol{\alpha}}$.}
\begin{equation}
\left[\sum_{j=1}^{N-1} f\left(\frac{\alpha_j}{k} \right) 
+ f \left(\alpha_N-\frac{1}{k} \sum_{j=1}^{N-1} \alpha_j \right) \right] \delta_{\v{\alpha}' \v{\alpha}}
\label{finUmatelt}
\end{equation}
To determine the splitting of the ground states, we need to understand the eigenvalue spectrum of the above matrix. Let us
assume that $f$ has a unique minimum at some $\alpha_j = q$ --- which is what we expect generically. Then, as long as the system size 
is commensurate with this value in the sense that $N q$ is a multiple of $k$, we can see that the above matrix has a unique
ground state $|\v{\alpha}\>$ with $\alpha_1 = ... = \alpha_{N-1} = q$ and $\alpha_N = Nq/k$ (mod $2$). Furthermore,
this ground state is separated from the lowest excited states by a finite gap, which is set 
by the function $f$. Thus, in this case, the finite $U$ corrections completely split the ground state degeneracy leading to 
a unique ground state with an energy gap.

Likewise, we can consider the opposite scenario where the magnetic impurities have finite $U$ while the superconducting
impurities have infinite $U$. Again, similar analysis shows that the corrections favor a unique ground state which 
is separated from the excited states by a finite gap. The main difference from the previous case is that
the matrix elements of the finite $U$ corrections are \emph{off-diagonal} in the $|\v{\alpha}\>$ basis so the ground state
is a superposition of many different $|\v{\alpha}\>$ states. 

To complete the discussion, let us consider the case where \emph{all} the impurities, magnetic and superconducting,
have finite $U$. If the magnetic impurities are at much stronger coupling
than the superconducting impurities or vice-versa then presumably the finite $U$ corrections drive the system to one
of the two gapped phases discussed above. On the other hand, if the two types of impurities have comparable values of 
$U$, then the low energy physics is more delicate since the finite
$U$ corrections associated with the two types of impurities do not commute with one other, i.e.
$[e^{i\Gamma_{2j}}, e^{i\Gamma_{2j \pm 1}}] \neq 0$. In this case, a more quantitative analysis is required to determine the fate of the
low energy spectrum.

\section{Conclusion}

In this paper we have presented a general recipe for computing the low energy spectrum of 
Hamiltonians of the form (\ref{genHamc}) in the limit $U \rightarrow \infty$. This
recipe is based on the construction of an effective Hamiltonian $H_{\text{eff}}$ and an effective
Hilbert space $\mathcal{H}_{\text{eff}}$ describing the low energy properties of our system in the
infinite $U$ limit. The key reason that our approach works is that this effective 
Hamiltonian is quadratic, so there is a simple procedure for diagonalizing it.

While our recipe gives exact results in the infinite $U$ limit, it provides only
approximate results when $U$ is finite; in order to obtain the exact spectrum in the finite $U$ case, we need to include 
additional (non-quadratic) terms in $H_{\text{eff}}$. As part of this work, we have discussed the general form of 
these finite $U$ corrections and how they scale with $U$. However, we have not discussed how
to actually compute these corrections. One direction for future research would be to develop
quantitative approaches for obtaining these corrections --- for example using the instanton approach outlined
in Ref.~\onlinecite{coleman1988aspects}. 

Some of the most promising directions for future work involve applications of our formalism to different
physical systems. In this paper, we have focused on the application to Abelian fractional quantum Hall
edges, but there are several other systems where our formalism could be useful. 
For example, it would be interesting to apply our methods to superconducting circuits --- quantum circuits
built out of inductors, capacitors, and Josephson junctions. In particular, several authors have identified
superconducting circuits with protected ground state degeneracies that could
be used as qubits.\cite{KitaevCurrentMirror,Gladchenko,BrooksKitaevPreskill,Dempster} The formalism developed here might be useful for finding other circuits with protected degeneracies.

\begin{acknowledgments}
We thank Chris Heinrich for stimulating discussions. SG gratefully acknowledges support by NSF-JQI-PFC and LPS-MPO-CMTC.
ML was supported in part by the NSF under grant No. DMR-1254741.
\end{acknowledgments}

\appendix

\section{Derivation of low energy effective theory}
\label{derivsect}
In this appendix, we derive an effective theory that describes the low energy spectrum of $H$ (\ref{genHamc})
in the limit $U \rightarrow \infty$. More specifically, we show that the low energy spectrum of $H$ in the infinite $U$ limit is described 
by the effective Hamiltonian $H_{\text{eff}}$ (\ref{Heffgenc}), which is defined within the effective Hilbert space 
$\mathcal{H}_{\text{eff}}$ (\ref{Hilbeff}). Before proving this result in generality, we first derive it for two illustrative examples in 
appendix \ref{Heffexsect} and a special case in appendix \ref{Heffgenxsect}. Finally, after this 
preparation, we work out the general case in appendix \ref{Heffgensect}.

\subsection{Two examples}
\label{Heffexsect}

\subsubsection{Harmonic oscillator with a cosine term}
\label{Heffex1sect}
To understand the basic ideas underlying the derivation, it is helpful to consider some simple examples. We start by studying 
a one dimensional harmonic oscillator with a cosine term:
\begin{equation}
H = \frac{p^2}{2m} + \frac{K x^2}{2}  - U \cos(2\pi x)
\end{equation}
In the following, we derive an effective Hamiltonian $H_{\text{eff}}$ and effective Hilbert space
$\mathcal{H}_{\text{eff}}$ that describe the low energy spectrum of $H$ in the limit $U \rightarrow \infty$. 

To begin, we decompose $H$ into two pieces, $H = H_1 + H_2$, where
\begin{align}
H_1 = \frac{p^2}{2m} - U \cos(2\pi x), \quad \quad H_2 = \frac{K x^2}{2}
\end{align}
Our strategy is as follows: first, we show that when $U$ is large, $H_1$ has a collection of nearly 
degenerate ground states which are separated from the lowest excited states by a large gap. Next we
argue that we can treat $H_2$ as a \emph{perturbation} which splits the ground state degeneracy
of $H_1$. Finally, using degenerate perturbation theory, we derive a low energy effective Hamiltonian 
for our system.

Following this plan, we start with the Hamiltonian $H_1$. This Hamiltonian describes a one dimensional particle 
moving in a cosine potential. The low energy physics of $H_1$ is especially simple when $U$ is large. In
this case, tunneling between the different cosine minima is 
suppressed so that $H_1$ has an infinite set of nearly degenerate ground states ---  one for each cosine minimum. 
We will label these states as $\{|\psi_q\>\}$ where $|\psi_q\>$ is localized around the minimum $x = q$ and 
$q=0, \pm 1, \pm 2,...$. 

We can estimate the energy gap of $H_1$ by expanding the cosine 
potential to quadratic order in $x$: $U \cos(2\pi x) \approx U (1 - 2 \pi^2 x^2)$. In this 
approximation, the cosine potential is equivalent to a harmonic oscillator with frequency 
$\omega = 2 \pi \sqrt{U/m}$. In particular, it follows that the ground states of $H_1$ are separated from 
the lowest excited states by an energy gap $\Delta$ of order $\Delta \sim \sqrt{U/m}$.

Now let us imagine adding $H_2$ to $H_1$. We would like to know how $H_2$ splits the degeneracy between 
the ground states $|\psi_q\>$. To answer this question, we will treat $H_2$ as a perturbation to 
$H_1$ and then we will compute the associated energy splitting using degenerate perturbation 
theory. However, before we do this, we first need to check that such a perturbative approach is justified 
in the infinite $U$ limit. To this end, we need to estimate the size of the matrix element $\<\psi_{ex}|H_2|\psi_q\>$ 
where $|\psi_q\>$ is an arbitrary ground state and $|\psi_{ex}\>$ is an arbitrary excited state of $H_1$. 
For large $U$ we can approximate $|\psi_q\>$ by a harmonic oscillator ground state centered at position $x=q$. 
Similarly, we can approximate $|\psi_{ex}\>$ by the $n$th excited state of the harmonic oscillator centered 
at $x=q$. Within this approximation, the matrix element $\<\psi_{ex}|H_2|\psi_q\>$ reduces to
\begin{equation}
\<\psi_{ex} |H_2 |\psi_q\> \approx \<n| K(x+q)^2 |0\>
\label{H2matapprox}
\end{equation} 
where $|0\>$ and $|n\>$ are the ground state and $n$th excited state of a harmonic oscillator with frequency 
$\omega = 2 \pi \sqrt{U/m}$, centered at $x=0$. The latter matrix element can be evaluated easily with the result
\begin{displaymath}
\<n| K(x+q)^2 |0\> \sim \begin{cases} \frac{Kq}{(mU)^{1/4}} & \mbox{if } n=1 \\
				      \frac{K}{\sqrt{mU}} & \mbox{if } n=2 \\
				      0 & \mbox{if } n \geq 3 \end{cases}
\end{displaymath}
Combining this expression with our formula for the energy gap $\Delta$, we obtain 
\begin{equation}
\sum_{|\psi_{ex}\>} \frac{|\<\psi_{ex}|H_2|\psi_q\>|^2}{\Delta} \sim \frac{K^2 q^2}{U}
\label{pterror}
\end{equation}
in the large $U$ limit.
This estimate is significant because the left hand side of (\ref{pterror}) 
is proportional to the \emph{second order} perturbative correction to the ground state energies. 
Evidently, this correction vanishes as $U \rightarrow \infty$, so we conclude that first order 
perturbation theory gives exact results in this limit.

With this justification, we now proceed with the perturbative calculation. According to first order degenerate 
perturbation theory, the energy splitting of the ground states can be determined by diagonalizing the matrix 
$\<\psi_{q'}| H_2 |\psi_q\>$. When $U$ is large, $|\psi_q\>$ can be approximated as a Gaussian wave function, 
centered at $x=q$. The width of this Gaussian is given by:
\begin{equation*}
(\Delta x)^2 \sim \frac{1}{m\omega} \sim \frac{1}{\sqrt{mU}}
\end{equation*}
We see that as $U \rightarrow \infty$, $(\Delta x)^2 \rightarrow 0$ so that $|\psi_q\>$ approaches a position eigenstate: 
$|\psi_q\> \rightarrow |x=q\>$. We conclude that the low energy spectrum of $H$ 
can be obtained by diagonalizing the matrix $\<q'| H_2 |q\>$. 

At this point, our calculation is essentially complete: the matrix elements $\<q'| H_2 |q\>$ define our low energy effective 
Hamiltonian, while the ground state subspace spanned by $\{|q\>\}$ defines our low energy Hilbert space. In other words, 
the low energy effective Hamiltonian is given by 
\begin{equation}
H_{\text{eff}} = H_2 = \frac{K x^2}{2}
\end{equation} 
while the low energy effective Hilbert space $\mathcal{H}_{\text{eff}}$ is the subspace spanned by position eigenstates $\{|q\>\}$ for 
$q = 0, \pm 1, \pm 2...$. Clearly this effective theory is valid for energies smaller than the gap of $H_1$, i.e. $E < 2\pi\sqrt{U/m}$.

Comparing these expressions with the effective Hamiltonian and Hilbert space from section \ref{effsummsect}, we see that they agree 
exactly. Thus we have successfully established equations (\ref{Heffgenc}) and (\ref{Hilbeff}) for this example.

\subsubsection{Harmonic oscillator with 2 cosine terms} \label{Heffex2sect}
Another important illustrative example is given by a one dimensional harmonic oscillator with \emph{two} cosine terms:
\begin{equation}
H = \frac{p^2}{2m} + \frac{K x^2}{2} - U \cos(d p) - U \cos(2\pi x)
\la{example2}
\end{equation}
Here, $d$ is a positive integer. 

Before analyzing $H$ we need to choose an appropriate basis in which to represent it. Because the arguments of the two cosine terms don't commute
with one another, neither the position basis nor the momentum basis are particularly convenient choices. Instead, we find it helpful to work in a third basis,
which consists of simultaneous eigenstates of the commuting operators $e^{i p}, e^{2\pi i x}$. We will denote these simultaneous eigenstates by
$|\theta, \varphi\>$ where
\begin{align}
e^{i p}|\theta, \varphi\> &= e^{i\theta} |\theta, \varphi\> \nonumber \\
e^{2\pi i x}|\theta, \varphi\> &= e^{i\varphi} |\theta, \varphi\>
\label{thetaphidef}
\end{align}
Here, the labels $\theta, \varphi$ take values in $0 \leq \theta, \varphi < 2\pi$. The explicit formula for
$|\theta, \varphi\>$ is
\begin{equation}
|\theta, \varphi\> = \frac{1}{\sqrt{2\pi}}\sum_{j} e^{i j \theta} |j + \frac{\varphi}{2\pi}\>
\end{equation}
where $|j + \frac{\varphi}{2\pi}\>$ denotes the position eigenstate at position $x = j + \frac{\varphi}{2\pi}$.

We now work out what the Hamiltonian $H$ looks like in the $\theta,\varphi$ representation. The first step is to express the $x,p$ operators in terms of $\theta, \varphi$. To this end, we observe that
\begin{eqnarray*}
e^{i a x} |\theta, \varphi\> &=& e^{i a \varphi/2\pi} |\theta+a, \varphi \> \nonumber \\
e^{i a p} |\theta, \varphi\> &=& |\theta, \varphi-2\pi a\>
\end{eqnarray*}
Differentiating these equations with respect to $a$, we derive
\begin{eqnarray*}
x |\theta, \varphi\> &=& \left(\frac{1}{i} \frac{\partial}{\partial \theta} + \frac{\varphi}{2\pi} \right) |\theta, \varphi\>  \nonumber \\
p |\theta, \varphi\> &=& - \frac{2\pi}{i} \frac{\partial}{\partial \varphi} |\theta, \varphi\>
\end{eqnarray*}
From these equations, we deduce that 
\begin{eqnarray*}
\<\theta, \varphi| x | \psi\> &=& \left(- \frac{1}{i}\frac{\partial}{\partial \theta} + \frac{\varphi}{2\pi} \right) \<\theta, \varphi|\psi\> \nonumber \\
\<\theta, \varphi| p | \psi\> &=&  \frac{2\pi}{i}\frac{\partial}{\partial \varphi} \<\theta, \varphi|\psi\> 
\end{eqnarray*}
for any state $|\psi\>$. We conclude that in the $\theta, \varphi$ representation, the operators $x, p$ take the form
\begin{align*}
x = - \frac{1}{i}\frac{\partial}{\partial \theta} + \frac{\varphi}{2\pi}, \quad p = \frac{2\pi}{i}\frac{\partial}{\partial \varphi}
\end{align*}
or equivalently
\begin{align}
x = -p_\theta +  \frac{\varphi}{2\pi}, \quad p = 2\pi p_\varphi
\label{prep}
\end{align}
where $p_\theta \equiv \frac{1}{i}\frac{\partial}{\partial \theta}$ and $p_\varphi \equiv \frac{1}{i}\frac{\partial}{\partial \varphi}$. 

The next step is to find expressions for $e^{2\pi i x}$ and $e^{ip}$ in terms of $\theta, \varphi$. One way to do this is to exponentiate (\ref{prep}):
\begin{equation*}
e^{2\pi i x} = e^{-2\pi i p_\theta +i\varphi}, \quad e^{ip} = e^{2\pi i p_\varphi}
\end{equation*}
We then make use of two operator identities which we will prove shortly:
\begin{equation}
e^{2\pi i p_\theta} = 1, \quad e^{2\pi i p_\varphi} =e^{i\theta}
\label{opid}
\end{equation}
With these identities, we can simplify the expressions for $e^{2\pi i x}$ and $e^{ip}$ to:
\begin{equation}
e^{2\pi i x} = e^{i\varphi}, \quad e^{ip} = e^{i\theta} 
\label{eiprep}
\end{equation}
(Alternatively, we could also have derived (\ref{eiprep}) directly from (\ref{thetaphidef})). The proof of the identities (\ref{opid}) relies on
the following observations:
\begin{align*}
e^{2\pi i p_\theta} |\theta, \varphi\> = |\theta-2\pi, \varphi\>, \quad e^{2\pi i p_\varphi}|\theta,\varphi\> = |\theta,\varphi-2\pi\>
\end{align*}
and 
\begin{align*}
|\theta-2\pi, \varphi\> = |\theta, \varphi\>, \quad |\theta,\varphi-2\pi\> = e^{i\theta} |\theta,\varphi\> 
\end{align*}
Putting these together, we deduce that
\begin{align*} 
e^{2\pi i p_\theta} |\theta, \varphi\> = |\theta, \varphi\>, \quad e^{2\pi i p_\varphi}|\theta,\varphi\> = e^{i\theta} |\theta,\varphi\>
\end{align*}
Since these relations hold for \emph{all} basis states, they imply the operator identities (\ref{opid}).

We now have all the ingredients to write the Hamiltonian $H$ in terms of $\theta,\varphi$: combining (\ref{prep}) and (\ref{eiprep}), we derive
\bea
H &=& \frac{4\pi^2 p_\varphi^2}{2m} + \frac{K (p_\theta - \varphi/2\pi)^2}{2} \nonumber \\
&-& U \cos(d \theta) - U \cos(\varphi)
\label{Hthetaphi}
\eea
This Hamiltonian is defined on a Hilbert space consisting of wave functions $\psi(\theta, \varphi)$ with
$0 \leq \theta, \varphi < 2\pi$. To find the boundary conditions for these wave functions, we use the identities
\begin{align*}
|\theta, 2\pi\> = e^{-i\theta} |\theta,0\>, \quad |2\pi,\varphi\> = |0,\varphi\>
\end{align*}
which follow from the definition of $|\theta,\varphi\>$. These identities imply that our wave functions $\psi$ satisfy the boundary conditions
\begin{align}
\psi(\theta,2\pi) = e^{i \theta}\psi(\theta,0), \quad  \psi(2\pi,\varphi) = \psi(0,\varphi)
\label{bcthetaphi}
\end{align}

So far, all we have done is derive the $\theta,\varphi$ representation (\ref{Hthetaphi}) of the Hamiltonian and the Hilbert space (\ref{bcthetaphi}). 
We now use this representation to find the low energy spectrum of $H$ in the large $U$ limit.
To begin, we note that $H$ can be thought of as describing a particle on a torus parameterized by
$\theta, \varphi$. This particle is coupled to a vector potential and two cosine potentials. Now consider the limit where
$U$ is large. In this limit, tunneling between the different cosine minima is suppressed, so we conclude that $H$ has a set of
nearly degenerate ground states, each of which is localized in a different minimum. There are $d$ different cosine minima 
located at positions $(\theta, \varphi) = (2\pi \alpha/d,0)$, with $\alpha = 0,1,...,d-1$, so $H$ has $d$ ground states. 
We will label these states by $|\psi_\alpha\>$.

To estimate the energy gap separating the ground states from the lowest excited states, we expand the cosine potentials to quadratic order. In this 
approximation, $H$ reduces to a sum of two decoupled harmonic oscillators with frequencies $\sqrt{U/m}$ and $\sqrt{UK}$. We conclude 
that the energy gap is of order $\Delta \sim \text{min}(\sqrt{U/m}, \sqrt{U K})$.

Let us now translate these results into the language of effective Hamiltonians. We have seen that $H$ has $d$ 
ground states $|\psi_\alpha\>$. We have also seen that these states are separated from the excited 
states by a large gap $\Delta$. Furthermore, it is easy to see that when $U \rightarrow \infty$, 
the width of $|\psi_\alpha\>$ becomes vanishingly small, so that $|\psi_\alpha\>$ approaches the state $|\theta = \frac{2\pi \alpha}{d}, \varphi = 0\>$:
\begin{equation*}
|\psi_\alpha\> \rightarrow |\theta = \frac{2\pi \alpha}{d}, \varphi = 0\>
\end{equation*}
Putting this all together, we conclude that the low energy spectrum of $H$ is described by an effective Hamiltonian, $H_{\text{eff}} = 0$
defined within an effective $d$-dimensional Hilbert space $\mathcal{H}_{\text{eff}}$ spanned by the states $|\theta = \frac{2\pi \alpha}{d}, \varphi = 0\>$, 
with $\alpha = 0,1,...,d-1$. This effective description is valid for energies $E < \Delta$. (Here, the reason we use $H_{\text{eff}} = 0$ rather than 
$H_{\text{eff}} = \text{const.}$ is that we only interested in energy \emph{differences} and therefore we are free to redefine the ground 
state energy to be $0$). Comparing these results with the effective Hamiltonian and Hilbert space from section \ref{effsummsect}, we can see that there is exact 
agreement. 

\subsection{Special case}
\label{Heffgenxsect}
We now generalize the example of appendix \ref{Heffex1sect} to a Hamiltonian of the form
\begin{align}
H = H_0 - U \sum_{i=1}^M \cos(2\pi x_i),
\label{genHamx}
\end{align}
defined on the $2N$ dimensional phase space $\{x_1,...,x_N, p_1,...,p_N\}$ with $[x_i, p_j] = i \delta_{ij}$.
Here $H_0$ is an arbitrary positive semidefinite quadratic function of $\{x_1,...,x_N, p_1,...,p_N\}$ with the only restriction being that the 
$M \times M$ matrix
\begin{equation}
\mathcal{N}_{ij} = - [x_i,[x_j,H_0]]
\end{equation}
is non-degenerate. 

Following the same outline as in the previous sections, we first derive an effective Hamiltonian and effective Hilbert space that
describe the low energy spectrum of $H$ in the infinite $U$ limit, and then we show that this effective theory agrees with
the general expressions from equations (\ref{Heffgenc}) and (\ref{Hilbeff}).

For the first step, we use the same strategy as in appendix \ref{Heffex1sect}: we decompose the Hamiltonian into two pieces, $H = H_1 + H_2$, where
\begin{align}
H_1 &= \sum_{i,j=1}^M \frac{(\mathcal{M}^{-1})_{ij}}{2} \cdot \Pi_i \Pi_j - U \sum_{i=1}^M \cos(2\pi x_i) \nonumber \\
H_2 &= H_0 - \sum_{i,j=1}^M \frac{(\mathcal{M}^{-1})_{ij}}{2} \cdot \Pi_i \Pi_j
\end{align}
Here $\mathcal{M}_{ij}$ is an $M \times M$ scalar matrix defined by $\mathcal{M} = \mathcal{N}^{-1}$
and $\Pi_1,..., \Pi_M$ are operators defined by
\begin{equation}
\Pi_{i} = -i \sum_{j=1}^M \mathcal{M}_{ij} [x_j, H_0]
\end{equation}
After making this decomposition, we will treat $H_2$ as a \emph{perturbation} to $H_1$ and then derive an effective Hamiltonian
using first order degenerate perturbation theory. 

Before executing this plan, we first make some preliminary observations. One observation is that
\begin{equation}
[x_i, \Pi_j] = i \delta_{ij}, \quad i = 1,...,M
\label{xpobs}
\end{equation}
Another observation is that we can assume without loss of generality that
\bea
[x_i, \Pi_j] = [p_i, \Pi_j] = 0, \quad i = M+1,...,N
\label{xpcassum}
\eea
The reason why we can assume (\ref{xpcassum}) is that we can always redefine the position and momentum operators $x_i, p_i$ for
$i = M+1,...,N$ according to
\begin{align*}
x_i \rightarrow x_i + \sum_{k=1}^M a_{ik} x_k, \quad p_i \rightarrow p_i + \sum_{k=1}^M b_{ik} x_k 
\end{align*}
where $a_{ik} = i [x_i, \Pi_k]$ and $b_{ik} = i [p_i, \Pi_k]$. After this redefinition, Eq. \ref{xpcassum} is automatically satisfied. A final observation is that $\Pi_j$ can be written in the form
\begin{equation}
\Pi_j = p_j + A_j
\label{pijexp}
\end{equation}
where $A_j$ is a linear combination of $\{x_1,...,x_M\}$. Indeed, this result follows immediately from (\ref{xpobs}) and (\ref{xpcassum}).

With these observations in mind, we now study the low energy spectrum of $H_1$. To begin we note that (\ref{pijexp}) implies that $H_1$ can be written in the form
\begin{equation*}
H_1 = \sum_{i,j=1}^M \frac{(\mathcal{M}^{-1})_{ij}}{2} \cdot (p_i + A_i)(p_j + A_j) - U \sum_{i=1}^M \cos(2\pi x_i) 
\end{equation*}
where $A_j$ is a linear function of $\{x_1,...,x_M\}$. Next, we note that $[x_i, H_1] = 0$ for $i = M+1,...,N$, which implies that 
$H_1$ can be diagonalized separately for 
each value of $\v{x}_\perp = (x_{M+1},...,x_N)$. Once we fix $\v{x}_\perp$, the Hamiltonian $H_1$ 
describes an $M$ dimensional particle with coordinates $(x_1,...,x_M)$, moving in a periodic potential $-U \sum_{i=1}^M \cos(x_i)$ and 
coupled to a vector potential $A$ that depends linearly on $\{x_1,...,x_M\}$. Let us consider
the low energy physics of this $M$ dimensional particle when $U$ is large. In this case, we can neglect 
tunneling between the different minima of the cosine potential, and treat each minimum in isolation. At the same time, it is easy to see that
the energy spectra associated with different cosine minima and different values of $\v{x}_\perp$ are all \emph{identical} 
since $H_1$ has discrete (magnetic) translational invariance in the $x_1,...,x_M$ directions as well as continuous translational 
invariance in the $x_{M+1},...,x_N$ directions. Putting these facts together, we conclude that the Hamiltonian $H_1$ has an infinite set 
of nearly degenerate ground states --- one ground state for each cosine minimum and each value of $\v{x}_\perp$. We label these ground states as 
$|\psi_{\v{q},\v{x}_\perp}\>$ where $\v{q} = (q_1,...,q_M)$ is an $M$ component integer vector describing the position of the cosine minimum
and $\v{x}_\perp = (x_{M+1},...,x_N)$ is an $N-M$ component real vector describing the position in the orthogonal directions.

We can estimate the energy gap between the ground states and excited states of $H_1$ by expanding the cosine potential to quadratic order in $\v{x}$. In this approximation, the cosine potential reduces to a multidimensional 
quadratic potential. Diagonalizing this potential gives a collection of harmonic oscillators with frequencies $\omega_i = \sqrt{U/m_i}$ 
where $m_i$ are the eigenvalues of the matrix $\mathcal{M}_{ij}$. The energy gap is determined by the smallest frequency and thus the 
largest eigenvalue $m_i$. We conclude that $H_1$ has an energy gap $\Delta \sim \sqrt{U/m}$ where $m$ is the maximum eigenvalue of $\mathcal{M}_{ij}$.

We now imagine adding $H_2$ to $H_1$, and we ask how the low energy spectrum changes. As in appendix \ref{Heffex1sect}, we will answer this question using a perturbative expansion in $H_2$. However, before we perform this calculation, we need to check that this perturbative approach is valid when $U$ is large. To this end, we need to estimate the size of the matrix elements $\<\psi_{ex} | H_2 | \psi_{\v{q},\v{x}_\perp}\>$ where $|\psi_{\v{q},\v{x}_\perp}\>$, $|\psi_{ex}\>$ are arbitrary ground states and excited states of $H_1$. The first step is to observe that $H_2$ has a special property: for any $i=1,...,M$, we have
\begin{eqnarray*}
[x_i, H_2] &=& [x_i, H_0] - i \sum_{j=1}^M (\mathcal{M}^{-1})_{ij} \cdot \Pi_j \nonumber \\
&=& [x_i,H_0] - [x_i,H_0] \nonumber \\
&=& 0
\end{eqnarray*}
It follows that the momentum operators $p_1,...,p_M$ do not appear in $H_2$. Thus, $H_2$ is a sum of three types of terms: $x_i x_j$, $p_{k} p_l$ and $x_i p_k$, where $i,j$ are arbitrary and $k,l \geq M+1$. We need to estimate the matrix elements corresponding to each of these terms. We can do this using the same argument as in appendix \ref{Heffex1sect}. First, we note that when $U$ is large, the states $|\psi_{\v{q},\v{x}_\perp}\>$ and $|\psi_{ex}\>$ can be approximated as harmonic oscillator eigenstates centered at some appropriate positions in space. The matrix elements of interest can then be related to harmonic oscillator matrix elements as in Eq. \ref{H2matapprox}. Omitting details, a straightforward calculation
shows that all three types of matrix elements fall off like $U^{-1/4}$ or faster as $U \rightarrow \infty$.
Combining this scaling law with the expression for $\Delta$, we see that
\begin{equation*}
\sum_{|\psi_{ex}\>} \frac{|\<\psi_{ex}|H_2|\psi_q\>|^2}{\Delta} = O \left(\frac{1}{U} \right)
\end{equation*}
As in appendix \ref{Heffex1sect}, this estimate implies that the second order perturbative corrections to the ground state energies vanish in the limit $U \rightarrow \infty$. Therefore, first order perturbation theory is exact in this limit.

We now proceed with the perturbative calculation.
According to first order degenerate perturbation theory, the energy splitting of the ground states can be determined by diagonalizing the matrix 
$\<\psi_{\v{q}',\v{x}'_\perp}| H_2 |\psi_{\v{q},\v{x}_\perp}\>$. These matrix elements are easy to compute. Indeed, for large $U$, the ground states $|\psi_{\v{q},\v{x}_\perp}\>$ can be approximated as harmonic oscillator ground states with a width of order $(\Delta x)^2  \sim \frac{1}{\sqrt{mU}}$.
In the limit $U \rightarrow \infty$, the width $\Delta x \rightarrow 0$ so $|\psi_{\v{q},\v{x}_\perp}\>$ approaches a position eigenstate: 
$|\psi_{\v{q},\v{x}_\perp}\> \rightarrow |\v{q},\v{x}_\perp\>$ where $|\v{q}, \v{x}_\perp\>$ denotes the position eigenstate located at $\v{x} = (\v{q}, \v{x}_\perp)$.
Thus, in this limit, the matrix we need to diagonalize is $\< \v{q}', \v{x}'_\perp | H_2 | \v{q}, \v{x}_\perp\>$.

Our calculation is now complete: we have shown that the low energy spectrum of $H$ in the limit $U \rightarrow \infty$ can be obtained by 
diagonalizing the operator $H_2$ within the subspace spanned
by $\{|\v{q},\v{x}_\perp\>\}$. In other words, the effective Hamiltonian for our system is
\begin{eqnarray}
H_{\text{eff}} &=& H_2 \nonumber \\
&=& H_0 - \sum_{i,j=1}^M \frac{(\mathcal{M}^{-1})_{ij}}{2} \cdot \Pi_i \Pi_j
\label{Heffx}
\end{eqnarray}
while the effective Hilbert space $\mathcal{H}_{\text{eff}}$ is spanned by position eigenstates $\{|\v{q},\v{x}_\perp\>\}$ for which
the first $M$ components $q_1,...,q_M$ are integers and the last $N-M$ components $x_{M+1},...,x_N$ are real valued. 
This effective theory is valid for energies $E \lesssim \sqrt{U/m}$. 

Comparing the above effective Hamiltonian and Hilbert space with the general expressions from equations (\ref{Heffgenc}) and 
(\ref{Hilbeff}), we can see that
they match exactly. This completes our proof of equations (\ref{Heffgenc}), (\ref{Hilbeff}) for the special case (\ref{genHamx}).
 
\subsection{General case}
\label{Heffgensect}
To complete the derivation of equations (\ref{Heffgenc}) and (\ref{Hilbeff}), we now consider the most general case:
\begin{align}
H = H_0 - U \sum_{i=1}^M \cos(C_i), \ \ C_i = \sum_{j=1}^N (\gamma_{ij} x_j + \gamma_{ij}' p_j+\delta_i)
\label{genHamc3}
\end{align}
Here $H$ is defined on a $2N$ dimensional phase space $\{x_1,...,x_N, p_1,...,p_N\}$ with $[x_i, p_j] = i \delta_{ij}$.
The first term $H_0$ is a positive semidefinite quadratic function of $\{x_1,...,x_N, p_1,...,p_N\}$ and $\gamma_{ij},\gamma_{ij}',\delta_i$
are arbitrary real numbers with the only constraints being that (1) $\{C_1,...,C_M\}$ are linearly independent,
(2) $[C_i, C_j]$ is an integer multiple of $2\pi i$ for all $i,j$ so that the different cosine terms commute with one another, 
and (3) the matrix
\begin{equation}
\mathcal{N}_{ij} = - \frac{1}{4\pi^2} [C_i, [C_j, H_0]]
\label{Ndefderiv}
\end{equation}
is non-degenerate. Our task is to derive an effective Hamiltonian and effective Hilbert space that describes the low energy spectrum of $H$ in the 
limit $U \rightarrow \infty$. We will then show that this effective theory is exactly the one defined in 
equations (\ref{Heffgenc}) and (\ref{Hilbeff}).

Our basic strategy is simple: we will map $H$ onto the Hamiltonian studied in the previous section and then we will derive the effective theory using our
previous results. While this is a simple plan at a conceptual level, there are some technical obstacles that make it difficult to define the desired mapping in one step. 
Therefore, we will build up the mapping in stages by making several successive changes of variables. 

In the first change of variables, we replace $x_i, p_i$ by new coordinates $x_i', p_i'$ which are chosen so that the constraints $C_i$ can be written as 
\emph{integer} linear combinations of $2\pi x_i'$ and $p_i'$. To this end, it is helpful to consider the $M \times M$ matrix $\mathcal{Z}_{ij}$ defined by
\begin{equation}
\mathcal{Z}_{ij} = \frac{1}{2\pi i} [C_i, C_j]
\end{equation}
Clearly the matrix $\mathcal{Z}_{ij}$ is integer and skew-symmetric. Therefore, 
there exists an integer matrix $\mathcal{V}$ with determinant $\pm 1$ such that $\mathcal{V} \mathcal{Z} \mathcal{V}^T = \mathcal{Z}'$ 
where $\mathcal{Z}'$ is in skew-normal form\cite{NewmanBook}:
\begin{equation}
\mathcal{Z}' = \bpm 0_I & -\mathcal{D} & 0 \\
	  \mathcal{D} & 0_I & 0 \\
	  0 & 0 & 0_{M-2I} \epm  , \ \ \mathcal{D} = \bpm d_1 & 0 & \dots & 0 \\
						  0 & d_2 & \dots & 0 \\
						  \vdots & \vdots & \vdots & \vdots \\
						   0 & 0 & \dots & d_I \epm
\end{equation}
Here $0 \leq I \leq M/2$ and $0_I$ denotes an $I \times I$ matrix of zeros, and similarly for $0_{M-2I}$. Let
\begin{align}
C_i' = \sum_{j=1}^M \mathcal{V}_{ij} C_j \textcolor{black}{+ \chi_i}
\label{cprimedef}
\end{align}
\textcolor{black}{where 
\begin{equation}
\chi_i = \pi \cdot \sum_{j < k} \mathcal{V}_{ij} \mathcal{V}_{ik} \mathcal{Z}_{jk} \pmod{2\pi}
\end{equation}
By construction, $[C_i', C_j'] = 2\pi i \mathcal{Z}_{ij}'$. Furthermore, because we chose the offset $\chi_i$ of the above form, we have the identity
\begin{equation}
e^{i C_i'} = \prod_{j=1}^M e^{i \mathcal{V}_{ij} C_j}
\label{ciprimeci1}
\end{equation}
as one can check using the Campbell-Baker-Hausdorff formula. Equivalently, using the fact that $\mathcal{V}_{ij}$ is an integer matrix with determinant $\pm 1$, the above identity 
can be inverted and rewritten as
\begin{equation}
e^{i C_i} = \prod_{j=1}^M e^{i \mathcal{V}_{ij}^{-1} C_j'}
\label{ciprimeci2}
\end{equation}
The two identities (\ref{ciprimeci1}-\ref{ciprimeci2}) will be useful below.}

We are now in a position to construct the new variables $x_i', p_i'$ that we seek. Specifically, we define
\begin{align*}
p_i' &= \frac{1}{d_i} C_i' \ \ \ \text{for } i = 1,...,I \nonumber \\
x_{i}' &= \frac{1}{2\pi} C_{i+I}' \ \ \ \text{for } i = 1,...,M-I
\end{align*}
and we define $p_{I+1}',...,p_N'$ and $x_{M-I+1}',...,x_N'$ to be some arbitrary linear combination of $x_i, p_i$ with the only
constraint being that they obey the correct commutation relations. In the new variables, the Hamiltonian takes the form
\begin{equation*}
H = H_0 - U \sum_{i=1}^M \cos \left(\sum_{j=1}^M (\mathcal{V}^{-1})_{ij} \textcolor{black}{[C_j'-\chi_j]} \right)
\end{equation*}
Note that the argument of the cosines are now \emph{integer} linear combinations of $2\pi x_i'$ and $p_i'$, just as we wanted.

In the next step we find an alternative representation of $H$ in which the arguments of the cosine terms all commute
with each other. We accomplish this by working in an unusual basis, which is similar to the one discussed in appendix \ref{Heffex2sect}. Specifically,
let us consider the basis of simultaneous eigenstates of the $N+I$ commuting operators
\begin{displaymath}
\{e^{i p_1'},...,e^{i p_I'}, e^{2\pi i x_1'},...,e^{2\pi i x_I'}, x_{I+1}',...,x_{N}'\}
\end{displaymath}
We denote these simultaneous eigenstates by 
\begin{displaymath}
|\v{\theta}, \v{\varphi}, \v{x}'\> \equiv |\theta_1,..,\theta_I,\varphi_1,...,\varphi_I,x_{I+1}',...,x_N'\>
\end{displaymath}
with
\begin{align*}
e^{i \hat{p}_i'} |\v{\theta}, \v{\varphi}, \v{x}'\> &=  e^{i \theta_i}  |\v{\theta}, \v{\varphi}, \v{x}'\>,   \ \ \ \text{$i=1,...,I$} \nonumber \\
e^{2\pi i \hat{x}_i'} |\v{\theta}, \v{\varphi}, \v{x}'\> &= e^{i \varphi_i}  |\v{\theta}, \v{\varphi}, \v{x}'\>,  \ \ \ \text{$i=1,...,I$} \nonumber \\
\hat{x}_i'  |\v{\theta}, \v{\varphi}, \v{x}'\> &= x_i' |\v{\theta}, \v{\varphi}, \v{x}'\>, \ \ \ \text{$i=I+1,...,N$}
\end{align*}
Here $0 \leq \theta_i, \varphi_i < 2\pi$ while $x_i'$ can be arbitrary real numbers. The formal definition of 
$|\v{\theta}, \v{\varphi}, \v{x}'\>$ is
\begin{eqnarray}
|\v{\theta}, \v{\varphi}, \v{x}'\> &=& \frac{1}{(2\pi)^{I/2}} \cdot \sum_{k_i} e^{i \sum_{i=1}^I k_i \theta_{i}} \nonumber \\
&\cdot& |k_1 + \frac{\varphi_{1}}{2\pi},...,k_I + \frac{\varphi_I}{2\pi},x_{I+1}',...,x_{N}'\>
\label{thetaphixdef}
\end{eqnarray}
where the ket on the right hand side denotes a position eigenstate localized at position 
$x_1' = k_1 + \frac{\varphi_1}{2\pi},...$ etc.

We now re-express the Hamiltonian $H$ in the $\v{\theta}, \v{\varphi}, \v{x}'$ representation. The first step is to find expressions for
the $x_i',p_i'$ operators in terms of $\theta_i, \varphi_i$. Following the same logic as in appendix \ref{Heffex2sect}, it is easy to show that $x_i', p_i'$ take the form
\begin{equation}
x_i' = -\frac{1}{i}\frac{\partial}{\partial \theta_i} + \frac{\varphi_i}{2\pi}, \quad p_i' =  \frac{2\pi}{i}\frac{\partial}{\partial \varphi_i}
\label{xp}
\end{equation}
for $i=1,...,I$. Equivalently, we can write this as
\begin{equation}
x_i' = -p_{\theta_i} + \frac{\varphi_i}{2\pi}, \quad p_i' = 2\pi p_{\varphi_i}
\label{xpirel}
\end{equation}
where $p_{\theta_i} \equiv \frac{1}{i}\frac{\partial}{\partial \theta_i}$ and $p_{\varphi_i} \equiv \frac{1}{i}\frac{\partial}{\partial \varphi_i}$
Likewise, when $i \geq I+1$, it is easy to see that $x_i', p_i'$ take the usual form, i.e. $p_i' = \frac{1}{i} \frac{\partial}{\partial x_i'}$, etc.

In addition to $x_i', p_i'$, we also need expressions for $e^{2\pi i x_i'}$ and $e^{i p_i'}$ for $i = 1,...,I$. We can obtain these 
expressions by exponentiating (\ref{xpirel}):
\begin{equation*}
e^{2\pi i x_i'} = e^{-2\pi i p_{\theta_i} +i\varphi_i}, \quad e^{ip_i'} = e^{2\pi i p_{\varphi_i}}
\end{equation*}
As in appendix \ref{Heffex2sect}, these equations can be simplified to
\begin{equation}
e^{2\pi i x_i'} = e^{i\varphi_i}, \quad e^{ip_i'} = e^{i\theta_i} 
\label{eixprep}
\end{equation}
using the operator identities
\begin{equation*}
e^{2\pi i p_{\theta_i}} = 1, \quad e^{2\pi i p_{\varphi_i}} =e^{i\theta_i}
\end{equation*}
(For an explanation of where these identities come from, see the discussion in appendix \ref{Heffex2sect}).

With this preparation, we are now ready to express the Hamiltonian $H$ in terms of $\v{\theta}, \v{\varphi}, \v{x}'$. Using (\ref{eixprep}) \textcolor{black}{together with the
identity (\ref{ciprimeci2})}, we can rewrite the cosine terms as
\begin{equation*}
\cos \left(\sum_{j=1}^M (\mathcal{V}^{-1})_{ij} [C_j' \textcolor{black}{-\chi_j]} \right) = \cos \left(2\pi \sum_{j=1}^M (\mathcal{V}^{-1})_{ij} \xi_j \right)
\end{equation*}
where 
\begin{align}
&(\xi_1,...,\xi_{M}) = \nonumber \\
&\left(\frac{d_1 \theta_{1}}{2\pi},...,\frac{d_I \theta_{I}}{2\pi},\frac{\varphi_1}{2\pi},...,\frac{\varphi_I}{2\pi},x_{I+1}',...,x_{M-I}' \right)
\label{xidef}
\end{align}
Similarly, using (\ref{xpirel}) we can write the quadratic term $H_0$ as a function of $\{\theta_1,...,\theta_I,\varphi_1,...\varphi_I,x_{I+1}',...,x_N'\}$
and $\{p_{\theta_1},...,p_{\theta_I}, p_{\varphi_1},...,p_{\varphi_I}, p_{I+1}',...,p_N'\}$. Thus, all together, the Hamiltonian is given by
\begin{equation}
H = H_0 - U \sum_{i=1}^M \cos \left(2\pi \sum_{j=1}^M (\textcolor{black}{\mathcal{V}^{-1})_{ij}} \xi_j \right)
\label{Htilde}
\end{equation}
This Hamiltonian is defined on a Hilbert space consisting of wave functions 
$\psi(\v{\theta},\v{\varphi},\v{x}')$ with $0 \leq \theta_i, \varphi_i < 2\pi$ and $x_{I+1}',...,x_{N}'$ arbitrary real numbers. 
As in appendix \ref{Heffex2sect}, these wave functions obey boundary conditions of the form
\begin{align}
\psi |_{\varphi_i = 2\pi} &= e^{i \theta_i} \psi |_{\varphi_i = 0} \nonumber \\
\psi |_{\theta_i = 2\pi} &= \psi |_{\theta_i = 0}
 \ \ \ \text{for $i = 1,...,I$}
\label{pbc}
\end{align}

To proceed further we make an approximation in which we temporarily ignore the periodic boundary conditions on $\varphi_i, \theta_i$. That is, we 
treat the Hamiltonian (\ref{Htilde}) as if it were defined in a Hilbert space consisting of wave functions 
$\psi(\v{\theta},\v{\varphi},\v{x}')$ 
where $\theta_i, \varphi_i$ range over $-\infty < \theta_i, \varphi_i < \infty$. Then, at the very end of our calculation we will reincorporate the 
fact that $\theta_i, \varphi_i$ are actually angular variables which range from $0$ to $2\pi$. The justification for this approximation is that the low 
energy eigenstates of $H$ are localized near the minima of the cosine potential and are therefore insensitive to the global boundary conditions in the 
large $U$ limit. Thus, the only effect of the periodic boundary conditions on the low energy theory is to identify certain states in the low energy 
Hilbert space $\mathcal{H}_{\text{eff}}$ --- an effect we can take account of at the end of our derivation. 

Once we make this approximation, the variables 
$\theta_i, \varphi_i$ and $x_{I+1}',...,x_N'$ are all on an equal footing. Our Hamiltonian is then defined in a $2N+2I$ dimensional phase space 
consisting of the (real valued) position variables $\{\theta_1,...,\theta_I,\varphi_1,...\varphi_I,x_{I+1}',...,x_N'\}$ together with their canonically conjugate 
(real valued) momenta $\{p_{\theta_1},...,p_{\theta_I}, p_{\varphi_1},...,p_{\varphi_I}, p_{I+1}',...,p_N'\}$.
This completes the second step of our derivation: we have successfully rewritten the Hamiltonian in a form (\ref{Htilde}) in which the arguments
of the cosine terms commute with one another.

In the final step, we make yet another change of variables, defining new position operators $\tilde{x}_1,...,\tilde{x}_{N+I}$ and new momenta 
$\tilde{p}_1,...,\tilde{p}_{N+I}$ which are linear combinations of the previous position and momenta operators. The goal of this transformation
is to simplify the arguments of the cosine terms even further. More specifically, we define
\begin{equation}
\tilde{x}_i = \sum_{j=1}^M (\textcolor{black}{\mathcal{V}^{-1})_{ij}} \xi_j, \quad i=1,...,M
\label{txdef}
\end{equation}
and we define $\tilde{x}_{M+1},...\tilde{x}_{N+I}$ and $\tilde{p}_1,...,\tilde{p}_{N+I}$ arbitrarily as long as they obey the canonical 
commutation relations. After this change of variables, our Hamiltonian takes the form
\begin{equation}
H = H_0 - U \sum_{i=1}^M \cos \left(2\pi \tilde{x}_i \right)
\label{Htilde2}
\end{equation}

At this point we have achieved the desired mapping: we can see that the above Hamiltonian (\ref{Htilde2}) is of the same form as (\ref{genHamx}). 
This is very convenient because it means we can write down a low energy effective theory in the limit $U \rightarrow \infty$ 
using our previous results without doing additional work. Indeed,
according to Eq. \ref{Heffx}, we have
\begin{equation}
H_{\text{eff}} =  H_0 - \sum_{i,j=1}^{M} \frac{(\tilde{\mathcal{M}}^{-1})_{ij}}{2} \cdot \tilde{\Pi}_i \tilde{\Pi}_j
\label{Heff2}
\end{equation}
where $\tilde{\mathcal{M}}_{ij}$ and $\tilde{\Pi}_i$ are defined by
\begin{align*}
\tilde{\mathcal{M}} = \tilde{\mathcal{N}}^{-1}, \quad \tilde{\mathcal{N}}_{ij} = - [\tilde{x}_i,[\tilde{x}_j, H_0]]
\end{align*}
and
\begin{equation*}
\tilde{\Pi}_{i} = - i \sum_{j=1}^{M} \tilde{\mathcal{M}}_{ij} [\tilde{x}_j, H_0]
\end{equation*}

To complete the derivation, we need to express the effective Hamiltonian (\ref{Heff2}) in terms of the original variables $x_i, p_i$. 
We will do this by relating the commutator $[\tilde{x}_i,H_0]$ to $[C_i,H_0]$. To this end, we make a few observations. First, we note that
\begin{eqnarray}
[2\pi p_{\varphi_i} - \theta_i, H_0] &=& 0 \label{pphithe} \\
\left[p_{\theta_i}, H_0 \right] &=& 0 \label{pthe}
\end{eqnarray}
Here the first commutator vanishes because $\varphi_i$ only appears in $H_0$ in the combination $(-p_{\theta_i} + \frac{1}{2\pi}\varphi_i)$,
while the second commutator vanishes because $\theta_i$ doesn't appear in $H_0$ at all (see (\ref{xpirel})). Next, we observe that
\begin{align}
[\theta_i, H_0] =  2\pi \left[p_{\varphi_i}, H_0 \right] = [p_i', H_0]
\label{piQ}
\end{align}
where the first equality follows from (\ref{pphithe}) and the second equality follows from (\ref{xpirel}). Similarly, we have
\begin{align}
\frac{1}{2\pi}[\varphi_i, H_0] = [-p_{\theta_i} + \frac{1}{2\pi}\varphi_i, H_0] = [x_i', H_0]
\label{xiQ}
\end{align}
where the first equality follows from (\ref{pthe}) and the second equality follows from (\ref{xpirel}). 
If we now multiply (\ref{xiQ}) by $2\pi$ and (\ref{piQ}) by $d_i$, we derive
\begin{align*}
2\pi [\xi_i, H_0] = [C_i', H_0] 
\end{align*}
for $i =1,...,M$. It follows that
\begin{equation}
2\pi [\tilde{x}_i, H_0] = [C_i, H_0]
\label{xcrel}
\end{equation}

With the help of (\ref{xcrel}), it is now straightforward to rewrite $H_{\text{eff}}$ in terms
of $x_i, p_i$. First, we take the commutator of (\ref{xcrel}) with $\tilde{x}_j$, and apply the Jacobi identity to derive
$\tilde{\mathcal{N}} = \mathcal{N}$ where $\mathcal{N}$ is defined in Eq. \ref{Ndefderiv}. Likewise, we have 
$\tilde{\mathcal{M}} = \mathcal{M}$ where $\mathcal{M} = \mathcal{N}^{-1}$. 
Substituting these relations into (\ref{Heff2}), we see that our effective Hamiltonian can be equivalently 
written as
\begin{equation}
H_{\text{eff}} =  H_0 - \sum_{i,j=1}^{M} \frac{(\mathcal{M}^{-1})_{ij}}{2} \cdot \Pi_i \Pi_j
\label{Heffgenxp}
\end{equation}
where
\begin{equation*}
\Pi_{i} = \frac{1}{2\pi i} \sum_{j=1}^{M} \mathcal{M}_{ij} [C_j, H_0]
\end{equation*}
Here we think of $H_0, \Pi_i$, etc, as functions of the original variables $x_i, p_i$.

So far we have focused on the low energy effective Hamiltonian $H_{\text{eff}}$, but we also need to discuss the low energy effective Hilbert space $\mathcal{H}_{\text{eff}}$ 
in which this Hamiltonian is defined. If we apply the results of the previous section, we see that $\mathcal{H}_{\text{eff}}$ consists 
of all states satisfying $\tilde{x}_i = \text{(integer)}$ for $i = 1,...,M$. Let us try to translate this back into our original variables.
First, from Eq. \ref{txdef} we see that $\mathcal{H}_{\text{eff}}$ can be equivalently described as consisting of all states with 
$\xi_i = \text{(integer)}$ for $i=1,...,M$
(here we use the fact that $\textcolor{black}{\mathcal{V}_{ij}}$ is an integer matrix with determinant $\pm 1$). Next, we convert to the 
$\v{\theta}, \v{\varphi}, \v{x}'$ description, using
the definition (\ref{xidef}). In the $\v{\theta}, \v{\varphi}, \v{x}'$ language, $\mathcal{H}_{\text{eff}}$ consists of all states 
$|\v{\theta}, \v{\varphi}, \v{x}'\>$ with $\theta_i = 2\pi/d_i \cdot \text{(integer)}$, $\varphi_i = 2\pi \cdot \text{(integer)}$, and 
$x_i' = \text{(integer)}$ for $i = I+1,...,M-I$.

It is at this point we should remember that $\theta_i$ and $\varphi_i$ are actually \emph{angular} variables (see the discussion below (\ref{pbc})); that is we should identify 
$\theta_i +2\pi$ with $\theta_i$ and $\varphi_i + 2\pi$ with $\varphi_i$. Thus, the basis states 
$|\v{\theta}, \v{\varphi}, \v{x}'\>$ for the Hilbert space $\mathcal{H}_{\text{eff}}$ can be parameterized in a non-redundant fashion 
by fixing $\varphi_i = 0$ and letting $\theta_i$ range over the values $\theta_i = 2\pi \alpha_i/d_i$ where 
$\alpha_i = 0,1,...,d_i-1$. Meanwhile, since $x_i'$ is \emph{not} angular valued, it can range over arbitrary integers for 
$i = I+1,...,M-I$ and arbitrary real numbers for $i = M-I+1,...,N$.

The above description in terms of the $\v{\theta}, \v{\varphi}, \v{x}'$ variables is perhaps the most explicit way to parameterize the low energy effective Hilbert space
$\mathcal{H}_{\text{eff}}$. However, it is also useful to describe $\mathcal{H}_{\text{eff}}$ in terms of our original variables $x_i, p_i$. 
One way to do this is to note that the above basis states $|\v{\theta}, \v{\varphi}, \v{x}_\perp'\>$ are precisely the states 
that satisfy the constraints $\cos(C_i) = 1$ for all $i$. 
Thus we can equivalently describe $\mathcal{H}_{\text{eff}}$ as the set of states satisfying $\cos(C_i) = 1$ for all $i$. 

This completes our derivation: we can see that our effective Hamiltonian $H_{\text{eff}}$ and effective Hilbert space $\mathcal{H}_{\text{eff}}$ exactly match the expressions in
equations (\ref{Heffgenc}) and (\ref{Hilbeff}). Thus we have derived these results in complete generality. 

\section{Diagonalizing the effective theory}
\label{diagsect}
In this appendix, we \textcolor{black}{derive a general recipe for diagonalizing the effective Hamiltonian 
$H_{\text{eff}}$ (\ref{Heffgenc}). This recipe is the one outlined in section \ref{diagsummsect}.} Our analysis can be divided into three steps.

\subsection{Step 1: Creation and annihilation operators}
\label{diagcreatsect}
In general, the key to diagonalizing quadratic Hamiltonians is to find appropriate creation and annihilation
operators. For the case of $H_{\text{eff}}$, this can be accomplished by searching for all operators $a$ that are 
linear combinations of $\{x_1,...,x_N, p_1,...,p_N\}$ and that satisfy
\begin{equation}
[a,H_{\text{eff}}] = E a,
\label{commuteq}
\end{equation}
for some scalar $E \neq 0 $, as well as
\begin{equation}
[a, C_i] = 0, \quad i=1,...,M
\label{concomm}
\end{equation}
While the first condition (\ref{commuteq}) is the usual definition of creation and annihilation
operators, the second condition is less standard; the motivation for this condition originates from the fact that $H_{\text{eff}}$ obeys
\begin{equation*}
[C_i, H_{\text{eff}}] = 0, \quad i = 1,...,M
\end{equation*}
(One can verify this identity with straightforward algebra). As a result, we can restrict to $a$ operators that 
commute with the $C_i$'s, and we will still have
enough quantum numbers to completely diagonalize $H_{\text{eff}}$. 

We will call the $a$ operators with $E > 0$, ``annihilation operators'', and those with $E <0$ ``creation operators.'' 
We will denote the annihilation operators by $a_1,...,a_K$, the creation operators by $a_1^\dagger,...,a_K^\dagger$, 
and the corresponding $E$'s by $E_1,...,E_K$ ($E_i > 0$).

The creation and annihilation operators have several important properties. To explain these properties, we 
need to recall some notation from section \ref{effsummsect}. There, we noted that
there exists a change of variables $C_i' = \sum_j \mathcal{V}_{ij} C_j \textcolor{black}{+ \chi_i}$, with $\mathcal{V}$ an integer matrix
with determinant $\pm 1$ \textcolor{black}{and $\chi$ defined by (\ref{chicond})}, such that the $M \times M$ matrix of commutators $[C_i', C_j']$ takes the form
\begin{equation}
[C_i', C_j'] = 2\pi i \bpm 0_I & -\mathcal{D} & 0 \\
          \mathcal{D} & 0_I & 0 \\
          0 & 0 & 0_{M-2I} \epm, 
\label{ciprimecomm}
\end{equation}
with
\begin{equation}
\mathcal{D} = \bpm d_1 & 0 & \dots & 0 \\
						  0 & d_2 & \dots & 0 \\
						  \vdots & \vdots & \vdots & \vdots \\
						   0 & 0 & \dots & d_I \epm
\end{equation}
Here $I$ is an integer with $0 \leq I \leq M/2$ and $0_I$ denotes an $I \times I$ matrix of zeros, and similarly for $0_{M-2I}$.

With this notation, we can state the first property of the creation/annihilation operators:
the number of linearly independent annihilation operators is exactly 
\begin{equation}
K = N-M+I
\label{numa}
\end{equation}
where $I$ is defined as above. The second property of these operators is that they can always be chosen so that
\begin{align}
[a_k, a_{k'}^\dagger] =\delta_{kk'} \ \ \ , \ \ \ [a_k, a_{k'}] = [a_k^\dagger, a_{k'}^\dagger] = 0
\label{aicomm}
\end{align}
The third property of these operators (which is a consequence of the previous two) is that 
each of the $\{x_1,...,x_N, p_1,...,p_N\}$ operators can be written as a linear combination of 
$\{a_1,...,a_{K},a_1^\dagger,...,a_{K}^\dagger,C_{2I+1}',...,C_M',\Pi_1,..,\Pi_M\}$.

In \textcolor{black}{appendix \ref{propcreapp}} we show that the above properties are guaranteed to hold provided that we make one (rather technical) assumption:
we assume that there is no operator $R$ which is a linear combination of $\{x_1,...,x_N, p_1,...,p_N\}$, is linearly independent 
from $\{C_1,...,C_M\}$, and satisfies 
\begin{equation}
[R, H_{\text{eff}}] = [R,C_i] = 0
\label{Rassumption}
\end{equation}
for all $i$.
To understand the physical meaning of this assumption, note that if such an $R$ operator existed, that would 
imply the existence of a continuous real-valued quantum number that
we could use to label the low energy eigenstates of $H$. Such continuous quantum numbers cannot occur in finite-sized systems
(and they don't occur in any of the examples discussed in this paper) so we do not sacrifice much generality in making this assumption. 

Using these properties we can derive an important result: the Hamiltonian $H_{\text{eff}}$ can be written in the form
\begin{equation}
H_{\text{eff}} = \sum_{k=1}^{K} E_k a_k^\dagger a_k +F(C_{2I+1}',...,C_M')
\label{Heffdiag}
\end{equation}
where $F$ is some quadratic function. One way to prove this result is to observe that $(H_{\text{eff}}-\sum_k E_k a^{\dagger}_k a_k)$
commutes with many other operators. For example,
\begin{equation}
[a_{k'}, H_{\text{eff}}-\sum_k E_k a^{\dagger}_k a_k] = 0
\label{ahcom}
\end{equation}
as one can see from the commutation relations between the $a_k$ operators and $H_{\text{eff}}$. Likewise, one can see that
\begin{equation}
[a_{k'}^\dagger, H_{\text{eff}}-\sum_k E_k a^{\dagger}_k a_k] = 0
\label{adaghcom}
\end{equation}
At the same time, we have
\begin{equation}
[C_j, H_{\text{eff}}-\sum_k E_k a^{\dagger}_k a_k] = 0
\label{chcom}
\end{equation}
since $[C_j, H_{\text{eff}}] = [C_j, a_k] = [C_j, a_k^\dagger] = 0$. 
To derive the consequences of these identities, we use the fact that the $x_i, p_i$ operators can be written as 
a linear combination of $a_k, a_k^\dagger, \Pi_j$, and $C_{2I+1}',...,C_M'$. Clearly, this result implies that
$(H_{\text{eff}} - \sum_k E_k a_k^\dagger a_k)$ can be written as a \emph{quadratic} 
function of these operators. Examining the commutators (\ref{chcom}) we see that this quadratic function cannot contain any
$\Pi_j$ operators since these would make the commutators with $C_j$ nonzero. Similarly, from (\ref{ahcom}) and (\ref{adaghcom}), 
we can see that this quadratic function cannot contain any $a_k, a_k^\dagger$ operators since these would make the commutators 
with $a_k^\dagger$ and $a_k$ nonzero. We conclude that 
$(H_{\text{eff}} - \sum_k E_k a_k^\dagger a_k)$ must depend only on $C_{2I+1}',...,C_M'$. The expansion in Eq. \ref{Heffdiag} follows 
immediately.

\subsection{Step 2: Occupation number basis}
\label{occnumsect}
In the next step we construct a basis for the Hilbert space $\mathcal{H}$ that is analogous to the conventional occupation
number basis for a harmonic oscillator. To this end, we note that the following operators all commute with each other (see Eq. \ref{ciprimecomm}):
\begin{equation}
\{e^{i C_{1}'/d_1},...,e^{i C_{I}'/d_I},e^{i C_{I+1}'},...,e^{i C_{2I}'},C_{2I+1}',...,C_M'\}
\label{commop2}
\end{equation}
Furthermore, these operators commute with the occupation number operators $\{a_1^\dagger a_1,...,a_{K}^\dagger a_{K}\}$. Therefore,
we can simultaneously diagonalize (\ref{commop2}) along with $\{a_k^\dagger a_k\}$. We denote the simultaneous eigenstates by
\begin{align*}
|\theta_1,...,\theta_I, \varphi_1,...,\varphi_I, x'_{I+1},...,x'_{M-I}, n_1,...,n_{K}\>
\end{align*}
or in more abbreviated form, $|\v{\theta}, \v{\varphi},\v{x}',\v{n}\>$. Here the quantum numbers are defined by
\begin{align}
&e^{i C_i'/d_{i}} |\v{\theta}, \v{\varphi},\v{x}',\v{n}\> = e^{i\theta_{i}} |\v{\theta}, \v{\varphi},\v{x}',\v{n}\>, \  i = 1,...,I \nonumber \\
&e^{i C_i'} |\v{\theta}, \v{\varphi},\v{x}',\v{n}\> = e^{i\varphi_{i-I}} |\v{\theta}, \v{\varphi},\v{x}',\v{n}\>, \ i = I+1,...,2I \nonumber\\
&C_i' |\v{\theta}, \v{\varphi},\v{x}',\v{n}\> = 2\pi x_{i-I}'  |\v{\theta}, \v{\varphi},\v{x}',\v{n}\>, \ i = 2I+1,...,M \nonumber \\
&a_k^\dagger a_k |\v{\theta}, \v{\varphi},\v{x}',\v{n}\> = n_k  |\v{\theta}, \v{\varphi},\v{x}',\v{n}\>, \  k = 1,...,K   
\label{geneigdef}
\end{align}
where $0 \leq \theta_i, \varphi_i < 2\pi$, while $x_i'$ is real valued and $n_i$ ranges over 
non-negative integers. Importantly, one can show that there is \emph{exactly one} simultaneous eigenstate $|\v{\theta}, \v{\varphi},\v{x}',\v{n}\>$
for each choice of $\v{\theta}, \v{\varphi}, \v{x}', \v{n}$ so our labeling scheme is well-defined (see appendix \ref{uniqueapp} for a proof).
By construction the $|\v{\theta}, \v{\varphi},\v{x}',\v{n}\>$ states form a complete orthonormal basis for
the Hilbert space $\mathcal{H}$: these are the basis states that we seek. 

In fact, not only do the $|\v{\theta}, \v{\varphi},\v{x}',\v{n}\>$ states form a basis for $\mathcal{H}$, but a \emph{subset} of these states form a basis for
the low energy Hilbert space $\mathcal{H}_{\text{eff}}$. \textcolor{black}{To see this, recall that $\mathcal{H}_{\text{eff}}$ consists of all states $|\psi\>$ that satisfy
$\cos(C_i) |\psi\> = |\psi\>$ for all $i$. Applying the two identities (\ref{ciprimeci1}-\ref{ciprimeci2}), we can see that $\mathcal{H}_{\text{eff}}$
can be equivalently defined as the set of all states $|\psi\>$ satisfying}
\begin{equation}
\cos(C_1') |\psi\> = ... = \cos(C_M') |\psi\> = |\psi\>
\label{heffdef2}
\end{equation}
Substituting $|\v{\theta}, \v{\varphi},\v{x}',\v{n}\>$ into the above
definition, we conclude that $|\v{\theta}, \v{\varphi},\v{x}',\v{n}\>$ belongs to $\mathcal{H}_{\text{eff}}$ if and only if
\begin{equation*}
\v{\theta} = (2\pi \alpha_1/d_1,..., 2\pi \alpha_I/d_I),
\end{equation*}
for some $\alpha_i =0,1,...,d_i-1$, and 
\begin{align*}
\v{\varphi} = (0,...,0), \quad \v{x}' = (q_1,...,q_{M-2I})
\end{align*}
with the $q_i$'s being \emph{integers}. 
It follows that the above states form a basis for $\mathcal{H}_{\text{eff}}$. We will denote these states using the abbreviated notation $\{|\v{\alpha},\v{q},\v{n}\>\}$,
where $\v{\alpha}$ and $\v{q}$ are defined above, and $\v{n} =(n_1,...,n_K)$ is the usual set of occupation numbers.

\subsection{Step 3: Eigenstates and energies}
\label{eigensect}
We now have everything we need to diagonalize $H_{\text{eff}}$. Indeed, we've already seen that the $|\v{\alpha},\v{q},\v{n}\>$ states
form a basis for the low energy Hilbert space $\mathcal{H}_{\text{eff}}$. At the same time, from Eq. \ref{Heffdiag} we can see that the $|\v{\alpha},\v{q},\v{n}\>$ states are 
eigenstates of $H_{\text{eff}}$ with energies
\begin{equation}
E(\v{\alpha},\v{q},\v{n}) = \sum_{k=1}^K n_k E_k + F(2\pi q_1,...,2\pi q_{M-2I})
\label{energyspect}
\end{equation}
We therefore have found all the eigenstates and energies of $H_{\text{eff}}$.

Now that we have the energy spectrum, we should point out one of its most important features: the energy $E$ is independent of the quantum numbers 
$\v{\alpha} =(\alpha_1,...,\alpha_I)$. Since $\alpha_i$ ranges from $0 \leq \alpha_i < d_i - 1$, it follows that every eigenvalue of $H_{\text{eff}}$ has a degeneracy of at least
\begin{equation}
D = \prod_{i=1}^I d_i
\end{equation}

Before concluding, there is one more issue we need to discuss: we haven't yet explained how to determine the quadratic function $F$. One approach for
computing $F$ is to define a new set of operators $\Phi_{2I+1},...,\Phi_{M}$ by
\begin{equation}
\Phi_i = \sum_{j=1}^M w_{ij} \Pi_j + \sum_{k=1}^K (x_{ik} a_k + x_{ik}^* a_k^\dagger)
\label{Phidef}
\end{equation}
where the coefficients $w_{ij}$ and $x_{ik}$ are defined by $w_{ij} = -(\mathcal{V}^{-1})_{ji}$, and
$x_{ik} = \sum_{j=1}^M (\mathcal{V}^{-1})_{ji} \cdot [\Pi_j,a_k^\dagger]$. These coefficients have been chosen
so that the $\Phi_i$ operators have simple commutators with other key operators. In particular, they have been chosen so that
\begin{align}
[\Phi_i, C_j'] = 2\pi i \delta_{ij}, \quad [\Phi_i, a_k] = [\Phi_i, a_k^\dagger] = 0
\label{Phicom}
\end{align}
Thus, the $\Phi_{2I+1},...,\Phi_M$ are \emph{conjugate} variables to $C_{2I+1}',...,C_M'$. 

Once we have defined these operators, we can compute $F$ by \textcolor{black}{considering the commutators 
$[\Phi_i, H_{\text{eff}}]$ and $[\Phi_i, [\Phi_j, H_{\text{eff}}]]$. Indeed, from} the expression 
(\ref{Heffdiag}), we \textcolor{black}{derive
\begin{align}
[\Phi_i, H_{\text{eff}}]] &=  2\pi i \frac{\partial F}{\partial q_i} \nonumber \\
[\Phi_i,[\Phi_j, H_{\text{eff}}]] &=  -4\pi^2 \frac{\partial^2 F}{\partial q_i \partial q_j}
\end{align}
These relations completely determine} $F$ since it is a quadratic function of the $q_i$'s. (Alternatively, one can often compute $F$ using 
problem-specific approaches, as in section \ref{examsect}).

\subsection{Another method for finding creation and annihilation operators}
\label{altmethsect}
We have seen that one of the key steps in computing the energy spectrum of $H_{\text{eff}}$ is finding creation and 
annihilation operators $a$ --- that is, finding solutions to Eqs. \ref{commuteq},\ref{concomm}. 
Here we point out that there is an another, more convenient, way to formulate these equations. 
In this alternative formulation, we search for all operators $a$ that 
obey
\begin{align}
[a,H_0] &= E a + \sum_{j=1}^M\lambda_{j} [C_j,H_0] \label{auxeq} \\
[a, C_i] &= 0, \quad i=1,...,M \label{auxeq2}
\end{align}
for some scalars $E, \lambda_{j}$ with $E \neq 0$. Here the $\lambda_j$ can be thought of as a kind of Lagrangian multiplier. We will
show below that the above equations are mathematically equivalent to the previous equations 
(\ref{commuteq}) and (\ref{concomm}) in the sense that every solution to 
(\ref{auxeq}) and (\ref{auxeq2}) is a solution to (\ref{commuteq}) and (\ref{concomm}) and vice versa. The above
equations are often more convenient than (\ref{commuteq}) and (\ref{concomm}) because they are written in terms of $H_0$ and thus 
do not require us to compute $H_{\text{eff}}$.

To prove the equivalence between the two approaches, let us suppose that $a$ obeys (\ref{auxeq}) and (\ref{auxeq2}) for some $E, \lambda_j$. We wish to 
show that $a$ also obeys (\ref{commuteq}). To prove this, we take the commutator of (\ref{auxeq}) with $C_i$. The result is:
\begin{equation*}
[C_i,[a,H_0]] = \sum_{j=1}^M \lambda_{j} [C_i,[C_j,H_0]]
\end{equation*}
We deduce that
\begin{eqnarray}
\lambda_{j} &=& - \frac{1}{4\pi^2} \sum_{i=1}^M \mathcal{M}_{ji} [C_i,[a,H_0]] \nonumber \\
&=& - \frac{1}{4\pi^2} \sum_{i=1}^M \mathcal{M}_{ji} [a,[C_i,H_0]] \label{lambdaj}
\end{eqnarray}
where in the second step, we used the Jacobi identity.
Substituting the above formula for $\lambda_j$ into (\ref{auxeq}), we derive
\begin{equation*}
[a,H_0] +  \frac{1}{4\pi^2} \sum_{j,i=1}^M \mathcal{M}_{ji} [C_j,H_0] \cdot [a,[C_i,H_0]]  = E a
\end{equation*}
which can be rewritten as
\begin{equation*}
[a,H_0] - \sum_{j,i=1}^M \mathcal{M}^{-1}_{ji} \Pi_j\cdot [a, \Pi_i] = E a
\end{equation*}
This is exactly (\ref{commuteq}). Conversely, one can check that any solution to (\ref{commuteq}), (\ref{concomm})
provides a solution to (\ref{auxeq}) with $\lambda_{j}$ given by Eq. \ref{lambdaj}. 

\section{Finite \texorpdfstring{$U$}{U} corrections}
\label{finUsect}
When $U$ is large but finite, $H_{\text{eff}}$ only gives the \emph{approximate} low energy spectrum of $H$.
It is natural to wonder: what kinds of corrections do we need to add to $H_{\text{eff}}$ if we want
to obtain the exact effective theory at finite $U$? The goal of this appendix is to address this question.

\subsection{Simple example}
We begin by discussing the example from section \ref{effsummsect}:
$H = \frac{p^2}{2m} + \frac{K x^2}{2}  - U \cos(2\pi x)$. For this example,
the low energy effective Hamiltonian in the infinite $U$ limit
is $H_{\text{eff}} = \frac{K x^2}{2}$ while the low energy Hilbert space $\mathcal{H}_{\text{eff}}$
is spanned by position eigenstates $\{|q\>\}$ where $q$ is an integer.

To understand the finite $U$ case, let us
imagine repeating the derivation of $H_{\text{eff}}$ from appendix \ref{Heffex1sect}, but without taking the
limit $U \rightarrow \infty$ or making any other approximations. In such a hypothetical exact calculation, we would first write $H$ as a sum
$H = H_1 + H_2$ where $H_1 = \frac{p^2}{2m} - U \cos(2\pi x)$ and $H_2 = \frac{K x^2}{2}$. Next, we would find
find the exact eigenstates and energies of $H_1$. According to Bloch's theorem, these states can be labeled
as $|k,n\>$ where $n=0,1,...$ is the band index and $k$ is the crystal momentum, $-\pi \leq k \leq \pi$.
To complete the calculation, we would construct a low energy effective theory describing the energy spectrum
of $H$ below the first band gap by treating $H_2$ as a perturbation to $H_1$, and including terms to all orders
in perturbation theory. In this way, we can imagine deriving
an effective Hamiltonian $H_{\text{eff}}$ giving the exact low energy spectrum of $H$ in the finite $U$ case. This
effective Hamiltonian $H_{\text{eff}}$ would be defined in an effective Hilbert
space $\mathcal{H}_{\text{eff}}$ spanned by the states in the lowest band, $\{|k,0\>\}$. Equivalently, we can describe
the effective Hilbert space, $\mathcal{H}_{\text{eff}}$ as the span of the Wannier states $\{|q\>\}$ defined by
$|q\> = \int_{-\pi}^\pi e^{-ikq} |k,0\> dk$. Here, $|q\>$ denotes the Wannier state localized near the cosine minimum at
$x=q$.

Although we will not perform the above calculation, we can still make some qualitative statements about the structure of the final result.
Indeed, from our analysis of the infinite $U$ limit, we know that the resulting $H_{\text{eff}}$ must take the form
\begin{equation}
H_{\text{eff}} = \sum_q \frac{K q^2}{2} |q\> \<q| + \sum_{q q'} \epsilon_{qq'}|q\>\<q'|
\label{HeffexfinU1}
\end{equation}
where $\epsilon_{qq'} \rightarrow 0$ as $U \rightarrow \infty$. Furthermore,
we can estimate the size of $\epsilon_{qq'}$ for finite $U$. There are two cases to consider: $q \neq q'$ and $q = q'$. When
$q \neq q'$, we expect that
$\epsilon_{qq'} \sim \sqrt{\frac{U}{m}} e^{-\text{const.} \cdot \sqrt{m U}}$ since these off-diagonal terms are generated by
tunneling between different cosine minima of $H_1$.
As for the $q = q'$ case, these terms can be estimated as $\epsilon_{qq} \sim \frac{K^2 q^2}{U}$ since the leading order
contribution to $\epsilon_{qq}$ comes from the second order terms in the perturbative expansion in $H_2$, computed in
Eq. \ref{pterror}.

It is instructive to rewrite the above expression (\ref{HeffexfinU1}) for $H_{\text{eff}}$ in terms of the operators $x,p$. In this
language, we have
\begin{equation}
H_{\text{eff}} = \frac{K x^2}{2} + \sum_{n=-\infty}^\infty e^{inp} \cdot \epsilon_n(x)
\label{HeffexfinU2}
\end{equation}
where $\epsilon_n$ is a function satisfying $\epsilon_{n}(q') = \epsilon_{(q'-n)q'}$ for all integers $q'$. Here, the equivalence between
(\ref{HeffexfinU1}) and
(\ref{HeffexfinU2}) follows from the fact that $e^{inp} |q'\> = |q'-n\>$. Translating our results about $\epsilon_{qq'}$ into this alternative
language, we see that $\epsilon_{n}$ has an exponential dependence on $\sqrt{U}$ for $n \neq 0$, and scales like $1/U$ for $n = 0$.

Equations (\ref{HeffexfinU1}, \ref{HeffexfinU2}) give the \emph{qualitative} structure of $H_{\text{eff}}$ in the finite $U$ case. To obtain 
more \emph{quantitative}
results, we would need to explicitly compute the coefficients $\epsilon_{qq'}$ or the function $\epsilon_n(x)$ as a function of $U,K,m$. In
principle it should be possible to compute these quantities in a large $U$ expansion --- for example using the instanton approach outlined
in section 7.2.3 of Ref.~\onlinecite{coleman1988aspects}.

\subsection{General case}
We now discuss the finite $U$ corrections for more general systems. First, we consider a scenario in which
only one of the cosine terms has a finite coefficient while the others are taken to be infinitely large.
In other words, we consider Hamiltonians of the form $H = H_0 - U \cos(C_{i}) - U'\sum_{j \neq i} \cos(C_j)$ in the limit
where $U$ is finite but $U' \rightarrow \infty$. In this case, the finite $U$ corrections only generate tunneling processes of the form
$C_{i} \rightarrow C_{i} - 2\pi n$; other tunneling processes, $C_j \rightarrow C_j - 2\pi n$, are suppressed by $U'$.
It follows that the finite $U$ corrections must commute with $\{C_1,...,C_{i-1},C_{i+1},...,C_M\}$ but don't have to commute with $C_i$. At
the same time we know that the most general low energy operator is of the form shown in Eq. \ref{loweop2}. Combining
these two facts, we conclude that the finite $U$ corrections can be written in the form
\begin{equation}
\sum_{n=-\infty}^\infty e^{in \Pi_i} \cdot
\epsilon_{n}(\{a_k, a_k^\dagger, C_{2I+i}'\})
\label{finU1}
\end{equation}
for some functions $\epsilon_n(a_1,...,a_k,a_1^\dagger,...,a_k^\dagger,C_{2I+1'},...,C_M')$.
Note that these terms are generalizations of the finite $U$ corrections (\ref{HeffexfinU2})
that we discussed for the above example.  As in the example, the functions $\epsilon_{n}$ have
a $U$ dependence which we have not shown explicitly. In particular, the $\epsilon_{n}$ with
$n \neq 0$ vanish exponentially in $\sqrt{U}$ as $U \rightarrow \infty$ since these terms are generated by (non-perturbative)
instanton effects,
while $\epsilon_{0}$ vanishes like $1/U$ since this term originates from perturbative corrections.

Next we consider the case where all the cosine terms have finite coefficients. In this case, all tunneling processes
$C_i \rightarrow C_i - 2\pi m_i$ are allowed so the finite $U$ corrections take the form
\begin{equation}
\sum_{\v{m}} e^{i \sum_{j=1}^M m_j \Pi_j} \cdot
\epsilon_{\v{m}}(\{a_k, a_k^\dagger, C_{2I+i}'\})
\label{finU2}
\end{equation}
with the sum running over $M$ component integer vectors $\v{m} = (m_1,...,m_M)$. As above, the $\epsilon_{\v{m}}$ are
unknown functions of $\{a_1,...,a_k,a_1^\dagger,...,a_k^\dagger,C_{2I+1'},...,C_M'\}$ which also depend on $U$. The $\epsilon_{\v{m}}$
with $\v{m} \neq 0$ vanish exponentially in $\sqrt{U}$ as $U \rightarrow \infty$, while $\epsilon_{0}$ vanishes like $1/U$. In principle,
it should be possible to compute $\epsilon_{\v{m}}$ using instanton methods~\cite{coleman1988aspects}, but we won't discuss this computation here.

\subsection{Splitting of ground state degeneracy}
One application of this formalism is that we can use it to analyze how the $D$-fold ground state degeneracy of $H_{\text{eff}}$ splits at finite $U$.
Indeed, according to lowest order perturbation theory, we can determine the splitting of the ground state
degeneracy by projecting the finite $U$ corrections onto the ground state subspace and then diagonalizing the resulting
$D \times D$ matrix. This diagonalization problem is system dependent so we cannot say much about it in general, but we would like to mention a 
result that is useful for setting up the computation. This result applies to any system in which the commutator matrix
$\mathcal{Z}_{ij}$ is non-degenerate, i.e. it applies to the case where $M = 2I$. The result states that the matrix elements of the
finite $U$ corrections are proportional to the matrix elements of $e^{i \sum_{j=1}^M m_j \Gamma_j}$ \textcolor{black}{(where $\Gamma_j$ is defined as in Eq. \ref{Gammadef}}): that is,
\begin{equation}
\<\alpha'|e^{i \sum_{j=1}^M m_j \Pi_j} \cdot \epsilon_{\v{m}} |\alpha\>
= u_{\v{m}} \cdot \<\alpha'|e^{i \sum_{j=1}^M m_j \Gamma_j}|\alpha\>
\label{finUid}
\end{equation}
where $|\alpha\>, |\alpha'\>$ are ground states and $u_{\v{m}}$ is some unknown proportionality constant.
Although this relation does not tell us the value of $u_{\v{m}}$, it is still useful since it tells us
the form of the matrix that we need to diagonalize.

To derive equation (\ref{finUid}), consider the difference $\Pi_j - \Gamma_j$. This difference is a linear function of
$x_i$'s and $p_i$'s so we know it can be written as a linear combination of $\{a_k, a_k^\dagger, C_i'\}$ as in Eq. \ref{Aexp2}.
\footnote{There are no $\Phi_{2I+i}$ operators in this expansion due to our assumption that $M = 2I$.}
At the same time, it is easy to see that $\Pi_j-\Gamma_j$ commutes with
the $C_i$ operators and hence also the $C_i'$ operators. It follows that the expansion of $\Pi_j -\Gamma_j$ in terms of
$\{a_k, a_k^\dagger, C_i'\}$ cannot contain any $C_i'$ operators: that is,
\begin{displaymath}
\Pi_j = \Gamma_j + (\text{linear combination of $a_k$ and $a_k^\dagger$}).
\end{displaymath}
If we exponentiate this relation, Eq. \ref{finUid} follows easily. Note that the constant prefactor $u_{\v{m}}$ comes from the
$a_k$ and $a_k^\dagger$ operators appearing in this expression.

\section{Matrix elements and operators}
\label{mateltopsect}
\subsection{Matrix elements}
\label{mateltsect}
If we wish to have a complete low energy theory, we need to do more than just find the energies of the 
low-lying states: we also need to be able to compute matrix elements of operators between these states. 
To address this issue, we now outline a general 
procedure for computing matrix elements $\<\v{\alpha},\v{q},\v{n}| \mathcal{O}| \v{\alpha},\v{q},\v{n}\>$. 

Our basic strategy is as follows: suppose we are given a general operator $\mathcal{O}$ which is some function of $\{x_1,...,x_N,p_1,...,p_N\}$.
What we will do is express $\mathcal{O}$ as a function of the operators $a_k, a_k^\dagger, C_i,\Pi_i$, etc. Then we will use the known matrix elements
of $a_k, a_k^\dagger, C_i, \Pi_i$, etc to compute the matrix elements of $\mathcal{O}$. 

We now demonstrate how this approach works in more detail. We focus on two cases which are particularly
important: (1) $\mathcal{O} = \mathcal{A}$ and (2) $\mathcal{O} = e^{i\mathcal{A}}$ where $\mathcal{A}$ is some \emph{linear} combination of 
$x_i$'s and $p_i$'s. We begin with the first case, $\mathcal{O} = \mathcal{A}$. In this case, our strategy will be to express $\mathcal{A}$ as a linear 
combination of $\{a_1,...,a_{K},a_1^\dagger,...,a_{K}^\dagger,C_{2I+1}',...,C_M',\Pi_1,..,\Pi_M\}$:
\begin{align}
\mathcal{A} &= \sum_{k=1}^{K} (\kappa_k a_k + \lambda_k a_k^\dagger) + \sum_{i=2I+1}^M \mu_i C_i' + \sum_{i=1}^M \nu_i \Pi_i 
\la{Aexp}
\end{align}
We know such an expression must exist, since the $x_i$'s and $p_i$'s can be written as linear combinations of these operators, as shown in 
appendix \ref{diagcreatsect}. 

Our task is now to find the expansion coefficients, $\kappa_k, \lambda_k, \mu_i, \nu_i$. This can be accomplished by considering appropriate commutators.
For example, if we take the commutator of Eq. \ref{Aexp} 
with $C_j$, we derive $\nu_j = -\frac{i}{2\pi} [C_j, \mathcal{A}]$. Similarly, taking the commutator with $a_k^\dagger$ and $a_k$, we derive
\begin{eqnarray*}
\kappa_k &=&  -[a_k^\dagger, \mathcal{A}] + \sum_{i=1}^M \nu_i [a_k^\dagger, \Pi_i] \nonumber \\
\lambda_k &=&  [a_k, \mathcal{A}] - \sum_{i=1}^M \nu_i [a_k, \Pi_i] 
\end{eqnarray*}
Finally, taking the commutator with $\Pi_j$, we find
\begin{eqnarray*}
-2\pi i\sum_{i=1}^M \mathcal{V}_{ij} \mu_i &=&  [\Pi_j,\mathcal{A}] - \sum_{k=1}^K \kappa_k [\Pi_j, a_k] \nonumber \\
&-& \sum_{k=1}^K \lambda_k [\Pi_j, a_k^\dagger]) - \sum_{i=1}^M \nu_i [\Pi_j, \Pi_i] 
\end{eqnarray*}
which we can invert to find $\mu_i$.

Once we have the expansion (\ref{Aexp}), our problem reduces to finding the matrix elements of the operators $a_k, a_k^\dagger, \Pi_i$ and 
$C_{2I+1}',...,C_M'$. The matrix elements for $a_k, a_k^\dagger, C_i'$ can be written down without any work:
\begin{eqnarray}
a_k|\v{\alpha},\v{q},\v{n}\> &=& \sqrt{n_k} |\v{\alpha},\v{q},\v{n}-\v{e_i}\> \nonumber \\
a_k^\dagger |\v{\alpha},\v{q},\v{n}\> &=& \sqrt{n_k+1} |\v{\alpha},\v{q},\v{n}+\v{e_i}\> \label{akmat} \\
C_i'|\v{\alpha},\v{q},\v{n}\> &=& \textcolor{black}{2\pi}q_{i-2I} |\v{\alpha},\v{q},\v{n}\>, \ i=2I+1,...,M \label{ciprimemat}
\end{eqnarray}
Here
$\v{e_i}$ denotes the $K$-component vector $\v{e_i} = (0,...,1,...,0)$ with a ``1'' in the $i$th entry and $0$ everywhere else;
thus the first two equations encode the fact that the $a_k, a_k^\dagger$ act as raising and lower operators in the occupation numbers $\v{n}$.
As for the last equation, this follows from the definition of $|\v{\alpha},\v{q},\v{n}\>$ in 
\textcolor{black}{section \ref{diagsummsect}.}

All that remains are the $\Pi_i$ operators. We now argue that the matrix elements of these operators 
vanish \emph{identically}:
\footnote{Note that Eq. \ref{pimat} does \emph{not} imply that 
\unexpanded{$\Pi_i|\v{\alpha},\v{q},\v{n}\> = 0$} since $\Pi_i$ has nonzero matrix elements between low energy
states \unexpanded{$|\v{\alpha},\v{q},\v{n}\>$} that belong to $\mathcal{H}_{\text{eff}}$ and higher energy states 
that do not belong to $\mathcal{H}_{\text{eff}}$.}
\begin{equation}
\<\v{\alpha}',\v{q}',\v{n}'| \Pi_i|\v{\alpha},\v{q},\v{n}\> = 0
\label{pimat}
\end{equation}
The derivation of Eq. \ref{pimat} follows from two observations. First, we recall from our
derivation of $H_{\text{eff}}$ that the low energy states $|\v{\alpha},\v{q},\v{n}\>$ are degenerate ground states of the Hamiltonian 
$H_1 = \sum_{i,j=1}^{M} \frac{(\mathcal{M}^{-1})_{ij}}{2} \cdot \Pi_i \Pi_j - U \sum_{i=1}^M \cos(C_i)$ in the limit $U \rightarrow \infty$. Thus,
we know that $H_1 |\v{\alpha},\v{q},\v{n}\> = E |\v{\alpha},\v{q},\v{n}\>$ for some scalar $E$. Second, we note that 
$\Pi_i = \frac{1}{2\pi i} \sum_{j=1}^{M} \mathcal{M}_{ij} [C_j, H_1]$, so in particular $\Pi_i$ is a linear combination of commutators $[C_j, H_1]$.
Equation (\ref{pimat}) now follows from the identity
\begin{align*}
&\<\v{\alpha}',\v{q}',\v{n}'| [C_j, H_1] |\v{\alpha},\v{q},\v{n}\> = \\
& \<\v{\alpha}',\v{q}',\v{n}'| C_j E - E C_j  |\v{\alpha},\v{q},\v{n}\> = 0
\end{align*}
Putting together equations (\ref{Aexp},\ref{akmat},\ref{ciprimemat},\ref{pimat}), our task is complete: we can compute matrix elements of any operator 
$\mathcal{A}$ that is a linear combination of $\{x_1,...,x_N,p_1,...,p_N\}$. 

We now move on to the second case: operators of the form $e^{i\mathcal{A}}$ where $\mathcal{A}$ is a linear combination of $x_i$'s and $p_i$'s.
For this case, our strategy is to express
$\mathcal{A}$ as a linear combination of the operators $\{a_1,...,a_K, a_1^\dagger,...,a_K^\dagger, C_1',...,C_M'\}$
along with $\Phi_{2I+1},...,\Phi_M$ (where the $\Phi_i$ operators are defined in Eq. \ref{Phidef}):
\begin{align}
\mathcal{A} &= \sum_{k=1}^{K} (\kappa_k a_k + \lambda_k a_k^\dagger) + \sum_{i=2I+1}^M \mu_i \Phi_i +  \sum_{i=1}^M \nu_i C_i' 
\la{Aexp2}
\end{align}
We know such an expression must exist because (1) the above set contains $2N$ operators, as can be seen from Eq. \ref{numa}, and (2) these operators are 
linearly independent, as one can show using the same arguments as in appendix \ref{propcreapp}. 

Just as in the previous case, we can find the expansion coefficients $\kappa_k, \lambda_k, \mu_i, \nu_i$ by taking appropriate commutators. 
After finding the coefficients, our problem reduces to computing the matrix elements of 
$e^{i \kappa_k a_k}, e^{i \lambda_k a_k^\dagger}, e^{i \mu_i \Phi_i}, e^{i \nu_i C_i'}$. 
The matrix elements of $e^{i \kappa_k a_k}, e^{i \lambda_k a_k^\dagger}$ can be found straightforwardly using (\ref{akmat}). As for $e^{i \mu_i \Phi_i}$,
one can show, using the commutation relations (\ref{Phicom}), that $e^{i \Phi_i}$ acts as
\begin{equation}
e^{\pm i \Phi_i} |\v{\alpha},\v{q},\v{n}\> = |\v{\alpha},\v{q} \pm \v{e_{i-2I}},\v{n}\>
\label{Phimat}
\end{equation}
for $i=2I+1,...,M$. Thus, $e^{\pm i \Phi_i}$ act as raising and lowering operators on the $\v{q}$ quantum numbers. 

Finally, we need to discuss the matrix elements of $e^{i \nu_i C_i'}$. There are three cases, each of which needs to be treated differently: 
(1) $1 \leq i \leq I$, (2) $I\textcolor{black}{+1} \leq i \leq 2I$, and (3) $2I+1 \leq i \leq M$. For the first case, Eq. \ref{geneigdef} implies that
\begin{equation}
e^{\pm i C_i'/d_{i}} |\v{\alpha},\v{q},\v{n}\> = e^{\pm 2\pi i\alpha_{i}/d_i} |\v{\alpha},\v{q},\v{n}\>, \ i = 1,...,I 
\label{eciprimemat1}
\end{equation}
For the second case, one can show using the commutation relations $[C_i', C_j'] = 2\pi i \mathcal{Z}_{ij}'$ that
\begin{equation}
e^{\pm i C_i'/d_{i-I}} |\v{\alpha},\v{q},\v{n}\> = |\v{\alpha} \pm \v{e_{i-I}},\v{q},\v{n}\>, \ i = I+1,...,2I 
\label{eciprimemat2}
\end{equation}
where the addition is performed modulo $d_i$. Note that equations (\ref{eciprimemat1}-{\ref{eciprimemat2}) imply that the
operators $e^{\pm i C_i'/d_{i}}$ act like ``clock'' matrices for $i=1,...,I$, while the operators $e^{\pm i C_i'/d_{i-I}}$ act like ``shift'' matrices
for $i=I+1,...,2I$; thus these operators generate a generalized Pauli algebra. Finally, for the third case, the matrix elements of $e^{i \nu_i C_i'}$ can be obtained 
by exponentiating Eq. \ref{ciprimemat}. 

The above equations tell us everything we need to compute $\<\v{\alpha}',\v{q}',\v{n}'|e^{i\mathcal{A}} |\v{\alpha},\v{q},\v{n}\>$ 
for the special case where the coefficient $\mu_i$ in (\ref{Aexp2}) is an integer for every $i$, 
and $\nu_i$ is an integer multiple of $1/d_i$ for $i=1,...,I$, and a multiple of $1/d_{i-I}$ for $i=I+1,...,2I$. To complete the story
we need to explain how to evaluate matrix elements if the $\mu_i$ and $\nu_i$ coefficients \emph{aren't} quantized in this manner. 
To understand this case, suppose that
$\mu_i$ is not an integer. It then follows that the commutator $[\textcolor{black}{\mathcal{A}},C_i']$ is not an integer multiple 
of $2\pi i$, which implies that the state $e^{i\mathcal{A}} |\v{\alpha},\v{q},\v{n}\>$ is an eigenstate of $C_i'$ with
non-integer eigenvalue. But this means that  $e^{i\mathcal{A}} |\v{\alpha},\v{q},\v{n}\>$ is
orthogonal to the low energy Hilbert space, $\mathcal{H}_{\text{eff}}$ so that all the matrix elements 
$\<\v{\alpha}',\v{q}',\v{n}'| e^{i\mathcal{A}} |\v{\alpha},\v{q},\v{n}\>$ vanish identically.
Similar reasoning shows that these matrix elements also vanish identically 
if the $\nu_i$ coefficients aren't quantized as above. Hence, the above special case is actually the only case where the matrix
elements are nonzero.

\subsection{Low energy operators}
We now turn to the question of how to describe the most general operators in the low energy theory --- that is, 
the most general operators that act within the low energy Hilbert space $\mathcal{H}_{\text{eff}}$. The simplest 
way to do this is to write the operators as linear combinations of outer products of the form
$|\v{\alpha}',\v{q}',\v{n}'\> \<\v{\alpha},\v{q},\v{n}|$. This approach is straightforward, but it can be unwieldy since many 
operators look complicated in this representation.

Alternatively, we can represent any operator $\mathcal{O}$ in the low energy theory as
an infinite power series in the operators 
\begin{align*}
&\{a_1,...,a_K, a_1^\dagger,...,a_K^\dagger\}, \\ 
&\{ C_{2I+1}',...,C_{M}',e^{\pm i\Phi_{2I+1}},...,e^{\pm i\Phi_{M}}\}, \\ 
&\{e^{\pm iC_1'/d_1},...,e^{\pm iC_I'/d_I}, e^{\pm iC_{I+1}'/d_1},...,e^{\pm iC_{2I}'/d_I}\}
\end{align*}
That is,
\begin{equation}
\mathcal{O} = f(\{a_k,a_k^\dagger,C_{2I+i}',e^{\pm i\Phi_{2I+i}}, e^{\pm iC_i'/d_i},e^{\pm iC_{I+i}'/d_i}\})
\label{loweop1}
\end{equation}
for some function $f$ that can be expanded as a power series. Indeed, to prove this result, we need
to establish two facts: (1) we need to show that the above operator $\mathcal{O}$ maps 
$\mathcal{H}_{\text{eff}} \rightarrow \mathcal{H}_{\text{eff}}$, 
and (2) we need to show that the functional form for $\mathcal{O}$ is sufficiently general that it can
reproduce \emph{any} linear map from $\mathcal{H}_{\text{eff}} \rightarrow \mathcal{H}_{\text{eff}}$. The first fact is easy to prove:
we can see that all of the operators 
$\{a_k, a_k^\dagger, C_{2I+i}', ,e^{\pm i\Phi_{2I+i}}, e^{\pm iC_i'/d_i},e^{\pm iC_{I+i}'/d_i}\}$
commute with $\{\cos(C_1),...,\cos(C_M)\}$ and therefore $\mathcal{O}$ also has this property. 
The second fact is harder to prove, but can be established straightforwardly by examining
the matrix elements for $\{a_k,a_k^\dagger,C_{2I+i}',e^{\pm i\Phi_{2I+i}},e^{\pm iC_i'/d_i},e^{\pm iC_{I+i}'/d_i}\}$ given in
equations (\ref{akmat}, \ref{ciprimemat},\ref{Phimat},\ref{eciprimemat1},\ref{eciprimemat2}).

To get more intuition about the above representation, we note that the operators 
$\{a_k, a_k^\dagger, e^{i\Phi_{2I+i}}, C_{2I+i}', e^{iC_{i}'/d_i},\textcolor{black}{e^{iC_{I+i}'/d_i}}\}$
have a natural interpretation in terms of the phase space of the low energy theory: the $\{a_k, a_k^\dagger\}$ operators are conjugate
variables that describe \emph{real} degrees of freedom at low energies; the $\{e^{i\Phi_{2I+i}}\}$ operators describe \emph{angular-valued} 
degrees of freedom and the $\{C_{2I+i}'\}$ describe the corresponding conjugate \emph{discrete} degrees of freedom; finally, the 
$\{e^{iC_{i}'/d_i},\textcolor{black}{e^{iC_{I+i}'/d_i}}\}$ operators
are conjugate variables that describe the \emph{finite discrete} degrees of freedom. The latter operators can be thought of as generalized
Pauli operators, similar to $\sigma^z, \sigma^x$. 

In fact, there is yet another way to parameterize the low energy operators which is often more convenient to use: every low 
energy operator $\mathcal{O}$ can be
written as an infinite power series in $\{a_k,a_k^\dagger\}$, $\{C_{2I+1}',...,C_{M}'\}$ and 
$\{e^{\pm i\Pi_1},...,e^{ \pm i\Pi_M}\}$. That is,
\begin{equation}
\mathcal{O} = f(\{a_k,a_k^\dagger,C_{2I+i}', e^{\pm i \Pi_i}\})
\label{loweop2}
\end{equation}
To derive this parameterization, it suffices to show that we can express the operators $\{e^{\pm i\Phi_i}\}$, 
$\{e^{\pm iC_i'/d_i}\}$ and $\textcolor{black}{\{e^{\pm i C_{I+i}'/d_i}\}}$ in this fashion; once we establish this
property then this parameterization follows immediately from the previous one (\ref{loweop1}). To derive the latter property, we use the definition of
$\Phi_i$ (\ref{Phidef}) which states that $\Phi_i$ is a linear combination of the $\Pi_j, a_k, a_k^\dagger$ operators.
Importantly, if we examine this expansion, we can see that the coefficients of the $\Pi_j$'s are \emph{integers}. Therefore if we 
exponentiate (\ref{Phidef}), we immediately see that $e^{\pm i \Phi_i}$ can be written as a monomial in $\{e^{\pm i\Pi_1},...,e^{ \pm i\Pi_M}\}$
multiplying a power series in $\{a_k,a_k^\dagger\}$. 
Identical reasoning shows that \textcolor{black}{$e^{\pm iC_i'/d_i}$ and $e^{\pm i C_{I+i}'/d_i}$} can also be written in this way; this 
completes the proof.

\subsection{Ground state operators}
So far we have focused on operators acting within the low energy Hilbert space $\mathcal{H}_{\text{eff}}$. However, in some cases we
may be interested in even \emph{lower} energy scales in which we will need to think about the $D$-dimensional subspace spanned 
by the set of degenerate ground states. In our notation, these ground states take the form $|\v{\alpha},\v{q},\v{n}\>$ 
with $\v{n} = \v{q} = 0$. We will label them by $|\v{\alpha}\> \equiv |\v{\alpha},0,0\>$. 

Once we consider the ground state subspace, we face a similar question: how do we describe the operators that act within this
subspace? The simplest way to do this is to write the operators as linear combinations of outer products, 
$|\v{\alpha}'\>\<\v{\alpha}|$ but this representation is not always the most convenient one.

Another way to represent the operators $\mathcal{O}$ that act within the ground state subspace is to write them as polynomials in
\begin{align*}
\{e^{\pm iC_1'/d_1},...,e^{\pm iC_I'/d_I},e^{\pm iC_{I+1}'/d_1},...,e^{\pm iC_{2I}'/d_I}\}. 
\end{align*}
That is,
\begin{equation}
\mathcal{O} = f(\{e^{\pm i C_i'/d_i},e^{\pm i C_{I+i}'/d_i}\})
\label{gsop1}
\end{equation}
for some polynomial function $f$. Indeed, to prove this result, we need to establish two facts: (1) we need to show that the 
above operator $\mathcal{O}$ maps the ground state subspace to itself, and (2)
we need to show that the functional form for $\mathcal{O}$ is sufficiently general that it can
reproduce \emph{any} linear map from the ground state subspace to itself. To prove the first fact we note that the operators
$\{e^{\pm i C_i'/d_i},e^{\pm i C_{I+i}'/d_i}\}$ commute with $\{\cos(C_1),...,\cos(C_M), H_{\text{eff}}\}$ and therefore $\mathcal{O}$ also has this property.
The second fact is harder to prove but can be established straightforwardly using
the matrix elements for $\{e^{\pm i C_i'/d_i},e^{\pm i C_{I+i}'/d_i}\}$ given in Eqs. \ref{eciprimemat1},\ref{eciprimemat2}.

There is also a third way to parameterize the ground state subspace operators which 
applies to the case where the commutator matrix $\mathcal{Z}_{ij} = \frac{1}{2\pi i} [C_i, C_j]$ is 
non-degenerate. This alternative parameterization involves the operators $\Gamma_1,...,\Gamma_M$, defined in Eq. \ref{Gammadef}.
More specifically, this parameterization states that general ground state operators $\mathcal{O}$ can be 
written as polynomials in $\{e^{\pm i\Gamma_1},...,e^{ \pm i\Gamma_M}\}$. That is,
\begin{equation}
\mathcal{O} = f(\{e^{\pm i \Gamma_i}\})
\label{gsop2}
\end{equation}
for some polynomial $f$.
 To derive this result, it suffices to show that we can write $e^{\pm i C_i'/d_i}$
and $e^{\pm i C_{i+I}'/d_i}$ as such polynomials; once we establish this fact, the parameterization follows immediately
from the previous one (\ref{gsop1}). To prove the latter property, we note that the operators $e^{\pm i C_i'/d_i}$
and $e^{\pm i C_{i+I}'/d_i}$ can be equivalently written as $\exp(\pm i \sum_{j=1}^{M} (\mathcal{Z}'^{-1})_{ji} C_j')$. We then
use the following identity, which can be derived easily from the definition of $C_j', \Gamma_j$:
\begin{align*}
\sum_{j=1}^{M} (\mathcal{Z}'^{-1})_{ji} C_j' =  \sum_{j=1}^{M}  (\mathcal{V}^{-1})_{ji} \Gamma_j \textcolor{black}{+ (\mathcal{Z}'^{-1})_{ji} \chi_j}
\end{align*}
The important point about this identity is that $(\mathcal{V}^{-1})_{ji}$ is an integer matrix, so we deduce that the
left hand side can be written as a linear combination of $\Gamma_j$'s with \emph{integer} coefficients. It follows
that $\exp(\pm i \sum_{j=1}^{M} (\mathcal{Z}'^{-1})_{ji} C_j')$ can be written as a monomial in 
$e^{\pm i \Gamma_1},...,e^{\pm i \Gamma_M}$,
which is of course a special kind of polynomial. This completes the proof.

\section{Properties of creation and annihilation operators}
\label{propcreapp}
In this appendix, we derive three properties of creation/annihilation operators that were used in appendix \ref{diagcreatsect}.
These properties are guaranteed hold as long as we assume that there is no operator $R$ which is a linear combination 
of $\{x_1,...,x_N, p_1,...,p_N\}$, is linearly independent from $\{C_1,...,C_M\}$, and satisfies 
\begin{equation}
[R,C_1] = ... = [R, C_M] = [R, H_{\text{eff}}] = 0
\label{assumapp}
\end{equation}
(See appendix \ref{diagcreatsect} for more discussion about this assumption).

The first property that we will derive is that the number $K$ of linearly independent annihilation operators is exactly 
\begin{equation}
K = N-M+I
\label{numaapp}
\end{equation}
We prove this result by establishing two opposing inequalities: $K \leq N-M+I$ and $K \geq N-M+I$. 
We start by showing the inequality $K \leq N-M+I$.

To begin, consider the set of all operators that are linear combinations 
of $\{x_1,...,x_N, p_1,..., p_N\}$ and that commute with $\{C_1,...,C_M\}$. This set forms a vector space
since it is closed under addition and scalar multiplication. We will call this vector space $V$. 

Next, observe that if $A$ is an operator that belongs to $V$, then the commutator $[A, H_{\text{eff}}]$ 
also belongs to $V$: to see this, it suffices to show that $[C_i, A] = 0$ implies $[C_i, [A, H_{\text{eff}}]] = 0$.
The latter result is a simple application of the Jacobi identity: 
\begin{equation*}
[C_i, [A, H_{\text{eff}}]] = -[A,[H_{\text{eff}},C_i]]-[H_{\text{eff}},[C_i,A]] = 0
\end{equation*}
where the second equality follows from the fact that $[H_{\text{eff}},C_i] = 0$. In view of this property, 
we can think of the commutator map $A \rightarrow [A, H_{\text{eff}}]$ as defining 
a linear mapping from $V \rightarrow V$. We will denote this linear mapping by $\mathcal{S}$. 

In this language, the annihilation operators $a$ correspond to eigenvectors of $\mathcal{S}$ with 
positive eigenvalue. Our task is thus to bound the number of these eigenvectors. To this end, we note 
that the total dimension of $V$ is $2N-M$. At the same time, we know that the 
operators $C_{2I+1}',...,C_M'$ belong to the space $V$ and are eigenvectors of $\mathcal{S}$ with eigenvalue $0$. 
It follows by dimension counting that $\mathcal{S}$ has at most $(2N-M)-(M-2I) = 2N - 2M + 2I$ linearly
independent eigenvectors with nonzero eigenvalues. 

To complete the derivation of the first inequality, we note that the eigenvectors of $\mathcal{S}$ with 
nonzero eigenvalues come in $\pm E$ pairs: if $[A, H_{\text{eff}}] = E \cdot A$ then $[A^\dagger, H_{\text{eff}}] = - E \cdot A^\dagger$. 
It follows that $\mathcal{S}$ has at most $N-M+I$ linearly independent eigenvectors with positive eigenvalues. In other words, 
$K \leq N-M+I$.

 
Now we prove the second inequality, i.e. $K \geq N-M+I$. The first step is to observe that physical
considerations guarantee that $H_{\text{eff}}$ is positive 
semi-definite. It follows that 
we can write $H_{\text{eff}}$ as
\begin{equation}
H_{\text{eff}} = \sum_{i=1}^L \frac{B_i^2}{2},
\end{equation}
where each $B_i$ is a real linear combination of $\{x_1,...,x_N, p_1,..., p_N\}$, and where each $B_i$ is linearly 
independent from the others. Here $L$ is some integer with $L \leq N$. 

Next, we observe that the $B_i$ operators have the property that $[B_i, C_j] = 0$ for
all $i,j$: one way to derive this fact is to expand the commutator $[C_j, H_{\text{eff}}]$ as 
$\sum_{i=1}^L [C_j, B_i] B_i$, and then use the fact that the $B_i$ operators are linearly independent 
as well as the fact that $[C_j, H_{\text{eff}}] = 0$.

To proceed further, consider the $L \times L$ matrix $\mathcal{Y}_{ij} = [B_j, B_i]$. This matrix is 
imaginary and skew-symmetric, \textcolor{black}{which implies} that it is \emph{diagonalizable}. Thus, we can find $L$ 
linearly independent eigenvectors $v_1,..., v_L$ with eigenvalues $E_1,..., E_L$. The corresponding operators 
$a_i \equiv \sum_{j=1}^L v_i^j B_j$ obey $[a_i, H_{\text{eff}}] = E_i a_i$. Furthermore, we know that 
$[a_i, C_j] = 0$ since $[B_i, C_j] = 0$ for all $i,j$.

Clearly the $a_i$ operators obey almost all of the conditions for annihilation operators. All we 
have to show is that we can find a subset of $N-M+I$ linearly independent $a_i$'s with eigenvalues $E_i > 0$. To this end, 
consider the set of all operators that are linear combinations of $\{x_1,...,x_N,p_1,...,p_N\}$, and 
commute with all the $C_i$ and $a_i$ operators. This set is a vector space since it is 
closed under addition and scalar multiplication. We will call this vector space $W$. Let us try to find the dimension of $W$.
To do this, note that any operator in $W$ commutes with all the $B_i$ operators, since the $B_i$ operators
can be written as linear combinations of the $a_i$'s. It then follows that any operator in $W$ must
also commute with $H_{\text{eff}}$. But by our assumption (\ref{assumapp}), the only operators that commute with both $H_{\text{eff}}$
and the $C_i$ operators can be written as linear combinations of $C_i$, or equivalently, linear combinations of 
$\{C_{2I+1}',..., C_M'\}$. We conclude that the dimension of $W$ is at most $M-2I$.

Given that the dimension of $W$ is at most $M-2I$, it follows that there must be at least $2N-M+2I$ linearly independent 
$a_i, C_i$ operators. Hence, at least $2N-2M+2I$ of the $a_i$ operators are linearly independent from the $C_i$ 
operators. It then follows from the assumption (\ref{assumapp}) that at least $2N-2M+2I$ of the $a_i$'s have $E_i \neq 0$.
At the same time, we know that the $E_i$ eigenvalues come in pairs of opposite sign since $\mathcal{Y}_{ij}$ is skew-symmetric and imaginary, 
so we conclude that at least $N-M+I$ of the $a_i$'s have positive eigenvalue. This establishes that $K \geq N-M+I$ and completes
the proof of Eq. \ref{numaapp}.

We now prove the second property of the creation and annihilation operators. This property states that the $a_k$'s can always be 
chosen so that
\begin{align}
[a_k, a_{k'}^\dagger] =\delta_{kk'}, \quad [a_k, a_{k'}] = [a_k^\dagger, a_{k'}^\dagger] = 0 
\label{aicommapp}
\end{align}
We begin with the first relation. To prove this, we observe that $[a_k, a_{k'}^\dagger] = 0$ unless $E_k = E_k'$.
Indeed, this follows from the Jacobi identity:
\begin{equation*}
[a_{k},[a_{k'}^\dagger,H_{\text{eff}}]] = -[a_{k'}^\dagger,[H_{\text{eff}}, a_{k}]]- [H_{\text{eff}},[a_{k},a_{k'}^\dagger]]
\end{equation*}
Now fix a particular $E_k$. There are two cases to consider: $E_k$ may be degenerate or non-degenerate. If $E_k$ is 
non-degenerate then we can simply normalize the corresponding
$a_k$ such that $[a_k, a_k^\dagger] = 1$; the vanishing of the commutators between $a_k$ and other $a_{k'}^\dagger$'s is 
guaranteed. On the other hand if $E_k$ has some degeneracy, then we can use the Gram-Schmidt 
procedure to choose the associated $a_k$'s so that they obey $[a_k, a_{k'}^\dagger] = \delta_{kk'}$.
Again, all the other commutators vanish automatically. Thus, in all cases we can choose the $a_k$ operators
to satisfy the first relation. As for the second relation, this follows from the Jacobi identity by similar reasoning.

Finally, we move on to prove the third property of the creation and annihilation operators. This property states that 
each of the $\{x_1,...,x_N, p_1,...,p_N\}$ operators can be written as a linear combination of 
$\{a_1,...,a_{K},a_1^\dagger,...,a_{K}^\dagger,C_{2I+1}',...,C_M',\Pi_1,..,\Pi_M\}$.
To prove this, we only need to show that the operators in this set are
\emph{linearly independent} since (1) there are $2N$ of them all together according to Eq. \ref{numaapp}, 
and (2) they are all linear combinations of $\{x_1,...,x_N,p_1,...,p_N\}$. 
The fact that they are linearly independent can be seen as follows: suppose that
\begin{equation*}
\sum_{k=1}^{K} (\kappa_k a_k + \lambda_k a_k^\dagger) + \sum_{i=2I+1}^M \mu_i C_i' + \sum_{i=1}^M \nu_i \Pi_i = 0
\end{equation*}
for some scalars $\kappa_k, \lambda_k, \mu_i, \nu_i$. Then, if we take the commutator of both sides with $C_j$, we immediately see that $\nu_j =0$
since $[a_k, C_j] = [a_k^\dagger, C_j] = [C_i',C_j] = 0$ while $[\Pi_i, C_j] = \delta_{ij}$. We therefore have
\begin{equation*}
\sum_{k=1}^{K} (\kappa_k a_k + \lambda_k a_k^\dagger) + \sum_{i=2I+1}^M \mu_i C_i' = 0
\end{equation*}
Next, we take the commutator of both sides with $a_k$ and we deduce that $\lambda_k = 0$ since
$[a_k, C_i'] = 0$ while $[a_k, a_{k'}^\dagger] = \delta_{kk'}$. Similarly, taking the commutator with $a_k^\dagger$ shows that
$\kappa_k = 0$. Our relation now becomes
\begin{equation*}
\sum_{i=2I+1}^M \mu_i C_i' = 0
\end{equation*}
Finally we note that the $C_i'$ are all linearly independent by construction, so all the $\mu_i$'s must vanish. 
Hence, all the coefficients in our original
expansion must vanish, which implies that the above operators are linearly independent, as claimed.

\section{Uniqueness of simultaneous eigenstates}
\label{uniqueapp}
In this appendix, we show that there is exactly one simultaneous eigenstate $|\v{\theta}, \v{\varphi},\v{x}',\v{n}\>$ 
for each choice of $\v{\theta}, \v{\varphi}, \v{x}', \v{n}$. Here $\v{\theta}, \v{\phi}$ are $I$ component angular-valued vectors,
while $\v{n}$ is a $K$ component non-negative integer vector, and $\v{x}'$ is a $M-2I$ component real valued vector.

To see that there is \emph{at least} one eigenstate for each choice of these quantum numbers, we observe that the explicit formula in 
Eq. \ref{thetaphixdef} implies the weaker 
result that there is at least one state for each choice of $\v{\theta},\v{\varphi}, \v{x}'$. Now consider the subspace of states
with fixed values of $\v{\theta},\v{\varphi}, \v{x}'$. It is clear that the creation and annihilation operators $a_k^\dagger, a_k$ map this 
subspace onto itself. Therefore, using the same arguments as in the algebraic analysis of the harmonic oscillators, it is easy to see that
this subspace contains at least one state for each choice of occupation number $\v{n}$.

Conversely, to see that there is \emph{at most} one state  $|\v{\theta}, \v{\varphi},\v{x}',\v{n}\>$ 
for each choice of $\v{\theta}, \v{\varphi}, \v{x}', \v{n}$, 
we recall that all the $x_i, p_i$ operators can be expressed as a linear combination of 
 $\{a_1,...,a_{K},a_1^\dagger,...,a_{K}^\dagger,C_{2I+1}',...,C_M',\Pi_1,..,\Pi_M\}$. One application of
these expressions is that we can use them to compute the expectation value of any operator $\mathcal{O}_{\kappa, \lambda}$ of the form 
$\mathcal{O}_{\v{\kappa},\v{\lambda}} = \exp(i \sum_i (\kappa_i x_i + \lambda_i p_i))$ in any state with quantum numbers
$\v{\theta}, \v{\varphi},\v{x}',\v{n}$. This computation is completely algebraic and depends only on the quantum numbers
$\v{\theta}, \v{\varphi},\v{x}',\v{n}$ as well as the parameters $\v{\kappa}, \v{\lambda}$. Thus we conclude that the 
quantum numbers $\v{\theta}, \v{\varphi},\v{x}',\v{n}$ completely fix 
the expectation values of the $\mathcal{O}_{\v{\kappa},\v{\lambda}}$ operators. But the above operators 
$\mathcal{O}_{\v{\kappa},\v{\lambda}}$ are sufficiently 
general that any operator $f(x_i,p_i)$ can be constructed by taking an appropriate linear combination of them. Hence, if two states share the same 
quantum numbers  $\v{\theta}, \v{\varphi},\v{x}',\v{n}$, then they must have the same expectation values with respect to \emph{all} operators
in the Hilbert space and hence must be equivalent to one another up to a phase. 

\section{Regularizing the cosine term}
\label{regapp}
In this appendix, we \textcolor{black}{revisit} the problem of a fractional quantum spin Hall edge with a single magnetic impurity:
\begin{equation}
H = H_0 -U \cos(C)
\end{equation}
The new element in our discussion is that we \emph{regularize} the argument of the cosine term, replacing $C = k (\phi_\up(0) + \phi_\down(0))$ with
\begin{equation*}
C = \int_{-L/2}^{L/2} k ( \phi_\up(x) + \phi_\down(x)) \tilde{\delta}(x) dx
\end{equation*}
where $\tilde{\delta}$ is an approximation to a delta function --- i.e. a narrowly peaked function with $\int \tilde{\delta}(x) dx =1$. 
After making this replacement, we repeat the analysis in section \ref{fmsisect} in which we constructed creation and annihilation operators for the low
energy effective Hamiltonian $H_{\text{eff}}$. Our main result is that we find that when the cosine term is properly regularized, the condition
$[a,C] = 0$ translates to the constraint in Eq. \ref{constraintAB2}; otherwise the regularization doesn't change much.

As in section \ref{fmsisect}, our task is to find all $a$ operators such that (1) $a$ is a linear combination of 
our fundamental phase space operators $\{\partial_y \phi_\up, \partial_y \phi_\down, \phi_\up(y_0), \phi_\down(y_0)\}$ and (2) $a$ obeys
\begin{align}
[a, H_{0}] &= E a + \lambda [C, H_0] + \lambda_\up [Q_\up, H_0] + \lambda_\down [Q_\down, H_0] \nonumber \\
[a, C] &= [a,Q_\up] = [a,Q_\down] = 0 
\label{auxeqsiapp}
\end{align}
for some scalars $E, \lambda, \lambda_\up, \lambda_\down$ with $E \neq 0$.

Given that $[a, Q_\up] = [a, Q_\down] = 0$, we deduce that $\phi_\up(y_0), \phi_\down(y_0)$ cannot appear in the expression 
for $a$. Hence, $a$ can be written in the general form
\begin{equation}
a=\int_{-L/2}^{L/2} [f_{\up}(y)\partial_{y}\phi_{\up}(y)+f_{\down}(y)\partial_{y}\phi_{\down}(y)] dy
\la{genmodeapp}
\end{equation}
Next, from the first line of Eq.~\ref{auxeqsiapp} we obtain the following set of differential equations,
\begin{align*}
-ivf_{\up}^{'}(y) & =Ef_{\up}(y)+\lambda kiv\tilde{\delta}(y)\\
ivf_{\down}^{'}(y) & = Ef_{\down}(y)-\lambda kiv\tilde{\delta}(y)
\end{align*}
(The $\lambda_\up, \lambda_\down$ terms drop out of these equations since $Q_\up, Q_\down$ commute with $H_0$).
Solving these equations, we get
\begin{align}
f_{\up}(y) & = e^{ipy}[A_{1}(1-\tilde{\Theta}_\up(y))+A_{2}\tilde{\Theta}_\up(y)]\\
f_{\down}(y) & = e^{-ipy}[B_{1}(1-\tilde{\Theta}_\down(y))+B_{2}\tilde{\Theta}_\down(y)]
\la{solsiapp}
\end{align} 
where $p = E/v$ and
\begin{equation}
A_{2}=A_{1}-\lambda k, B_{2}=B_{1}-\lambda k
\label{bcsiapp}
\end{equation}
and
\begin{eqnarray*}
\tilde{\Theta}_\up(y) &=& \int_{-L/2}^y e^{-ipx} \tilde{\delta}(x) dx \nonumber \\
\tilde{\Theta}_\down(y) &=& \int_{-L/2}^y e^{ipx} \tilde{\delta}(x) dx 
\end{eqnarray*}
Note that both $\tilde{\Theta}_{\up}(x),\tilde{\Theta}_\down(x)$ reduce to the Heaviside step function $\Theta(x)$ in the limit
that $\tilde{\delta}(x)$ is a delta function. 
Eliminating $\lambda$ from (\ref{bcsiapp}) we see that
\begin{equation}
A_2 - A_1 = B_2 - B_1
\label{constraintAB1app}
\end{equation}

So far the regularization hasn't taught us anything new: all of our results are similar to what we found in equations (\ref{solsi},\ref{bcsi},\ref{constraintAB1}), when
we analyzed the unregularized cosine term with $C = k(\phi_\up(0)+\phi_\down(0))$. So why is the regularization
important? The reason is that it allows us to properly treat the constraint $[a,C] = 0$, as we now demonstrate. The 
first step is to use (\ref{genmodeapp}) to translate the constraint to 
\begin{eqnarray*}
\int_{-L/2}^{L/2} [f_{\up}(y)-f_{\down}(y)] \tilde {\delta}(y) dy =0
\end{eqnarray*}
Next, substituting our expressions for $f_\up, f_\down$ (\ref{solsiapp}), we derive
\begin{align}
&\int_{-L/2}^{L/2} (A_1 e^{ipy} - B_1 e^{-ipy}) \tilde{\delta}(y) dy \nonumber \\ 
&+ \int_{-L/2}^{L/2} ((A_2-A_1) e^{ipy} \tilde{\Theta}_\up(y))\tilde{\delta}(y) dy  \nonumber \\
&-\int_{-L/2}^{L/2}(B_2-B_1) e^{-ipy} \tilde{\Theta}_\down(y)) \tilde{\delta}(y) dy = 0
\label{consint}
\end{align}
While the constraint (\ref{consint}) looks complicated, it simplifies considerably in the low energy, long wavelength limit, 
i.e. the limit where $p b \ll 1$ where $b$ is the width of the function $\tilde{\delta}$. In this limit, 
the first integral in (\ref{consint}) evaluates to $(A_1-B_1)$, but the second integral is a bit trickier. 
To compute this integral, we use the identity
\begin{equation}
\lim_{pb  \rightarrow 0} \int_{-L/2}^{L/2} \tilde{\delta}(y) \tilde{\Theta}_\sigma(y) e^{\pm ipy} = \frac{1}{2} 
\label{thetadelta}
\end{equation}
where $\sigma = \up,\down$. This identity can proved for the case $\sigma =\up$ by noting that
\begin{align*}
&\lim_{pb  \rightarrow 0} \int_{-L/2}^{L/2} \tilde{\delta}(y) \tilde{\Theta}_{\up}(y) e^{\pm ipy} dy \\
&=\lim_{pb  \rightarrow 0} \int_{-L/2}^{L/2} \int_{-L/2}^y \tilde{\delta}(y) \tilde{\delta}(x) e^{- ipx \pm ipy} 
dx dy \\ 
&=\lim_{pb  \rightarrow 0} \int_{-L/2}^{L/2} \int_{-L/2}^y \tilde{\delta}(y) \tilde{\delta}(x) dx dy \\ 
&=\lim_{pb  \rightarrow 0} \frac{1}{2} \int_{-L/2}^{L/2} \int_{-L/2}^{L/2} \tilde{\delta}(y) 
\tilde{\delta}(x) dx dy = \frac{1}{2}
\end{align*}
The proof for the case $\sigma=\down$ is similar. 

Applying the above identity (\ref{thetadelta}), the second integral in (\ref{consint}) evaluates to $\frac{1}{2}[(A_2-A_1)-(B_2-B_1)]$ 
so that (\ref{consint}) becomes 
\begin{equation*}
(A_1 -B_1) + \frac{(A_2-A_1) - (B_2-B_1)}{2} = 0
\end{equation*}
Simplifying, we obtain
\begin{equation}
\frac{A_1+A_2}{2} = \frac{B_1+B_2}{2}
\label{constraintAB2app}
\end{equation}
This is precisely the constraint from Eq. \ref{constraintAB2}. 

\section{Degeneracy from spontaneously broken time-reversal symmetry}
\label{ssb}

In this appendix, we consider a FQSH edge in a disk geometry 
with $N$ \emph{time-reversal invariant} impurities located at positions $x_1,...,x_N$. 
We show that when $U \rightarrow \infty$, the ground state is two-fold degenerate,
which is consistent with spontaneous time-reversal symmetry breaking.

Because we consider time reversal-invariant impurities, the dominant scattering process in this system involves \emph{two-particle}
backscattering. Thus, the Hamiltonian takes the form
\begin{align}
&H=H_0-U\sum_{i=1}^{N}\cos(C_i) \\
&C_{i}= 2 k (\phi_{\up}(x_i)+\phi_{\down}(x_i))
\la{fmmitr}
\end{align}
where $H_0$ is defined in Eq. \ref{Hclean}. Notice the factor of $2$ in the definition of $C_i$.

We can see that this system is identical to the magnetic impurity system studied in section \ref{fmmisect} except for the above factor of $2$. As a result, almost all of the results derived in section \ref{fmmisect} carry over to this case without change. In fact, the only difference comes when we compute the commutator matrix $\mathcal{Z}_{ij} = \frac{1}{2\pi i} [C_i, C_j]$. For the above case,
$\mathcal{Z}_{ij}$ takes the form
\begin{eqnarray*}
\mathcal{Z}_{ij}
&=& \bpm 0 & \cdots & 0 & 2 & -2 \\ \vdots & \vdots & \vdots & \vdots & \vdots \\ 0 & \cdots & 0 & 2 & -2  \\
-2 & \cdots & -2 & 0 & 0 \\ 2 & \cdots & 2 & 0 & 0 \epm
\end{eqnarray*}

In order to transform the commutator matrix into canonical form, we make the change of variables
\begin{align*}
&C_1' = C_1, \quad C_2' = -2\pi Q_\up, \quad C_3' = 2\pi Q_\up + 2\pi Q_\down \\
&C_m' = C_{m-2} - C_{m-3}, \quad m= 4,...,N+2 
\end{align*}
In this basis we obtain
\begin{eqnarray*}
\mathcal{Z}_{ij}' &=& \frac{1}{2\pi i} [C_i', C_j'] \nonumber \\
&=& \bpm 0 & -2 & 0 & \cdots & 0 \\ 2 & 0 & 0 & \cdots & 0 \\ 0 & 0 & 0 & \cdots & 0 \\ \vdots & \vdots & \vdots & \vdots & \vdots \\
0 & 0 & 0 & \cdots & 0 \epm
\end{eqnarray*}
We can see that $\mathcal{Z}'_{ij}$ is in the canonical skew-normal form shown in Eq. \ref{zprime}, with the parameters $M = N+2$, $I = 1$, $d_1 = 2$.

If we now compute the degeneracy using Eq.~\ref{Deg}, we obtain $D = d_1 = 2$. We conclude that the ground state is two-fold degenerate (in fact, every state shares this degeneracy). This degeneracy makes perfect sense physically since we expect that the edge will exhibit spontaneous time-reversal
symmetry breaking in the limit of large $U$~\cite{XuMoore,WuBernevigZhang,LevinStern}, and such symmetry breaking naturally leads to a two-fold degeneracy.

\bibliography{quadHamarxiv4}
\end{document}